\documentclass[]{article}
\usepackage{graphicx}
\usepackage{epstopdf}
\usepackage{cite}
\usepackage{amssymb}
\usepackage{amsthm}
\usepackage{mathtools}
\usepackage[font=small]{caption}
\usepackage{subcaption}
\usepackage{amsmath}
\usepackage{amsfonts}
\usepackage{tikz}
\usepackage{float}
\usepackage{relsize}
\usepackage[hmarginratio=2:3]{geometry}
\usetikzlibrary{decorations.pathmorphing}
\usepackage{pifont}
\newtheorem{prop}{Proposition} 
\usepackage[titletoc,title]{appendix} 
\usepackage{authblk}
\title{Stokes Phenomena in Discrete Painlev\'{e} II}
\author{N. Joshi, C. J. Lustri and S. Luu\footnote{Electronic address: S.Luu@maths.usyd.edu.au; Corresponding author}}
\date{}
\affil{\textit{School of Mathematics and Statistics, F07, The University of Sydney, New South Wales 2006, Australia}}

\begin{document}

\maketitle

\begin{abstract}
We consider the asymptotic behaviour of the second discrete Painlev\'{e} equation in the limit as the independent variable becomes large. Using asymptotic power series, we find solutions that are asymptotically pole-free within some region of the complex plane. These asymptotic solutions exhibit Stokes phenomena, which is typically invisible to classical power series methods. We subsequently apply exponential asymptotic techniques to investigate such phenomena, and obtain mathematical descriptions of the rapid switching behaviour associated with Stokes curves. Through this analysis, we determine the regions of the complex plane in which the asymptotic approximations are valid, and find that the behaviour of these asymptotic solutions shares a number of features with the \textit{tronqu\'{e}e} and \textit{tri-tronqu\'{e}e} solutions of the second continuous Painlev\'{e} equation.
\end{abstract}

\section{Introduction}\label{S:Introduction}
In this paper we consider the second discrete Painlev\'{e} equation ($\text{dP}_{\text{II}}$) 
\begin{equation}\label{dPII}
x_{n+1}+x_{n-1}=\frac{\left(\alpha  n+\beta\right)x_n+\gamma}{1-x_n^2}
\end{equation}
in the asymptotic limit $n\rightarrow\infty$. It is part of a class of integrable, second order nonlinear difference equations known as the discrete Painlev\'{e} equations that tend to the ordinary Painlev\'{e} equations in the continuum limit. This equation is of interest in mathematical physics \cite{Shohat1939,Forrester2003,Periwal1990,Magnus1995,Magnus1999,VanAssche2015} and also appears as a reduction of the discrete modified Korteweg-de Vries (mKdV) equation \cite{Nijhoff1991, Nijhoff1999}. Equation (\ref{dPII}) tends to the second continuous Painlev\'{e} equation ($\text{P}_{\text{II}}$) in the continuum limit $x_n=\epsilon w$, $z_n=\alpha  n +\beta=2+\epsilon^2t$ and $\gamma=\epsilon^3\mu$ as $\epsilon\rightarrow 0$, 
\begin{equation}\label{PII}
\text{P}_{\text{II}}: \qquad \frac{d^2w}{dt^2}=2w^3+tw+\mu.
\end{equation}
The continuous second Painlev\'{e} equation appears in similar contexts as its discrete version. It is obtainable via reductions of partial differential equations used in fluid dynamics \cite{Fokas1981,Gromak2002,Clarkson2006, Olver2010,Ablowitz1977,Claeys2010} and appears in the Tracy-Widom distribution which describes the limiting distribution of particular eigenvalues of a class of matrices \cite{Clarkson2006,Tracey1993} in the study of combinatorics. It has also been used as a model in electrodynamics \cite{kudray1997,Bass2010,Bass1964} and in mathematical physics \cite{Seib2005,Periwal1990,Tracy1999,Schiappa2014,Forrester2015}. Reviews of both continuous and discrete Painlev\'{e} equations can be found in \cite{Conte1999,Grammaticos2004}.

Another version of \eqref{dPII} exists for which the $\gamma$ term is replaced by $(-1)^n\gamma$ \cite{Grammaticos2004}. This version can be obtained by considering the singularity confinement of the McMillan map
\begin{equation*}
x_{n+1}+x_{n-1}=\frac{a_n x_n+b_n}{1-x_n^2}.
\end{equation*}
In particular, the authors of \cite{Grammaticos2004} show that $b_n=\gamma+\delta(-1)^n$ and $a_n=\alpha n+\beta$. The odd-even dependence can be removed by choosing $\delta=0$, producing (\ref{dPII}).

Motivated by these applications, the asymptotic study of the Painlev\'{e} equations have been pursued. However, there have been few corresponding studies for the discrete versions. Previous asymptotics studies for the first discrete Painlev\'{e} equation have been conducted in \cite{Joshi1997,Veresh1995,Chris2015} where the authors found solutions asymptotically free of poles in the large independent variable limit. However, to the best of our knowledge there appears to be no asymptotic studies for $\text{dP}_{\text{II}}$ in the literature. There is another class of discrete Painlev\'{e} equations known as $q$-difference Painlev\'{e} equations for which asymptotic studies have been considered. In particular, Joshi \cite{Joshi2015} investigated unstable solutions for the first $q$-difference Painlev\'{e} equation. 

Using techniques of exponential asymptotics we will find solutions which are asymptotically free of poles within certain regions of the complex plane. We begin the analysis by finding the formal series solutions containing exponentially-small terms, then study the Stokes phenomena present within these asymptotic solutions and use this to deduce their regions of validity. We will find that these asymptotic solutions share features with the \textit{tronqu\'{e}e} and \textit{tri-tronqu\'{e}e} solutions of the second continuous Painlev\'{e} equation ($\text{P}_{\text{II}}$)\cite{Boutroux1913}. 

Exponential asymptotic techniques for differential-difference equations were developed by King and Chapman \cite{King2001} in order to study a nonlinear model of atomic lattices based on the work of \cite{Olde1995,Chapman1998}. Joshi and Lustri \cite{Chris2015} applied the Stokes-smoothing technique described in \cite{King2001} to the first discrete Painlev\'{e} equation and obtained asymptotic approximations which contain exponentially-small contributions. Motivated by their work, we extend this to $\text{dP}_{\text{II}}$ in order to study asymptotic solutions with similar features.

We note that there have been other exponential asymptotic approaches used to study difference equations \cite{Olde2004,Olver2000,Immink1988}. In particular, Olde Daalhuis \cite{Olde2004} considered a particular class of second-order linear difference equations, and applied Borel summation techniques in order to obtain asymptotic expansions with exponentially-small error. The Stokes-smoothing technique described in \cite{King2001} may be performed directly on the nonlinear difference equation, rather than requiring the solution to be formulated in terms of an integral expression. While this method does not produce the integral expressions and controlled error estimates associated with Borel summation techniques, the direct applicability of the method permits the analysis to be easily extended to a wide range of problems.

\subsection{Background on $\textrm{P}_{\textrm{II}}$}\label{S:Background}
General solutions to the Painlev\'{e} equations are higher, transcendental functions, which cannot be expressed in terms of known functions. Therefore, many authors have studied their asymptotic behaviours or sought special solutions. Many investigations have considered the asymptotic behaviours of the Painlev\'{e} transcendents in the limit as the independent variable goes to infinity \cite{Boutroux1913,Deift1995,Joshi1988,Joshi2002,Clarkson2006,Kitaev1994}. The first known study was conducted by Boutroux \cite{Boutroux1913}, who considered both the first and second Painlev\'{e} equations in the limit $|t|\rightarrow\infty$.

Boutroux \cite{Boutroux1913} was able to show that the asymptotic behaviour of the solutions to \eqref{PII} are valid within certain sectors of the complex plane. This study subsequently identified particular solutions which are asymptotically free of poles, meaning that for a sufficiently large radius, these solutions are pole-free. These sectors in the complex plane have angular width $2\pi/3$, bounded by rays and were characterized by its bisector, or ray of symmetry\footnote{These special rays are examples of the Stokes curves and anti-Stokes curves described in Section \ref{S:Exponential asymptotics and Stokes curves}.}. As such, these special asymptotic solutions are known as the \textit{tronqu\'{e}e} (asymptotically pole free along a bisector) and \textit{tri-tronqu\'{e}e} (asymptotically pole free along three successive bisectors) solutions whose asymptotic behaviour is described by either $y\sim \sqrt{-t/2}$ or $y\sim -\mu/t$ as $|t|\rightarrow\infty$.

Studies on the discrete Painlev\'{e} equations have been guided by their continuous counterparts due to the analogous results between the discrete and continuous Painlev\'{e} equations. However, as mentioned previously, it appears that there has been no asymptotic study of $\text{dP}_{\text{II}}$. Since $\text{dP}_{\text{II}}$ is of interest in mathematical physics, we investigate the asymptotic behaviour of (\ref{dPII}).     

\subsection{Exponential asymptotics and Stokes curves}\label{S:Exponential asymptotics and Stokes curves}
The formal series studied in this paper will be shown to be divergent, indicating the presence of exponentially-small contributions to the solution behaviour. Conventional asymptotic power series methods fail to capture the presence of such terms, and therefore these terms are often described as lying \textit{beyond all orders}. In order to investigate these exponentially-small contributions, exponential asymptotic methods are used. The underlying principle of these methods is that divergent asymptotic series may be truncated so that the divergent tail, known as the remainder term, is exponentially-small in the asymptotic limit \cite{Boyd1999}. This is known as an optimally-truncated asymptotic series. Thereafter, the problem can be rescaled in order to directly study the behaviour of these exponentially-small remainder terms. This idea was introduced by Berry \cite{Berry1988,Berry1989,Berry1991}, and Berry and Howls \cite{Berry1990}, who used these methods to determine the behaviour of special functions such as the Airy function. 

The basis of this study uses techniques of exponential asymptotics developed by Olde Daalhuis et al. \cite{Olde1995} for linear differential equations, extended by Chapman et al. \cite{Chapman1998} for application to nonlinear differential equations, and further developed by King and Chapman \cite{King2001} for nonlinear differential-difference equations. A brief outline of the key steps of the process will be provided here, however more detailed explanation of the methodology may be found in these studies.

In order to optimally truncate an asymptotic series, the general form of the coefficients of the asymptotic series is needed. However, in many cases this is an algebraically intractable problem. Dingle \cite{Dingle1973} investigated singular perturbation problems and noted that the calculation of successive terms of the asymptotic series involves repeated differentiation of the earlier terms. Hence, the late-order terms, $a_m$, of the asymptotic series typically diverge as the ratio between a factorial and an increasing power of some function as $m\rightarrow \infty$. A typical form describing this is given by the expression 
\begin{equation}\label{general late order terms}
a_m\sim \frac{A\,\Gamma(m+\gamma)}{\chi^{m+\gamma}},
\end{equation}
as $m\rightarrow\infty$ where $\Gamma$ is the gamma function defined in \cite{Abram2012}, while $A$, and $\chi$ are functions of the independent variable and do not depend on $m$, known as the prefactor and singulant respectively. The singulant is subject to the condition that it vanishes at the singular points of the leading order behaviour, ensuring that the singularity is present in all higher-order terms. Chapman et al. \cite{Chapman1998} noted this behaviour in their investigations and utilize \eqref{general late order terms} as an ansatz for the late-order terms, which may then be used to optimally truncate the asymptotic expansion.

Following \cite{Olde1995} we substitute the optimally-truncated series back into the governing equation and study the exponentially-small remainder term. When investigating these terms we will discover two important curves known as Stokes and anti-Stokes curves \cite{Bender1999}. Stokes curves are curves on which the switching exponential is maximally subdominant compared to the leading-order behaviour. As Stokes curves are crossed, the exponentially-small behaviour experiences a smooth, rapid change in value in the neighbourhood of the curve; this is known as Stokes switching.  Anti-Stokes curves are curves along which the exponential term switches from being exponentially-small to exponentially-large (and vice-versa). We will use these definitions to determine the locations of the Stokes and anti-Stokes curves in this study.

By studying the switching behaviour of the exponentially-small remainder term in the neighbourhood of Stokes curves, it is possible to obtain an expression of the remainder term itself. The behaviour of the remainder associated with the late-order terms in \eqref{general late order terms} typically takes the form $\mathcal{S}A\exp(-\chi/\epsilon)$, where $\mathcal{S}$ is a Stokes multiplier that is constant away from Stokes curves, but varies rapidly between constant values as Stokes curves are crossed.  From this form, it can be shown that Stokes lines follow curves along which $\chi$ is real and positive, while anti-Stokes lines follow curves along which $\chi$ is imaginary. A more detailed discussion of the behaviour of Stokes curves is given in \cite{Bender1999}\footnote{Note that this book follows the American convention in switching the naming of Stokes and anti-Stokes curves.}. 
\subsection{Paper outline}\label{S:Paper outline}
In Section \ref{S:Asymptotic series expansions}, we find formal series expansions of the asymptotic solutions of $\text{dP}_{\text{II}}$, and provide the recurrence relations for the coefficients. In Section \ref{S:Exponential Asymptotics}, we determine the form of the late-order terms and use this to determine the Stokes structure of these asymptotic series expansions. We then calculate the behaviours of the exponentially-small contributions present in these solutions as Stokes curves are crossed. This is used to determine the regions in which these asymptotic approximations are valid. In Section \ref{S:Non-vanishing Asymptotics}, we consider solutions which grow in the asymptotic limit and finally, we discuss the results and conclusions of the paper in Section \ref{S:Conclusions}.  Appendices A-C contain detailed calculations needed in Section \ref{S:Exponential Asymptotics}.
\section{Asymptotic series expansions}\label{S:Asymptotic series expansions}
In this section, we expand the solution as a formal power series in the limit $n\rightarrow\infty$, obtain the recurrence relation for the coefficients of the series and determine the general expression of the late-order terms of the series.

In order to capture the far-field behaviour, we set $s=\epsilon n$. We also define the equation parameters so that they are retained in this scaling, giving
\begin{equation}\label{smallscalings}
x_n=\epsilon f(s), \qquad \alpha =\epsilon\hat{\alpha} , \qquad \beta=\hat{\beta} , \qquad \gamma=\epsilon\hat{\gamma} . 
\end{equation}
We drop the hat notation for simplicity and obtain the rescaled equation
\begin{equation}\label{rescaled dPII}
\left(f\left(s+\epsilon\right)+f\left(s-\epsilon\right)\right)\left(1-\epsilon^2f\left(s\right)^2\right)=\left(\alpha s+\beta \right)f\left(s\right)+\gamma,
\end{equation}
where we consider the limit $\epsilon \rightarrow 0$. We assume that $f(s)$ is an analytic function of $s$ so that we may expand the solutions in \eqref{rescaled dPII} to give
\begin{equation}\label{taylorexpanded rescaled dPII}
\sum_{j=0}^{\infty}\frac{2 \epsilon^{2j}f^{(2j)}(s)}{(2j)!}\bigg(1-\epsilon^2f(s)^2\bigg)=(\alpha s+\beta )f(s)+\gamma 
\end{equation}
We now expand the solution, $f(s)$, as an asymptotic power series in $\epsilon$ 
\begin{equation}\label{asymptotic series}
f(s)\sim \sum_{r=0}^{\infty}\epsilon^r f_r(s)
\end{equation}
as $\epsilon\rightarrow 0$. This allows us to rewrite equation \eqref{taylorexpanded rescaled dPII} as 
\begin{equation*}
\sum_{j=0}^{\infty}\frac{2 \epsilon^{2j}}{(2j)!}\sum_{k=0}^{\infty}\epsilon^k f_k^{(2j)}\bigg(1-\epsilon^2\sum_{l=0}^{\infty}\epsilon^l f_l\sum_{m=0}^{\infty}\epsilon^mf_m\bigg)=(\alpha s+\beta )\sum_{j=0}^{\infty}\epsilon^j f_j +\gamma .
\end{equation*}
Matching orders of $\epsilon$ as $\epsilon\rightarrow 0$, we obtain 
\begin{align*}
\mathcal{O}(\epsilon^0):& \qquad 2f_0=(\alpha s+\beta )f_0+\gamma , \\
\mathcal{O}(\epsilon^1):& \qquad 2f_1=(\alpha s+\beta )f_1.
\end{align*}
Solving these equations gives
\begin{equation}\label{leading order solutions}
f_0=\frac{-\gamma }{\alpha s+\beta -2}, \qquad f_1=0, \qquad f_2=\frac{-2\gamma (\alpha -\gamma )(\alpha +\gamma )}{(\alpha s+\beta -2)^4}.
\end{equation}
We see that the leading order solution contains a simple pole located at $s=(2-\beta )/\alpha$. 

In general, we find 
\begin{equation}\label{reccurence for coeffs}
\mathcal{O}(\epsilon^r): \qquad \sum_{j=0}^{\lfloor r/2 \rfloor}\frac{2f_{r-2j}^{(2j)}}{(2j)!}-\sum_{m=0}^{r-2}f_m\sum_{l=0}^{r-m-2}f_l\sum_{j=0}^{\lfloor (r-m-l-2)/2 \rfloor}\frac{2f_{r-m-l-2j-2}^{(2j)}}{(2j)!}=(\alpha s+\beta )f_r
\end{equation} 
for $r\geq 2$. Rearranging this equation to obtain an expression for $f_r$ gives 
\begin{equation}\label{recurrence relation for proposition}
(\alpha s+\beta -2)f_r=\sum_{j=1}^{\lfloor r/2 \rfloor}\frac{2f_{r-2j}^{(2j)}}{(2j)!}-\sum_{m=0}^{r-2}f_m\sum_{l=0}^{r-m-2}f_l\sum_{j=0}^{\lfloor (r-m-l-2)/2 \rfloor}\frac{2f_{r-m-l-2j-2}^{(2j)}}{(2j)!}.
\end{equation}
We can show that the coefficients $f_{2n+1}$ vanish as a consequence of the fact that $f_1=0$.
\begin{prop}
All the odd coefficients of the asymptotic series \eqref{asymptotic series} are zero. That is, $f_{2n+1}=0$ for all $n\geq 0$.
\end{prop}
\begin{proof}
We first apply $r\mapsto 2r+1$ to \eqref{recurrence relation for proposition} so that we are only dealing with the odd coefficients. The case $n=1$ is easy to show; a direct calculation can easily show that $f_3=0$. We then assume that $f_{2m+1}=0$ is true for $m=0,1,2,...,K$ where $K$ is arbitrary and show that it is also true for $m=K+1$. This is easy to see, because the first sum in \eqref{recurrence relation for proposition} has subscript $f_{2r-2j+1}$ which is always odd, so there will be no contributions from this term. The remaining triple sum involves the subscripts $f_{2r-m-l-2j-1}f_{l}f_{m}$. We will also show that this term produces no contributions.

The first subscript can be written has $f_{2(n-j)-m-l-1}$ and this is always odd provided that $m+l$ is even. In this case, $m+l$ can be a combination of either (odd+odd) or (even+even) but for either combination, the resulting term will always be zero, since there will always be at least one odd subscript. In order to obtain a non-zero contribution, we require the first subscript to be even, which means that $m+l$ must be odd. In this case, $m+l$ must be (odd+even), which ensures that one subscript is odd, and therefore the whole term is zero. Thus, our proposition is proved.
\end{proof}

From the recurrence relation \eqref{reccurence for coeffs} we observe that the calculation of $f_r$ requires two differentiations of $f_{r-2}$. Hence, if $f_0$ has a singularity of strength $k$ then $f_2$ will have the same singularity but with strength $k+2$. As such, our late-order coefficient terms will be described by \eqref{general late order terms}, causing the asymptotic series \eqref{asymptotic series} to diverge and exhibit the Stokes phenomenon.

We have determined the leading order asymptotic solution to \eqref{rescaled dPII} and the recurrence equation for the coefficients of \eqref{asymptotic series}. In the subsequent analysis we will optimally truncate the asymptotic series and this requires the general form of the coefficients to be known. In the next section, we will determine the general behaviour of $f_r$ as $r\rightarrow\infty$, enabling us to optimally truncate \eqref{asymptotic series} and investigate the Stokes phenomena present in these asymptotic solutions.

\section{Exponential Asymptotics}\label{S:Exponential Asymptotics}
In this section, we will completely determine the form of the late-order terms. This will allow us to optimally truncate \eqref{asymptotic series} and study the behaviour of the exponentially-small contribution. We will investigate how the Stokes phenomena affect these terms, and deduce the regions in the complex plane for which these asymptotic solutions are valid.

\subsection{Late-order terms}\label{S:Late-order terms}
As discussed in Section \ref{S:Exponential asymptotics and Stokes curves}, the late-order ansatz is given by a factorial-over-power form. Therefore, our late-order terms have the form 
\begin{equation}\label{late-order terms}
f_r \sim \frac{F(s)\Gamma(r+k)}{\chi(s)^{r+k}},
\end{equation}
as $r\rightarrow\infty$, where $\chi(s)$ is the singulant, $F(s)$ is the prefactor and $k$ is a constant. Recalling that the singulant vanishes at the singularities of the leading order solution we deduce that the singulant is subject to the condition $$\chi\left(\frac{2-\beta }{\alpha }\right)=0.$$ We apply this ansatz to equation \eqref{reccurence for coeffs} and match orders of $r$ as $r\rightarrow\infty$. The leading order equation as $r\rightarrow\infty$ is given by 
\begin{equation}\label{singulant equation}
\mathcal{O}(f_r): \qquad \sum_{j=0}^{\lfloor r/2 \rfloor}\frac{2(-\chi')^{2j}}{(2j)!}\frac{F\,\Gamma(r+k)}{\chi^{r+k}}=(\alpha s+\beta )\frac{F\,\Gamma(r+k)}{\chi^{r+k}}.
\end{equation}
Continuing to the next order as $r\rightarrow\infty$, we obtain the equation 
\begin{equation}\label{prefactor equation}
\mathcal{O}(f_{r-1}): \qquad \sum_{j=1}^{\lfloor r/2 \rfloor}\frac{2}{(2j)!}\bigg(2j(-\chi')^{2j-1}F'+(2j-1)(-\chi')^{2j-2}(-\chi'')F\bigg)=0,
\end{equation}
after simplification. In order to determine the singulant, $\chi(s)$, we consider \eqref{singulant equation} which can be reduced to
\begin{equation}\label{singulantsumequation}
\sum_{j=0}^{\lfloor r/2 \rfloor}\frac{2(-\chi')^{2j}}{(2j)!}=(\alpha s+\beta ).
\end{equation}
We replace the upper summation limit by infinity in \eqref{singulantsumequation}, introducing exponentially-small error to the singulant as $r\rightarrow \infty$ \cite{King2001}, which may be neglected here. This gives     
\begin{equation*}
\cosh(\chi')=\frac{\alpha s+\beta }{2}
\end{equation*}
which has solutions
\begin{equation}\label{singulant dash equation}
\chi'=\pm\cosh^{-1}\bigg(\frac{\alpha s+\beta }{2}\bigg)+2M\pi i,
\end{equation}
where $M \in \mathbb{Z}$. Noting that there are two different equations for the singulant, we name them $\chi_1(s)$ and $\chi_2(s)$ with the choice of the positive and negative signs respectively. In general, the behaviour of $f_r$ will be the sum of expressions \eqref{late-order terms}, with each value of $M$ and sign of the singulant \cite{Dingle1973}. However, this sum will be dominated by the two terms associated with $M=0$ as this is the value for which $|\chi|$ is smallest \cite{Chapman1998}. Thus, we consider the $M=0$ case in the subsequent analysis.

Recalling that the singulant must vanish at the singularity, $s_0=(2-\beta )/\alpha$, we integrate \eqref{singulant dash equation} to find that the singulants are given by 
\begin{align}\label{singulants}
\chi_1=&\frac{1}{\alpha }\Bigg(\sqrt{\bigg(\frac{\alpha s+\beta }{2}\bigg)^2-1}-\alpha s\cosh^{-1}\bigg(\frac{\alpha s+\beta }{2}\bigg) \nonumber \\
&\qquad +\beta \log(2)-\beta \log\bigg(\alpha s+\beta +\sqrt{\bigg(\frac{\alpha s+\beta }{2}\bigg)^2-1}\bigg)\Bigg), \\
\chi_2=&-\chi_1.
\end{align}
In order to find the prefactor associated with each singulant we solve equation \eqref{prefactor equation}. As before, we extend the summation terms to infinity, obtaining
\begin{equation}\label{explicit prefactor equation}
-F'\sinh(\chi')-F\chi''\bigg(\frac{1-\cosh(\chi')+\chi'\sinh(\chi')}{(\chi')^2}\bigg)=0.
\end{equation}
This equation is independent of the choice of $\chi_1$ or $\chi_2$. We also note that the parameter, $\gamma$, does not appear in either the singulant or prefactor equations. As a consequence, $\gamma$ will not play any role in the Stokes phenomena.

In order to completely determine the form of the late-order terms, we must also determine the value of $k$. This requires matching the late-order expression given in \eqref{late-order terms} to the leading-order behaviour in the neighbourhood of the singularity. The technical details of this process are given in Appendix \ref{S:Appendix Calculate late order}. From this analysis, we find that $k = 1/2$. 

Hence, in the neighbourhood of the singularity at $s = s_0$, the late-order terms are given by 
\begin{equation*}\label{local late order terms expression}
f_r\sim \frac{\Lambda_1\Gamma(r+1/2)}{(s-s_0)^{1/4}\big(-\tfrac{2}{3}\sqrt{\alpha }(s-s_0)^{3/2}\big)^{r+1/2}}+\frac{\Lambda_2\Gamma(r+1/2)}{(s-s_0)^{1/4}\big(\tfrac{2}{3}\sqrt{\alpha }(s-s_0)^{3/2}\big)^{r+1/2}},
\end{equation*}
in which $\Lambda_1$ and $\Lambda_2$ are arbitrary constants that may be determined numerically, illustrated in Appendix \ref{S:Appendix calculate prefactors}. These constants may also be used to determine an appropriate boundary condition at $s = s_0$ for the prefactor equation \eqref{explicit prefactor equation}, although the explicit evaluation of this term is unnecessary in the present analysis.

\subsection{Stokes smoothing}\label{S:Stokes smoothing}
In order to determine the behaviour of the exponentially-small contributions in the neighbourhood of the Stokes curve we need to optimally truncate \eqref{asymptotic series}. We truncate the asymptotic series as follows
\begin{equation}\label{optimal truncation}
f(s)=\sum_{r=0}^{N-1}\epsilon^rf_r(s) + R_N(s),
\end{equation}
where $N$ is the optimal truncation point and $R_N$ is the optimally-truncated error. We choose $N$ such that the series are truncated after their smallest terms. As the analysis is technical, we will summarize the key results in this section with the details provided in Appendix \ref{S:Appendix Stokes smoothing}.

In Appendix \ref{S:Appendix Stokes smoothing}, we show that the optimal truncation point is given by 
\begin{equation*}
N=\frac{|\chi|}{\epsilon}+\omega,
\end{equation*}
where $\omega\in[0,1)$ is chosen such that $N\in \mathbb{Z}$. The remainder terms can be shown to take the form
\begin{equation}\label{explicit R_N FORM}
R_N \sim \mathcal{S}_1F_1e^{-\chi_1/\epsilon}+\mathcal{S}_2F_2e^{-\chi_2/\epsilon},
\end{equation}
where $\mathcal{S}_i$ is the Stokes switching parameter which varies rapidly in the neighbourhood of Stokes curves. Substituting \eqref{optimal truncation} with \eqref{explicit R_N FORM} into \eqref{rescaled dPII} we obtain 
\begin{align}\label{governing equation for opt error MAIN TEXT}
\sum_{j=1}^{\infty}\frac{2\epsilon^{2j}R_N^{(2j)}}{(2j)!}&-\epsilon^2\sum_{r=0}^{N-1}\epsilon^r\sum_{j=0}^{\infty}\frac{2\epsilon^{2j}f_r^{(2j)}}{(2j)!}\bigg(2R_N\sum_{k=0}^{N-1}\epsilon^kf_k+R_N^2\bigg) \nonumber \\ 
&-\epsilon^2\sum_{j=0}^{\infty}\frac{2\epsilon^{2j}R_N^{(2j)}}{(2j)!}\Bigg(\bigg(\sum_{r=0}^{N-1}\epsilon^rf_r\bigg)^2-2R_N\sum_{r=0}^{N-1}\epsilon^rf_r+R_N^2\Bigg) \nonumber \\
&+\epsilon^Nf_N +\ldots \sim (\alpha s+\beta -2)R_N,
\end{align}
where the omitted terms are smaller than those which have been retained in the limit $\epsilon\rightarrow 0$.

In particular, it can be shown in Appendix \ref{S:Appendix Stokes smoothing} that the Stokes multiplier, $\mathcal{S}_i$, changes in value by 
\begin{equation}\label{SECTION3.2STOKESJUMP}
\big[\mathcal{S}\big]^+_-\sim \frac{i\pi}{\sqrt{\epsilon}H(|\chi|)},
\end{equation}
as Stokes curves are crossed, where $H$ is the function defined by $H(\chi)=\chi'\sinh\left(\chi'\right)$, where $\chi'$ is treated as a function of $s$, which in turn is treated as a function of $\chi$.

Consequently, the optimally-truncated asymptotic series \eqref{optimal truncation} can be rewritten explicitly as 
\begin{equation}\label{COMPLETE OPTIMALLY TRUNC ASYM SERIES}
f(s)\sim \sum_{r=0}^{N-1}\epsilon^rf_r(s)+\mathcal{S}_1F_1e^{-\chi_1/\epsilon}+\mathcal{S}_2F_2e^{-\chi_2/\epsilon},
\end{equation}
where $\mathcal{S}_i$ varies in value by \eqref{SECTION3.2STOKESJUMP} as Stokes curves are crossed, the leading orders are given in \eqref{leading order solutions}, and the late-order behaviour is given in \eqref{late-order terms}. This expression is therefore an accurate asymptotic approximation up to exponentially-small terms, valid in certain sectors of the complex $s$-plane. In particular, \eqref{COMPLETE OPTIMALLY TRUNC ASYM SERIES} contains one parameter of freedom; either $\mathcal{S}_1$ or $\mathcal{S}_2$ is free. This will be further explained in Section \ref{S:Stokes structure}.

We have successfully determined a family of asymptotic solutions to \eqref{rescaled dPII} which contains exponentially-small error. These exponentially-small terms exhibit Stokes switching and therefore the asymptotic solution \eqref{COMPLETE OPTIMALLY TRUNC ASYM SERIES} will be valid in certain regions of the complex plane. The regions of validity of \eqref{COMPLETE OPTIMALLY TRUNC ASYM SERIES} will be determined in the next section.

\subsection{Stokes structure}\label{S:Stokes structure}
With the results for $\chi_1$ and $\chi_2$ given by \eqref{singulants}, we can investigate the switching behaviour of the exponentially-small contributions. As discussed in the introduction, we know that these terms are proportional to $\exp(-\chi/\epsilon)$; this is explicitly shown in the Appendix \ref{S:Appendix Stokes smoothing} using a WKB ansatz on the homogeneous form of \eqref{governing equation for opt error MAIN TEXT}. This term is exponentially-small when Re$(\chi)>0$ and exponentially large when Re$(\chi)<0$. In order to investigate how these terms behave we consider the solution's Stokes structure. We recall that Stokes curves follow curves where Im$(\chi)=0$ while anti-Stokes curves follow curves where Re$(\chi)=0$. Additionally, we recall that exponentially-small terms may only switch across Stokes curves where Re$(\chi)>0$.

Without loss of generality, we demonstrate the case where $\alpha $ and $\beta $ are real valued parameters. In particular, we study the Stokes structure with parameter values $\alpha =1,\beta =1$. In the general case where $\alpha ,\beta \in\mathbb{C}$, we find that complex $\alpha$ rotates the Stokes structure, while complex $\beta$ translates it. These effects are illustrated in Figure 4.

\begin{figure}[H]
\begin{minipage}{0.5\linewidth} \centering
\scalebox{0.5}{\begin{tikzpicture}{center}
\draw[<->] (-5,0) -- (5,0);
\node [right] at (5,0) {Re(s)};
\draw[<->] (0,-5) -- (0,5);
\node [above] at (0,5) {Im(s)};
\draw [-,decorate,decoration=zigzag,very thick] (1,0) -- (-4.9,0); 
\draw [dashed,line width=2] (3.50734, 5.) -- (3.45431, 4.88288) -- (3.4165, 4.79778) -- (3.34515, 
4.64286) -- (3.3051, 4.55204) -- (3.26428, 4.46429) -- (3.21429, 4.3544) -- 
(3.19266, 4.30734) -- (3.18232, 4.28571) -- (3.15428, 4.22571) -- (3.07988, 
4.06298) -- (3.03571, 3.96857) -- (3.01677, 3.92857) -- (2.96638, 3.81933) -- 
(2.93297, 3.75) -- (2.85714, 3.58829) -- (2.85171, 3.57686) -- (2.84901, 
3.57143) -- (2.84138, 3.55567) -- (2.73701, 3.33442) -- (2.67742, 3.21429) -- 
(2.6212, 3.09309) -- (2.51146, 2.8686) -- (2.50602, 2.85714) -- (2.50413, 
2.85301) -- (2.5, 2.84446) -- (2.4182, 2.67857) -- (2.38714, 2.61286) -- 
(2.32851, 2.5) -- (2.26875, 2.37411) -- (2.16039, 2.16039) -- (2.15181, 
2.14286) -- (2.14898, 2.13673) -- (2.14286, 2.12443) -- (2.06076, 1.96429) -- 
(2.02937, 1.8992) -- (1.96998, 1.78571) -- (1.96429, 1.77408) -- (1.90815, 
1.66328) -- (1.87697, 1.60714) -- (1.78571, 1.43051) -- (1.78504, 1.42925) -- 
(1.78463, 1.42857) -- (1.7833, 1.42616) -- (1.66264, 1.19451) -- (1.60714, 
1.09245) -- (1.59498, 1.07143) -- (1.5807, 1.04498) -- (1.53727, 0.962729) -- 
(1.49747, 0.892857) -- (1.42857, 0.760484) -- (1.41154, 0.731314) -- 
(1.40127, 0.714286) -- (1.36742, 0.653139) -- (1.34807, 0.616212) -- 
(1.30078, 0.535714) -- (1.28484, 0.500871) -- (1.25336, 0.446429) -- (1.25, 
0.439713) -- (1.22097, 0.386171) -- (1.2023, 0.357143) -- (1.16071, 
0.280141) -- (1.15596, 0.272607) -- (1.15258, 0.267857) -- (1.14196, 
0.249098) -- (1.12375, 0.215534) -- (1.11607, 0.201678) -- (1.10189, 
0.178571) -- (1.09166, 0.158343) -- (1.07661, 0.133929) -- (1.07143, 
0.123167) -- (1.05933, 0.101384) -- (1.05018, 0.0892857) -- (1.02679, 
0.0460447) -- (1.02622, 0.0452116) -- (1.02378, 0.0416322) -- (1.01532, 
0.0233344); 
\draw [dashed,line width=2] (3.50734, -5.) -- (3.45431, -4.88288) -- (3.4165, -4.79778) -- (3.34515, 
-4.64286) -- (3.3051, -4.55204) -- (3.26428, -4.46429) -- (3.21429, 
-4.3544) -- (3.19266, -4.30734) -- (3.18232, -4.28571) -- (3.15428, 
-4.22571) -- (3.07988, -4.06298) -- (3.03571, -3.96857) -- (3.01677, 
-3.92857) -- (2.96638, -3.81933) -- (2.93297, -3.75) -- (2.85714, 
-3.58829) -- (2.85171, -3.57686) -- (2.84901, -3.57143) -- (2.84138, 
-3.55567) -- (2.73701, -3.33442) -- (2.67742, -3.21429) -- (2.6212, 
-3.09309) -- (2.51146, -2.8686) -- (2.50602, -2.85714) -- (2.50413, 
-2.85301) -- (2.5, -2.84446) -- (2.4182, -2.67857) -- (2.38714, -2.61286) -- 
(2.32851, -2.5) -- (2.26875, -2.37411) -- (2.16039, -2.16039) -- (2.15181, 
-2.14286) -- (2.14898, -2.13673) -- (2.14286, -2.12443) -- (2.06076, 
-1.96429) -- (2.02937, -1.8992) -- (1.96998, -1.78571) -- (1.96429, 
-1.77408) -- (1.90815, -1.66328) -- (1.87697, -1.60714) -- (1.78571, 
-1.43051) -- (1.78504, -1.42925) -- (1.78463, -1.42857) -- (1.7834, 
-1.42626) -- (1.72369, -1.31203) -- (1.68935, -1.25) -- (1.66189, 
-1.19525) -- (1.60714, -1.09245) -- (1.59498, -1.07143) -- (1.5807, 
-1.04498) -- (1.53727, -0.962729) -- (1.49747, -0.892857) -- (1.42857, 
-0.760484) -- (1.41154, -0.731314) -- (1.40127, -0.714286) -- (1.36742, 
-0.653139) -- (1.34807, -0.616212) -- (1.30078, -0.535714) -- (1.28484, 
-0.500871) -- (1.25336, -0.446429) -- (1.25, -0.439713) -- (1.22097, 
-0.386171) -- (1.2023, -0.357143) -- (1.16071, -0.280141) -- (1.15596, 
-0.272607) -- (1.15258, -0.267857) -- (1.14196, -0.249098) -- (1.12375, 
-0.215534) -- (1.11607, -0.201678) -- (1.10189, -0.178571) -- (1.09166, 
-0.158343) -- (1.07661, -0.133929) -- (1.07143, -0.123167) -- (1.05933, 
-0.101384) -- (1.05018, -0.0892857) -- (1.02679, -0.0460447) -- (1.02622, 
-0.0452116) -- (1.02541, -0.0445882) -- (1.02378, -0.0416355) -- (1.01536, 
-0.0251399); 
\draw[line width=2] (-2.6549, 5.) -- (-2.58148, 4.91852) -- (-2.5, 4.82719) -- (-2.49477, 
4.82143) -- (-2.41313, 4.72973) -- (-2.33654, 4.64286) -- (-2.32143, 
4.62536) -- (-2.24606, 4.53966) -- (-2.18039, 4.46429) -- (-2.14286, 
4.42024) -- (-2.08032, 4.34826) -- (-2.02645, 4.28571) -- (-1.96429, 
4.21179) -- (-1.91595, 4.15548) -- (-1.87479, 4.10714) -- (-1.78571, 
3.99981) -- (-1.75302, 3.96127) -- (-1.72476, 3.92857) -- (-1.59156, 
3.76558) -- (-1.45722, 3.60008) -- (-1.43402, 3.57143) -- (-1.43162, 
3.56838) -- (-1.42857, 3.56446) -- (-1.2912, 3.39286) -- (-1.27323, 
3.36963) -- (-1.25, 3.33944) -- (-1.15087, 3.21429) -- (-1.11649, 3.16923) -- 
(-1.07143, 3.11001) -- (-1.01311, 3.03571) -- (-0.961437, 2.96713) -- 
(-0.892857, 2.87615) -- (-0.878049, 2.85714) -- (-0.808112, 2.76332) -- 
(-0.744704, 2.67857) -- (-0.714286, 2.63591) -- (-0.65653, 2.55776) -- 
(-0.613507, 2.5) -- (-0.535714, 2.38995) -- (-0.50669, 2.35045) -- 
(-0.365074, 2.15079) -- (-0.35951, 2.14286) -- (-0.358571, 2.14143) -- 
(-0.357143, 2.13929) -- (-0.234589, 1.96429) -- (-0.212474, 1.93038) -- 
(-0.178571, 1.87962) -- (-0.112545, 1.78571) -- (-0.0682017, 1.71751) -- (0.00630805, 1.60714) -- (0.0743253, 
1.5029) -- (0.125046, 1.42857) -- (0.178571, 1.34219) -- (0.21553, 
1.28696) -- (0.241031, 1.25) -- (0.28536, 1.17822) -- (0.347271, 1.0813) -- 
(0.35353, 1.07143) -- (0.354819, 1.0691) -- (0.357143, 1.06529) -- 
(0.41007, 0.982143) -- (0.423592, 0.959306) -- (0.446429, 0.922488) -- 
(0.465711, 0.892857) -- (0.492004, 0.849147) -- (0.520984, 0.803571) -- 
(0.535714, 0.778328) -- (0.560119, 0.738691) -- (0.576189, 0.714286) -- 
(0.625, 0.632818) -- (0.628042, 0.628042) -- (0.63866, 0.61134) -- 
(0.684784, 0.535714) -- (0.695045, 0.516474) -- (0.714286, 0.483871) -- 
(0.738456, 0.446429) -- (0.762251, 0.405108) -- (0.77289, 0.387825) -- 
(0.791472, 0.357143) -- (0.803571, 0.333968) -- (0.827991, 0.292276) -- 
(0.845383, 0.267857) -- (0.892857, 0.183451) -- (0.894877, 0.180592) -- 
(0.89636, 0.178571) -- (0.902062, 0.169366) -- (0.927481, 0.123909) -- 
(0.952603, 0.0892857) -- (0.95952, 0.0666632) -- (0.976398, 0.0428371); 
\draw[line width=2] (-2.6549, -5.) -- (-2.58148, -4.91852) -- (-2.5, -4.82719) -- (-2.49477, 
-4.82143) -- (-2.41313, -4.72973) -- (-2.33654, -4.64286) -- (-2.32143, 
-4.62536) -- (-2.24606, -4.53966) -- (-2.18039, -4.46429) -- (-2.14286, 
-4.42024) -- (-2.08032, -4.34826) -- (-2.02645, -4.28571) -- (-1.96429, 
-4.21179) -- (-1.91595, -4.15548) -- (-1.87479, -4.10714) -- (-1.78571, 
-3.99981) -- (-1.75302, -3.96127) -- (-1.72476, -3.92857) -- (-1.59156, 
-3.76558) -- (-1.45722, -3.60008) -- (-1.43402, -3.57143) -- (-1.43162, 
-3.56838) -- (-1.42857, -3.56446) -- (-1.2912, -3.39286) -- (-1.27323, 
-3.36963) -- (-1.25, -3.33944) -- (-1.15087, -3.21429) -- (-1.11649, 
-3.16923) -- (-1.07143, -3.11001) -- (-1.01311, -3.03571) -- (-0.961437, 
-2.96713) -- (-0.892857, -2.87615) -- (-0.878049, -2.85714) -- (-0.808112, 
-2.76332) -- (-0.744704, -2.67857) -- (-0.714286, -2.63591) -- (-0.65653, 
-2.55776) -- (-0.613507, -2.5) -- (-0.535714, -2.38995) -- (-0.50669, 
-2.35045) -- (-0.365074, -2.15079) -- (-0.35951, -2.14286) -- (-0.358571, 
-2.14143) -- (-0.357143, -2.13929) -- (-0.234589, -1.96429) -- (-0.212474, 
-1.93038) -- (-0.178571, -1.87962) -- (-0.112545, -1.78571) -- (-0.0682017, 
-1.71751) -- (6.66134*10^-16, -1.61601) -- (0.00630805, -1.60714) -- 
(0.0743253, -1.5029) -- (0.125046, -1.42857) -- (0.178571, -1.34219) -- 
(0.21553, -1.28696) -- (0.241031, -1.25) -- (0.28536, -1.17822) -- 
(0.347271, -1.0813) -- (0.35353, -1.07143) -- (0.354819, -1.0691) -- 
(0.357143, -1.06529) -- (0.41007, -0.982143) -- (0.423592, -0.959306) -- 
(0.446429, -0.922488) -- (0.465711, -0.892857) -- (0.492004, -0.849147) -- 
(0.520984, -0.803571) -- (0.535714, -0.778328) -- (0.560119, -0.738691) -- 
(0.576189, -0.714286) -- (0.625, -0.632818) -- (0.628072, -0.628072) -- 
(0.630146, -0.625) -- (0.638065, -0.611935) -- (0.66173, -0.572444) -- 
(0.684784, -0.535714) -- (0.695045, -0.516474) -- (0.714286, -0.483871) -- 
(0.738456, -0.446429) -- (0.762251, -0.405108) -- (0.77289, -0.387825) -- 
(0.791133, -0.357143) -- (0.795487, -0.349058) -- (0.803571, -0.335187) -- 
(0.817658, -0.3125) -- (0.828571, -0.292856) -- (0.843867, -0.267857) -- 
(0.848214, -0.259794) -- (0.861662, -0.236662) -- (0.870496, -0.223214) -- 
(0.892857, -0.183834) -- (0.894877, -0.180592) -- (0.89636, -0.178571) -- 
(0.902062, -0.169366) -- (0.927481, -0.123909) -- (0.952603, -0.0892857) -- 
(0.95952, -0.0666624) -- (0.976384, -0.0428631); 
\draw[line width=2] (1,0) -- (4.8,0); 
\node [right,align=center] at (3,2) {Re$(\chi_1)<0$, Im$(\chi_1)<0$\\Re$(\chi_2)>0$, Im$(\chi_2)>0$};
\node [right,align=center] at (3,-2) {Re$(\chi_2)>0$, Im$(\chi_2)<0$\\Re$(\chi_1)<0$, Im$(\chi_1)>0$};
\node [below,align=center] at (0.4,-3.5) {Re$(\chi_2)<0$, Im$(\chi_2)<0$\\Re$(\chi_1)>0$, Im$(\chi_1)>0$};
\node [above,align=center] at (0.4,3.5) {Re$(\chi_1)>0$, Im$(\chi_1)<0$\\Re$(\chi_2)<0$, Im$(\chi_2)>0$};
\node [align=center] at (-3,2) {Re$(\chi_1)>0$, Im$(\chi_1)>0$\\Re$(\chi_2)<0$, Im$(\chi_2)<0$};
\node [align=center] at (-3,-2) {Re$(\chi_2)<0$, Im$(\chi_2)>0$\\Re$(\chi_1)>0$, Im$(\chi_1)<0$};
\draw [line width=2] (-5,7) -- (-4,7);

\node [right] at (-3.9,7) {Stokes Curve};
\draw [dashed,line width=2] (-5,6.5) -- (-4,6.5);
\node [right] at (-3.9,6.5) {Anti-Stokes Curve};
\draw [-,decorate,decoration=zigzag,very thick] (-5,6) -- (-4,6);
\node [right] at (-3.9,6) {Branch Cut};
\end{tikzpicture}}
\subcaption{Singulant behaviour.}\label{fig:1a}
\end{minipage}
\begin{minipage}{0.5\linewidth} 
\centering
\scalebox{0.5}{\begin{tikzpicture}{center}
\draw [white] (-5,7) -- (-4,7);
\draw[<->] (-5,0) -- (5,0);
\node [right] at (5,0) {Re(s)};
\draw[<->] (0,-5) -- (0,5);
\node [above] at (0,5) {Im(s)};
\draw [-,decorate,decoration=zigzag,very thick] (1,0) -- (-4.9,0); 
\draw [dashed,line width=2] (3.50734, 5.) -- (3.45431, 4.88288) -- (3.4165, 4.79778) -- (3.34515, 
4.64286) -- (3.3051, 4.55204) -- (3.26428, 4.46429) -- (3.21429, 4.3544) -- 
(3.19266, 4.30734) -- (3.18232, 4.28571) -- (3.15428, 4.22571) -- (3.07988, 
4.06298) -- (3.03571, 3.96857) -- (3.01677, 3.92857) -- (2.96638, 3.81933) -- 
(2.93297, 3.75) -- (2.85714, 3.58829) -- (2.85171, 3.57686) -- (2.84901, 
3.57143) -- (2.84138, 3.55567) -- (2.73701, 3.33442) -- (2.67742, 3.21429) -- 
(2.6212, 3.09309) -- (2.51146, 2.8686) -- (2.50602, 2.85714) -- (2.50413, 
2.85301) -- (2.5, 2.84446) -- (2.4182, 2.67857) -- (2.38714, 2.61286) -- 
(2.32851, 2.5) -- (2.26875, 2.37411) -- (2.16039, 2.16039) -- (2.15181, 
2.14286) -- (2.14898, 2.13673) -- (2.14286, 2.12443) -- (2.06076, 1.96429) -- 
(2.02937, 1.8992) -- (1.96998, 1.78571) -- (1.96429, 1.77408) -- (1.90815, 
1.66328) -- (1.87697, 1.60714) -- (1.78571, 1.43051) -- (1.78504, 1.42925) -- 
(1.78463, 1.42857) -- (1.7833, 1.42616) -- (1.66264, 1.19451) -- (1.60714, 
1.09245) -- (1.59498, 1.07143) -- (1.5807, 1.04498) -- (1.53727, 0.962729) -- 
(1.49747, 0.892857) -- (1.42857, 0.760484) -- (1.41154, 0.731314) -- 
(1.40127, 0.714286) -- (1.36742, 0.653139) -- (1.34807, 0.616212) -- 
(1.30078, 0.535714) -- (1.28484, 0.500871) -- (1.25336, 0.446429) -- (1.25, 
0.439713) -- (1.22097, 0.386171) -- (1.2023, 0.357143) -- (1.16071, 
0.280141) -- (1.15596, 0.272607) -- (1.15258, 0.267857) -- (1.14196, 
0.249098) -- (1.12375, 0.215534) -- (1.11607, 0.201678) -- (1.10189, 
0.178571) -- (1.09166, 0.158343) -- (1.07661, 0.133929) -- (1.07143, 
0.123167) -- (1.05933, 0.101384) -- (1.05018, 0.0892857) -- (1.02679, 
0.0460447) -- (1.02622, 0.0452116) -- (1.02378, 0.0416322) -- (1.01532, 
0.0233344); 
\draw [dashed,line width=2] (3.50734, -5.) -- (3.45431, -4.88288) -- (3.4165, -4.79778) -- (3.34515, 
-4.64286) -- (3.3051, -4.55204) -- (3.26428, -4.46429) -- (3.21429, 
-4.3544) -- (3.19266, -4.30734) -- (3.18232, -4.28571) -- (3.15428, 
-4.22571) -- (3.07988, -4.06298) -- (3.03571, -3.96857) -- (3.01677, 
-3.92857) -- (2.96638, -3.81933) -- (2.93297, -3.75) -- (2.85714, 
-3.58829) -- (2.85171, -3.57686) -- (2.84901, -3.57143) -- (2.84138, 
-3.55567) -- (2.73701, -3.33442) -- (2.67742, -3.21429) -- (2.6212, 
-3.09309) -- (2.51146, -2.8686) -- (2.50602, -2.85714) -- (2.50413, 
-2.85301) -- (2.5, -2.84446) -- (2.4182, -2.67857) -- (2.38714, -2.61286) -- 
(2.32851, -2.5) -- (2.26875, -2.37411) -- (2.16039, -2.16039) -- (2.15181, 
-2.14286) -- (2.14898, -2.13673) -- (2.14286, -2.12443) -- (2.06076, 
-1.96429) -- (2.02937, -1.8992) -- (1.96998, -1.78571) -- (1.96429, 
-1.77408) -- (1.90815, -1.66328) -- (1.87697, -1.60714) -- (1.78571, 
-1.43051) -- (1.78504, -1.42925) -- (1.78463, -1.42857) -- (1.7834, 
-1.42626) -- (1.72369, -1.31203) -- (1.68935, -1.25) -- (1.66189, 
-1.19525) -- (1.60714, -1.09245) -- (1.59498, -1.07143) -- (1.5807, 
-1.04498) -- (1.53727, -0.962729) -- (1.49747, -0.892857) -- (1.42857, 
-0.760484) -- (1.41154, -0.731314) -- (1.40127, -0.714286) -- (1.36742, 
-0.653139) -- (1.34807, -0.616212) -- (1.30078, -0.535714) -- (1.28484, 
-0.500871) -- (1.25336, -0.446429) -- (1.25, -0.439713) -- (1.22097, 
-0.386171) -- (1.2023, -0.357143) -- (1.16071, -0.280141) -- (1.15596, 
-0.272607) -- (1.15258, -0.267857) -- (1.14196, -0.249098) -- (1.12375, 
-0.215534) -- (1.11607, -0.201678) -- (1.10189, -0.178571) -- (1.09166, 
-0.158343) -- (1.07661, -0.133929) -- (1.07143, -0.123167) -- (1.05933, 
-0.101384) -- (1.05018, -0.0892857) -- (1.02679, -0.0460447) -- (1.02622, 
-0.0452116) -- (1.02541, -0.0445882) -- (1.02378, -0.0416355) -- (1.01536, 
-0.0251399); 
\draw[line width=2] (-2.6549, 5.) -- (-2.58148, 4.91852) -- (-2.5, 4.82719) -- (-2.49477, 
4.82143) -- (-2.41313, 4.72973) -- (-2.33654, 4.64286) -- (-2.32143, 
4.62536) -- (-2.24606, 4.53966) -- (-2.18039, 4.46429) -- (-2.14286, 
4.42024) -- (-2.08032, 4.34826) -- (-2.02645, 4.28571) -- (-1.96429, 
4.21179) -- (-1.91595, 4.15548) -- (-1.87479, 4.10714) -- (-1.78571, 
3.99981) -- (-1.75302, 3.96127) -- (-1.72476, 3.92857) -- (-1.59156, 
3.76558) -- (-1.45722, 3.60008) -- (-1.43402, 3.57143) -- (-1.43162, 
3.56838) -- (-1.42857, 3.56446) -- (-1.2912, 3.39286) -- (-1.27323, 
3.36963) -- (-1.25, 3.33944) -- (-1.15087, 3.21429) -- (-1.11649, 3.16923) -- 
(-1.07143, 3.11001) -- (-1.01311, 3.03571) -- (-0.961437, 2.96713) -- 
(-0.892857, 2.87615) -- (-0.878049, 2.85714) -- (-0.808112, 2.76332) -- 
(-0.744704, 2.67857) -- (-0.714286, 2.63591) -- (-0.65653, 2.55776) -- 
(-0.613507, 2.5) -- (-0.535714, 2.38995) -- (-0.50669, 2.35045) -- 
(-0.365074, 2.15079) -- (-0.35951, 2.14286) -- (-0.358571, 2.14143) -- 
(-0.357143, 2.13929) -- (-0.234589, 1.96429) -- (-0.212474, 1.93038) -- 
(-0.178571, 1.87962) -- (-0.112545, 1.78571) -- (-0.0682017, 1.71751) -- (0.00630805, 1.60714) -- (0.0743253, 
1.5029) -- (0.125046, 1.42857) -- (0.178571, 1.34219) -- (0.21553, 
1.28696) -- (0.241031, 1.25) -- (0.28536, 1.17822) -- (0.347271, 1.0813) -- 
(0.35353, 1.07143) -- (0.354819, 1.0691) -- (0.357143, 1.06529) -- 
(0.41007, 0.982143) -- (0.423592, 0.959306) -- (0.446429, 0.922488) -- 
(0.465711, 0.892857) -- (0.492004, 0.849147) -- (0.520984, 0.803571) -- 
(0.535714, 0.778328) -- (0.560119, 0.738691) -- (0.576189, 0.714286) -- 
(0.625, 0.632818) -- (0.628042, 0.628042) -- (0.63866, 0.61134) -- 
(0.684784, 0.535714) -- (0.695045, 0.516474) -- (0.714286, 0.483871) -- 
(0.738456, 0.446429) -- (0.762251, 0.405108) -- (0.77289, 0.387825) -- 
(0.791472, 0.357143) -- (0.803571, 0.333968) -- (0.827991, 0.292276) -- 
(0.845383, 0.267857) -- (0.892857, 0.183451) -- (0.894877, 0.180592) -- 
(0.89636, 0.178571) -- (0.902062, 0.169366) -- (0.927481, 0.123909) -- 
(0.952603, 0.0892857) -- (0.95952, 0.0666632) -- (0.976398, 0.0428371); 
\draw[line width=2] (-2.6549, -5.) -- (-2.58148, -4.91852) -- (-2.5, -4.82719) -- (-2.49477, 
-4.82143) -- (-2.41313, -4.72973) -- (-2.33654, -4.64286) -- (-2.32143, 
-4.62536) -- (-2.24606, -4.53966) -- (-2.18039, -4.46429) -- (-2.14286, 
-4.42024) -- (-2.08032, -4.34826) -- (-2.02645, -4.28571) -- (-1.96429, 
-4.21179) -- (-1.91595, -4.15548) -- (-1.87479, -4.10714) -- (-1.78571, 
-3.99981) -- (-1.75302, -3.96127) -- (-1.72476, -3.92857) -- (-1.59156, 
-3.76558) -- (-1.45722, -3.60008) -- (-1.43402, -3.57143) -- (-1.43162, 
-3.56838) -- (-1.42857, -3.56446) -- (-1.2912, -3.39286) -- (-1.27323, 
-3.36963) -- (-1.25, -3.33944) -- (-1.15087, -3.21429) -- (-1.11649, 
-3.16923) -- (-1.07143, -3.11001) -- (-1.01311, -3.03571) -- (-0.961437, 
-2.96713) -- (-0.892857, -2.87615) -- (-0.878049, -2.85714) -- (-0.808112, 
-2.76332) -- (-0.744704, -2.67857) -- (-0.714286, -2.63591) -- (-0.65653, 
-2.55776) -- (-0.613507, -2.5) -- (-0.535714, -2.38995) -- (-0.50669, 
-2.35045) -- (-0.365074, -2.15079) -- (-0.35951, -2.14286) -- (-0.358571, 
-2.14143) -- (-0.357143, -2.13929) -- (-0.234589, -1.96429) -- (-0.212474, 
-1.93038) -- (-0.178571, -1.87962) -- (-0.112545, -1.78571) -- (-0.0682017, 
-1.71751) -- (6.66134*10^-16, -1.61601) -- (0.00630805, -1.60714) -- 
(0.0743253, -1.5029) -- (0.125046, -1.42857) -- (0.178571, -1.34219) -- 
(0.21553, -1.28696) -- (0.241031, -1.25) -- (0.28536, -1.17822) -- 
(0.347271, -1.0813) -- (0.35353, -1.07143) -- (0.354819, -1.0691) -- 
(0.357143, -1.06529) -- (0.41007, -0.982143) -- (0.423592, -0.959306) -- 
(0.446429, -0.922488) -- (0.465711, -0.892857) -- (0.492004, -0.849147) -- 
(0.520984, -0.803571) -- (0.535714, -0.778328) -- (0.560119, -0.738691) -- 
(0.576189, -0.714286) -- (0.625, -0.632818) -- (0.628072, -0.628072) -- 
(0.630146, -0.625) -- (0.638065, -0.611935) -- (0.66173, -0.572444) -- 
(0.684784, -0.535714) -- (0.695045, -0.516474) -- (0.714286, -0.483871) -- 
(0.738456, -0.446429) -- (0.762251, -0.405108) -- (0.77289, -0.387825) -- 
(0.791133, -0.357143) -- (0.795487, -0.349058) -- (0.803571, -0.335187) -- 
(0.817658, -0.3125) -- (0.828571, -0.292856) -- (0.843867, -0.267857) -- 
(0.848214, -0.259794) -- (0.861662, -0.236662) -- (0.870496, -0.223214) -- 
(0.892857, -0.183834) -- (0.894877, -0.180592) -- (0.89636, -0.178571) -- 
(0.902062, -0.169366) -- (0.927481, -0.123909) -- (0.952603, -0.0892857) -- 
(0.95952, -0.0666624) -- (0.976384, -0.0428631); 
\draw[line width=2] (1,0) -- (4.8,0); 
\node [right,align=center] at (3,2) {$\chi_1$: Inactive\\$\chi_2$: Active in state 1 (exp. small)};
\node [right,align=center] at (3,-2) {$\chi_1$: Inactive\\$\chi_2$: Active in state 2 (exp. small)};
\node [below,align=center] at (0.4,-3.8) {$\chi_1$: Inactive\\$\chi_2$: Active in state 2 (exp. large)};
\node [above,align=center] at (0.4,3.9) {$\chi_2$: Active in state 1 (exp. large)\\$\chi_1$: Inactive};
\node [align=center] at (-3,2) {$\chi_1$: Active (exp. small)\\$\chi_2$: Active in state 1 (exp. large)};
\node [align=center] at (-3,-1) {$\chi_1$: Active (exp. small)\\$\chi_2$: Active in state 2 (exp. large)};
\draw [white,line width=2] (-5,7) -- (-4,7);
\node [draw,circle,scale=0.8] at (4.5,0.35) {\textbf{2}};
\node [draw,circle,scale=0.8] at (-2.7,4.5) {\textbf{1}};
\node [draw,circle,scale=0.8] at (-2.7,-4.5) {\textbf{1}};
\end{tikzpicture}}
\subcaption{Exponential behaviour.}\label{fig:1b}
\end{minipage}
\caption{These figures depict the Stokes structure for parameter values  $\alpha =1$ and $\beta =1$. Figure \ref{fig:1a} illustrates the behaviour of the singulants as Stokes and anti-Stokes curves (denoted by thick black curves and dashed curves respectively) are crossed. Figure \ref{fig:1b} illustrates the regions of the complex $s$-plane in which the exponential contributions associated with $\chi_1$ and $\chi_2$ are active. The exponential contribution associated with $\chi_1$ is switched across the Stokes curves denoted by \ding{172}, which the contribution associated with $\chi_2$ is switched when crossing the Stokes curve denoted by \ding{173}. This convention will be followed in subsequent figures.}
\end{figure}
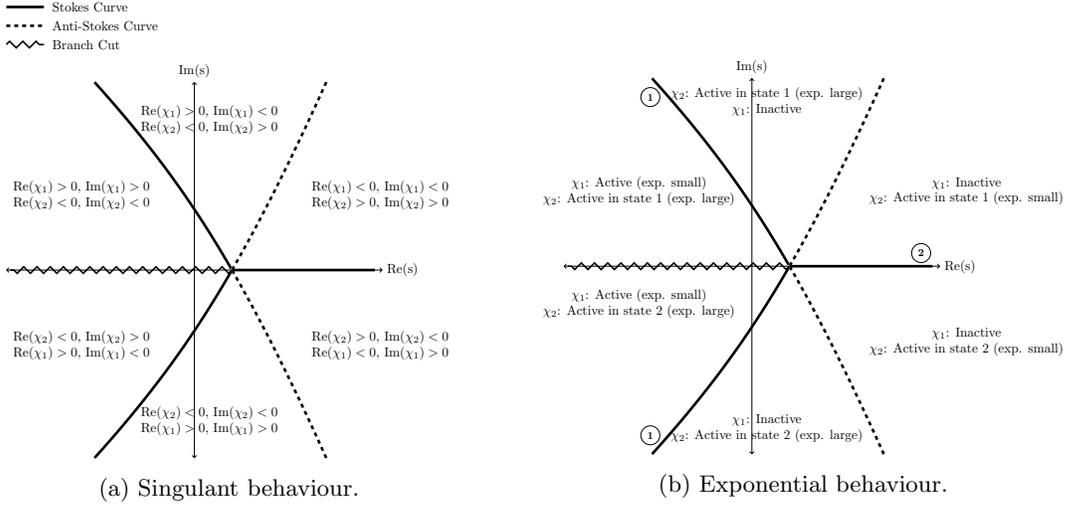

In Figure \ref{fig:1a} we see that there are three Stokes curves and two anti-Stokes curves emanating from the singularity in the complex $s$-plane. The Stokes curve located on the positive real axis switches the exponential contributions associated with $\chi_2$, while the remaining two Stokes curves switches the exponential associated with $\chi_1$. Additionally, there is a branch cut located along the negative real axis extending to the singularity, $s_0=1$. Using this knowledge, we can determine the switching behaviour as the Stokes curves are crossed. 

Since there are six critical curves (Stokes and anti-Stokes curves and a branch cut) in total, we have the freedom to choose within which region we wish to have a valid asymptotic solution. The most natural one to choose is the Stokes curve located on the positive real axis. Thus we see that the Stokes structure naturally separates the complex $s$-plane into separate regions.  
From Figure 1 we deduce that the remainder term associated with $\chi_1$ must not be present in the neighbourhood of this Stokes curve as it would exponentially dominate the leading order solution of \eqref{COMPLETE OPTIMALLY TRUNC ASYM SERIES}. In order for the asymptotic solution to be valid we require the remainder term associated with $\chi_1$ be absent on the positive real axis, and therefore $\mathcal{S}_1=0$. However, we see that the remainder term associated with $\chi_2$ is exponentially-small since Re$(\chi_2)>0$, and therefore the leading order solution of (\ref{COMPLETE OPTIMALLY TRUNC ASYM SERIES}) remains valid in the presence of these terms. Hence, the value of $\mathcal{S}_2$ about the real axis may be freely specified, and will therefore contain a free parameter. Since the remainder term associated with $\chi_2$ will exhibit Stokes switching, it will vary as it crosses a Stokes curve; say, from state one to state two. Consequently, we conclude that the exponentially-small contributions associated with $\chi_1$ is present in the regions bounded by the Stokes curves located in the upper and lower complex plane containing the branch cut. If we assume that
\begin{figure}[h!]
\begin{minipage}{0.5\linewidth} \centering
\scalebox{0.5}{\begin{tikzpicture}{center}
\draw[<->] (-5,0) -- (5,0);
\node [right] at (5,0) {Re(s)};
\draw[<->] (0,-5) -- (0,5);
\node [above] at (0,5) {Im(s)};
\draw [-,decorate,decoration=zigzag,very thick] (1,0) -- (-4.9,0); 
\draw [dashed,line width=2] (3.50734, 5.) -- (3.45431, 4.88288) -- (3.4165, 4.79778) -- (3.34515, 
4.64286) -- (3.3051, 4.55204) -- (3.26428, 4.46429) -- (3.21429, 4.3544) -- 
(3.19266, 4.30734) -- (3.18232, 4.28571) -- (3.15428, 4.22571) -- (3.07988, 
4.06298) -- (3.03571, 3.96857) -- (3.01677, 3.92857) -- (2.96638, 3.81933) -- 
(2.93297, 3.75) -- (2.85714, 3.58829) -- (2.85171, 3.57686) -- (2.84901, 
3.57143) -- (2.84138, 3.55567) -- (2.73701, 3.33442) -- (2.67742, 3.21429) -- 
(2.6212, 3.09309) -- (2.51146, 2.8686) -- (2.50602, 2.85714) -- (2.50413, 
2.85301) -- (2.5, 2.84446) -- (2.4182, 2.67857) -- (2.38714, 2.61286) -- 
(2.32851, 2.5) -- (2.26875, 2.37411) -- (2.16039, 2.16039) -- (2.15181, 
2.14286) -- (2.14898, 2.13673) -- (2.14286, 2.12443) -- (2.06076, 1.96429) -- 
(2.02937, 1.8992) -- (1.96998, 1.78571) -- (1.96429, 1.77408) -- (1.90815, 
1.66328) -- (1.87697, 1.60714) -- (1.78571, 1.43051) -- (1.78504, 1.42925) -- 
(1.78463, 1.42857) -- (1.7833, 1.42616) -- (1.66264, 1.19451) -- (1.60714, 
1.09245) -- (1.59498, 1.07143) -- (1.5807, 1.04498) -- (1.53727, 0.962729) -- 
(1.49747, 0.892857) -- (1.42857, 0.760484) -- (1.41154, 0.731314) -- 
(1.40127, 0.714286) -- (1.36742, 0.653139) -- (1.34807, 0.616212) -- 
(1.30078, 0.535714) -- (1.28484, 0.500871) -- (1.25336, 0.446429) -- (1.25, 
0.439713) -- (1.22097, 0.386171) -- (1.2023, 0.357143) -- (1.16071, 
0.280141) -- (1.15596, 0.272607) -- (1.15258, 0.267857) -- (1.14196, 
0.249098) -- (1.12375, 0.215534) -- (1.11607, 0.201678) -- (1.10189, 
0.178571) -- (1.09166, 0.158343) -- (1.07661, 0.133929) -- (1.07143, 
0.123167) -- (1.05933, 0.101384) -- (1.05018, 0.0892857) -- (1.02679, 
0.0460447) -- (1.02622, 0.0452116) -- (1.02378, 0.0416322) -- (1.01532, 
0.0233344); 
\draw [dashed,line width=2] (3.50734, -5.) -- (3.45431, -4.88288) -- (3.4165, -4.79778) -- (3.34515, 
-4.64286) -- (3.3051, -4.55204) -- (3.26428, -4.46429) -- (3.21429, 
-4.3544) -- (3.19266, -4.30734) -- (3.18232, -4.28571) -- (3.15428, 
-4.22571) -- (3.07988, -4.06298) -- (3.03571, -3.96857) -- (3.01677, 
-3.92857) -- (2.96638, -3.81933) -- (2.93297, -3.75) -- (2.85714, 
-3.58829) -- (2.85171, -3.57686) -- (2.84901, -3.57143) -- (2.84138, 
-3.55567) -- (2.73701, -3.33442) -- (2.67742, -3.21429) -- (2.6212, 
-3.09309) -- (2.51146, -2.8686) -- (2.50602, -2.85714) -- (2.50413, 
-2.85301) -- (2.5, -2.84446) -- (2.4182, -2.67857) -- (2.38714, -2.61286) -- 
(2.32851, -2.5) -- (2.26875, -2.37411) -- (2.16039, -2.16039) -- (2.15181, 
-2.14286) -- (2.14898, -2.13673) -- (2.14286, -2.12443) -- (2.06076, 
-1.96429) -- (2.02937, -1.8992) -- (1.96998, -1.78571) -- (1.96429, 
-1.77408) -- (1.90815, -1.66328) -- (1.87697, -1.60714) -- (1.78571, 
-1.43051) -- (1.78504, -1.42925) -- (1.78463, -1.42857) -- (1.7834, 
-1.42626) -- (1.72369, -1.31203) -- (1.68935, -1.25) -- (1.66189, 
-1.19525) -- (1.60714, -1.09245) -- (1.59498, -1.07143) -- (1.5807, 
-1.04498) -- (1.53727, -0.962729) -- (1.49747, -0.892857) -- (1.42857, 
-0.760484) -- (1.41154, -0.731314) -- (1.40127, -0.714286) -- (1.36742, 
-0.653139) -- (1.34807, -0.616212) -- (1.30078, -0.535714) -- (1.28484, 
-0.500871) -- (1.25336, -0.446429) -- (1.25, -0.439713) -- (1.22097, 
-0.386171) -- (1.2023, -0.357143) -- (1.16071, -0.280141) -- (1.15596, 
-0.272607) -- (1.15258, -0.267857) -- (1.14196, -0.249098) -- (1.12375, 
-0.215534) -- (1.11607, -0.201678) -- (1.10189, -0.178571) -- (1.09166, 
-0.158343) -- (1.07661, -0.133929) -- (1.07143, -0.123167) -- (1.05933, 
-0.101384) -- (1.05018, -0.0892857) -- (1.02679, -0.0460447) -- (1.02622, 
-0.0452116) -- (1.02541, -0.0445882) -- (1.02378, -0.0416355) -- (1.01536, 
-0.0251399); 
\draw[line width=2] (-2.6549, 5.) -- (-2.58148, 4.91852) -- (-2.5, 4.82719) -- (-2.49477, 
4.82143) -- (-2.41313, 4.72973) -- (-2.33654, 4.64286) -- (-2.32143, 
4.62536) -- (-2.24606, 4.53966) -- (-2.18039, 4.46429) -- (-2.14286, 
4.42024) -- (-2.08032, 4.34826) -- (-2.02645, 4.28571) -- (-1.96429, 
4.21179) -- (-1.91595, 4.15548) -- (-1.87479, 4.10714) -- (-1.78571, 
3.99981) -- (-1.75302, 3.96127) -- (-1.72476, 3.92857) -- (-1.59156, 
3.76558) -- (-1.45722, 3.60008) -- (-1.43402, 3.57143) -- (-1.43162, 
3.56838) -- (-1.42857, 3.56446) -- (-1.2912, 3.39286) -- (-1.27323, 
3.36963) -- (-1.25, 3.33944) -- (-1.15087, 3.21429) -- (-1.11649, 3.16923) -- 
(-1.07143, 3.11001) -- (-1.01311, 3.03571) -- (-0.961437, 2.96713) -- 
(-0.892857, 2.87615) -- (-0.878049, 2.85714) -- (-0.808112, 2.76332) -- 
(-0.744704, 2.67857) -- (-0.714286, 2.63591) -- (-0.65653, 2.55776) -- 
(-0.613507, 2.5) -- (-0.535714, 2.38995) -- (-0.50669, 2.35045) -- 
(-0.365074, 2.15079) -- (-0.35951, 2.14286) -- (-0.358571, 2.14143) -- 
(-0.357143, 2.13929) -- (-0.234589, 1.96429) -- (-0.212474, 1.93038) -- 
(-0.178571, 1.87962) -- (-0.112545, 1.78571) -- (-0.0682017, 1.71751) -- (0.00630805, 1.60714) -- (0.0743253, 
1.5029) -- (0.125046, 1.42857) -- (0.178571, 1.34219) -- (0.21553, 
1.28696) -- (0.241031, 1.25) -- (0.28536, 1.17822) -- (0.347271, 1.0813) -- 
(0.35353, 1.07143) -- (0.354819, 1.0691) -- (0.357143, 1.06529) -- 
(0.41007, 0.982143) -- (0.423592, 0.959306) -- (0.446429, 0.922488) -- 
(0.465711, 0.892857) -- (0.492004, 0.849147) -- (0.520984, 0.803571) -- 
(0.535714, 0.778328) -- (0.560119, 0.738691) -- (0.576189, 0.714286) -- 
(0.625, 0.632818) -- (0.628042, 0.628042) -- (0.63866, 0.61134) -- 
(0.684784, 0.535714) -- (0.695045, 0.516474) -- (0.714286, 0.483871) -- 
(0.738456, 0.446429) -- (0.762251, 0.405108) -- (0.77289, 0.387825) -- 
(0.791472, 0.357143) -- (0.803571, 0.333968) -- (0.827991, 0.292276) -- 
(0.845383, 0.267857) -- (0.892857, 0.183451) -- (0.894877, 0.180592) -- 
(0.89636, 0.178571) -- (0.902062, 0.169366) -- (0.927481, 0.123909) -- 
(0.952603, 0.0892857) -- (0.95952, 0.0666632) -- (0.976398, 0.0428371); 
\draw[line width=2] (-2.6549, -5.) -- (-2.58148, -4.91852) -- (-2.5, -4.82719) -- (-2.49477, 
-4.82143) -- (-2.41313, -4.72973) -- (-2.33654, -4.64286) -- (-2.32143, 
-4.62536) -- (-2.24606, -4.53966) -- (-2.18039, -4.46429) -- (-2.14286, 
-4.42024) -- (-2.08032, -4.34826) -- (-2.02645, -4.28571) -- (-1.96429, 
-4.21179) -- (-1.91595, -4.15548) -- (-1.87479, -4.10714) -- (-1.78571, 
-3.99981) -- (-1.75302, -3.96127) -- (-1.72476, -3.92857) -- (-1.59156, 
-3.76558) -- (-1.45722, -3.60008) -- (-1.43402, -3.57143) -- (-1.43162, 
-3.56838) -- (-1.42857, -3.56446) -- (-1.2912, -3.39286) -- (-1.27323, 
-3.36963) -- (-1.25, -3.33944) -- (-1.15087, -3.21429) -- (-1.11649, 
-3.16923) -- (-1.07143, -3.11001) -- (-1.01311, -3.03571) -- (-0.961437, 
-2.96713) -- (-0.892857, -2.87615) -- (-0.878049, -2.85714) -- (-0.808112, 
-2.76332) -- (-0.744704, -2.67857) -- (-0.714286, -2.63591) -- (-0.65653, 
-2.55776) -- (-0.613507, -2.5) -- (-0.535714, -2.38995) -- (-0.50669, 
-2.35045) -- (-0.365074, -2.15079) -- (-0.35951, -2.14286) -- (-0.358571, 
-2.14143) -- (-0.357143, -2.13929) -- (-0.234589, -1.96429) -- (-0.212474, 
-1.93038) -- (-0.178571, -1.87962) -- (-0.112545, -1.78571) -- (-0.0682017, 
-1.71751) -- (6.66134*10^-16, -1.61601) -- (0.00630805, -1.60714) -- 
(0.0743253, -1.5029) -- (0.125046, -1.42857) -- (0.178571, -1.34219) -- 
(0.21553, -1.28696) -- (0.241031, -1.25) -- (0.28536, -1.17822) -- 
(0.347271, -1.0813) -- (0.35353, -1.07143) -- (0.354819, -1.0691) -- 
(0.357143, -1.06529) -- (0.41007, -0.982143) -- (0.423592, -0.959306) -- 
(0.446429, -0.922488) -- (0.465711, -0.892857) -- (0.492004, -0.849147) -- 
(0.520984, -0.803571) -- (0.535714, -0.778328) -- (0.560119, -0.738691) -- 
(0.576189, -0.714286) -- (0.625, -0.632818) -- (0.628072, -0.628072) -- 
(0.630146, -0.625) -- (0.638065, -0.611935) -- (0.66173, -0.572444) -- 
(0.684784, -0.535714) -- (0.695045, -0.516474) -- (0.714286, -0.483871) -- 
(0.738456, -0.446429) -- (0.762251, -0.405108) -- (0.77289, -0.387825) -- 
(0.791133, -0.357143) -- (0.795487, -0.349058) -- (0.803571, -0.335187) -- 
(0.817658, -0.3125) -- (0.828571, -0.292856) -- (0.843867, -0.267857) -- 
(0.848214, -0.259794) -- (0.861662, -0.236662) -- (0.870496, -0.223214) -- 
(0.892857, -0.183834) -- (0.894877, -0.180592) -- (0.89636, -0.178571) -- 
(0.902062, -0.169366) -- (0.927481, -0.123909) -- (0.952603, -0.0892857) -- 
(0.95952, -0.0666624) -- (0.976384, -0.0428631); 
\draw[line width=2] (1,0) -- (4.8,0); 

\node [right,align=center] at (3,2) {$\begin{aligned}
\mathcal{S}_1&=0 \\ \mathcal{S}_2&=\mathcal{S}_2^+
\end{aligned}$};
\node [right,align=center] at (3,-1.5) {$\begin{aligned}
\mathcal{S}_1&=0\\\mathcal{S}_2&=\mathcal{S}_2^-
\end{aligned}$};
\node [right,align=center] at (0.5,-2.8) {$\begin{aligned}
\mathcal{S}_1&=0\\\mathcal{S}_2&=\mathcal{S}_2^-
\end{aligned}$};
\node [right,align=center] at (0.5,3.3) {$\begin{aligned}
\mathcal{S}_1&=0\\\mathcal{S}_2&=\mathcal{S}_2^+
\end{aligned}$};
\node [right,align=center] at (-3.5,2) {$\begin{aligned}
\mathcal{S}_1&=\mathcal{S}_1^+\\\mathcal{S}_2&=\mathcal{S}_2^+
\end{aligned}$};
\node [right,align=center] at (-3.5,-1.5) {$\begin{aligned}
\mathcal{S}_1&=\mathcal{S}_1^+\\\mathcal{S}_2&=\mathcal{S}_2^-
\end{aligned}$};
\node [draw,circle,scale=0.8] at (4.5,0.35) {\textbf{2}};
\node [draw,circle,scale=0.8] at (-2.7,4.5) {\textbf{1}};
\node [draw,circle,scale=0.8] at (-2.7,-4.5) {\textbf{1}};
\end{tikzpicture}}
\subcaption{Stokes multipliers.}\label{fig:2a}
\end{minipage}
\begin{minipage}{0.5\linewidth} 
\centering
\scalebox{0.5}{\begin{tikzpicture}{center}
\draw [fill,gray!30!white] (1,0) circle (3.5);
\begin{scope}
\clip (1,0) circle (3.5);
\clip (3.50734, 5.) -- (3.45431, 4.88288) -- (3.4165, 4.79778) -- (3.34515, 4.64286) -- (3.3051, 4.55204) -- (3.26428, 4.46429) -- (3.21429, 4.3544) -- (3.19266, 4.30734) -- (3.18232, 4.28571) -- (3.15428, 4.22571) -- (3.07988, 4.06298) -- (3.03571, 3.96857) -- (3.01677, 3.92857) -- (2.96638, 3.81933) -- (2.93297, 3.75) -- (2.85714, 3.58829) -- (2.85171, 3.57686) -- (2.84901, 3.57143) -- (2.84138, 3.55567) -- (2.73701, 3.33442) -- (2.67742, 3.21429) -- 
(2.6212, 3.09309) -- (2.51146, 2.8686) -- (2.50602, 2.85714) -- (2.50413, 2.85301) -- (2.5, 2.84446) -- (2.4182, 2.67857) -- (2.38714, 2.61286) -- (2.32851, 2.5) -- (2.26875, 2.37411) -- (2.16039, 2.16039) -- (2.15181, 2.14286) -- (2.14898, 2.13673) -- (2.14286, 2.12443) -- (2.06076, 1.96429) -- (2.02937, 1.8992) -- (1.96998, 1.78571) -- (1.96429, 1.77408) -- (1.90815, 1.66328) -- (1.87697, 1.60714) -- (1.78571, 1.43051) -- (1.78504, 1.42925) -- 
(1.78463, 1.42857) -- (1.7833, 1.42616) -- (1.66264, 1.19451) -- (1.60714, 1.09245) -- (1.59498, 1.07143) -- (1.5807, 1.04498) -- (1.53727, 0.962729) -- (1.49747, 0.892857) -- (1.42857, 0.760484) -- (1.41154, 0.731314) -- (1.40127, 0.714286) -- (1.36742,0.653139) -- (1.34807, 0.616212) -- (1.30078, 0.535714) -- (1.28484, 0.500871) -- (1.25336, 0.446429) -- (1.25, 0.439713) -- (1.22097, 0.386171) -- (1.2023, 0.357143) -- (1.16071,0.280141) -- (1.15596, 0.272607) -- (1.15258, 0.267857) -- (1.14196, 0.249098) -- (1.12375, 0.215534) -- (1.11607, 0.201678) -- (1.10189, 0.178571) -- (1.09166, 0.158343) -- (1.07661, 0.133929) -- (1.07143, 0.123167) -- (1.05933, 0.101384) -- (1.05018, 0.0892857) -- (1.02679, 0.0460447) -- (1.02622, 0.0452116) -- (1.02378, 0.0416322) -- (1.01532, 0.0233344) -- (1,0) -- (-3.5,0) -- (-3.5,3.5) -- cycle;
\draw [fill,gray!80!white] (-3.5,3.5) rectangle (3.5,0);
\end{scope}
\begin{scope}
\clip (1,0) circle (3.5);
\clip (3.50734, -5.) -- (3.45431, -4.88288) -- (3.4165, -4.79778) -- (3.34515, -4.64286) -- (3.3051, -4.55204) -- (3.26428,-4.46429) -- (3.21429, -4.3544) -- (3.19266, -4.30734) -- (3.18232, -4.28571) -- (3.15428, -4.22571) -- (3.07988, -4.06298) -- (3.03571, -3.96857) -- (3.01677, -3.92857) -- (2.96638, -3.81933) -- (2.93297, -3.75) -- (2.85714, -3.58829) -- (2.85171, -3.57686) -- (2.84901, -3.57143) -- (2.84138, 
-3.55567) -- (2.73701, -3.33442) -- (2.67742, -3.21429) -- (2.6212, -3.09309) -- (2.51146, -2.8686) -- (2.50602, -2.85714) -- (2.50413, -2.85301) -- (2.5, -2.84446) -- (2.4182, -2.67857) -- (2.38714, -2.61286) -- (2.32851, -2.5) -- (2.26875, -2.37411) -- (2.16039, -2.16039) -- (2.15181, -2.14286) -- (2.14898, -2.13673) -- (2.14286, -2.12443) -- (2.06076, -1.96429) -- (2.02937, -1.8992) -- (1.96998, -1.78571) -- (1.96429, 
-1.77408) -- (1.90815, -1.66328) -- (1.87697, -1.60714) -- (1.78571, -1.43051) -- (1.78504, -1.42925) -- (1.78463, -1.42857) -- (1.7834, -1.42626) -- (1.72369, -1.31203) -- (1.68935, -1.25) -- (1.66189, -1.19525) -- (1.60714, -1.09245) -- (1.59498, -1.07143) -- (1.5807, -1.04498) -- (1.53727, -0.962729) -- (1.49747, -0.892857) -- (1.42857, -0.760484) -- (1.41154, -0.731314) -- (1.40127, -0.714286) -- (1.36742, -0.653139) -- (1.34807, -0.616212) -- (1.30078, -0.535714) -- (1.28484, 
-0.500871) -- (1.25336, -0.446429) -- (1.25, -0.439713) -- (1.22097, -0.386171) -- (1.2023, -0.357143) -- (1.16071, -0.280141) -- (1.15596, -0.272607) -- (1.15258, -0.267857) -- (1.14196, -0.249098) -- (1.12375, -0.215534) -- (1.11607, -0.201678) -- (1.10189, -0.178571) -- (1.09166, -0.158343) -- (1.07661, -0.133929) -- (1.07143, -0.123167) -- (1.05933, 
-0.101384) -- (1.05018, -0.0892857) -- (1.02679, -0.0460447) -- (1.02622, -0.0452116) -- (1.02541, -0.0445882) -- (1.02378, -0.0416355) -- (1.01536, -0.0251399) -- (1,0) -- (-3.5,0) -- (-3.5,-3.5) -- cycle;
\draw [fill,gray!80!white] (-3.5,-3.5) rectangle (3.5,0);
\end{scope}
\draw[<->] (-5,0) -- (5,0);
\node [right] at (5,0) {Re(s)};
\draw[<->] (0,-5) -- (0,5);
\node [above] at (0,5) {Im(s)};
\draw [-,decorate,decoration=zigzag,very thick] (1,0) -- (-4.9,0); 
\draw [dashed,line width=2] (3.50734, 5.) -- (3.45431, 4.88288) -- (3.4165, 4.79778) -- (3.34515, 
4.64286) -- (3.3051, 4.55204) -- (3.26428, 4.46429) -- (3.21429, 4.3544) -- 
(3.19266, 4.30734) -- (3.18232, 4.28571) -- (3.15428, 4.22571) -- (3.07988, 
4.06298) -- (3.03571, 3.96857) -- (3.01677, 3.92857) -- (2.96638, 3.81933) -- 
(2.93297, 3.75) -- (2.85714, 3.58829) -- (2.85171, 3.57686) -- (2.84901, 
3.57143) -- (2.84138, 3.55567) -- (2.73701, 3.33442) -- (2.67742, 3.21429) -- 
(2.6212, 3.09309) -- (2.51146, 2.8686) -- (2.50602, 2.85714) -- (2.50413, 
2.85301) -- (2.5, 2.84446) -- (2.4182, 2.67857) -- (2.38714, 2.61286) -- 
(2.32851, 2.5) -- (2.26875, 2.37411) -- (2.16039, 2.16039) -- (2.15181, 
2.14286) -- (2.14898, 2.13673) -- (2.14286, 2.12443) -- (2.06076, 1.96429) -- 
(2.02937, 1.8992) -- (1.96998, 1.78571) -- (1.96429, 1.77408) -- (1.90815, 
1.66328) -- (1.87697, 1.60714) -- (1.78571, 1.43051) -- (1.78504, 1.42925) -- 
(1.78463, 1.42857) -- (1.7833, 1.42616) -- (1.66264, 1.19451) -- (1.60714, 
1.09245) -- (1.59498, 1.07143) -- (1.5807, 1.04498) -- (1.53727, 0.962729) -- 
(1.49747, 0.892857) -- (1.42857, 0.760484) -- (1.41154, 0.731314) -- 
(1.40127, 0.714286) -- (1.36742, 0.653139) -- (1.34807, 0.616212) -- 
(1.30078, 0.535714) -- (1.28484, 0.500871) -- (1.25336, 0.446429) -- (1.25, 
0.439713) -- (1.22097, 0.386171) -- (1.2023, 0.357143) -- (1.16071, 
0.280141) -- (1.15596, 0.272607) -- (1.15258, 0.267857) -- (1.14196, 
0.249098) -- (1.12375, 0.215534) -- (1.11607, 0.201678) -- (1.10189, 
0.178571) -- (1.09166, 0.158343) -- (1.07661, 0.133929) -- (1.07143, 
0.123167) -- (1.05933, 0.101384) -- (1.05018, 0.0892857) -- (1.02679, 
0.0460447) -- (1.02622, 0.0452116) -- (1.02378, 0.0416322) -- (1.01532, 
0.0233344); 
\draw [dashed,line width=2] (3.50734, -5.) -- (3.45431, -4.88288) -- (3.4165, -4.79778) -- (3.34515, 
-4.64286) -- (3.3051, -4.55204) -- (3.26428, -4.46429) -- (3.21429, 
-4.3544) -- (3.19266, -4.30734) -- (3.18232, -4.28571) -- (3.15428, 
-4.22571) -- (3.07988, -4.06298) -- (3.03571, -3.96857) -- (3.01677, 
-3.92857) -- (2.96638, -3.81933) -- (2.93297, -3.75) -- (2.85714, 
-3.58829) -- (2.85171, -3.57686) -- (2.84901, -3.57143) -- (2.84138, 
-3.55567) -- (2.73701, -3.33442) -- (2.67742, -3.21429) -- (2.6212, 
-3.09309) -- (2.51146, -2.8686) -- (2.50602, -2.85714) -- (2.50413, 
-2.85301) -- (2.5, -2.84446) -- (2.4182, -2.67857) -- (2.38714, -2.61286) -- 
(2.32851, -2.5) -- (2.26875, -2.37411) -- (2.16039, -2.16039) -- (2.15181, 
-2.14286) -- (2.14898, -2.13673) -- (2.14286, -2.12443) -- (2.06076, 
-1.96429) -- (2.02937, -1.8992) -- (1.96998, -1.78571) -- (1.96429, 
-1.77408) -- (1.90815, -1.66328) -- (1.87697, -1.60714) -- (1.78571, 
-1.43051) -- (1.78504, -1.42925) -- (1.78463, -1.42857) -- (1.7834, 
-1.42626) -- (1.72369, -1.31203) -- (1.68935, -1.25) -- (1.66189, 
-1.19525) -- (1.60714, -1.09245) -- (1.59498, -1.07143) -- (1.5807, 
-1.04498) -- (1.53727, -0.962729) -- (1.49747, -0.892857) -- (1.42857, 
-0.760484) -- (1.41154, -0.731314) -- (1.40127, -0.714286) -- (1.36742, 
-0.653139) -- (1.34807, -0.616212) -- (1.30078, -0.535714) -- (1.28484, 
-0.500871) -- (1.25336, -0.446429) -- (1.25, -0.439713) -- (1.22097, 
-0.386171) -- (1.2023, -0.357143) -- (1.16071, -0.280141) -- (1.15596, 
-0.272607) -- (1.15258, -0.267857) -- (1.14196, -0.249098) -- (1.12375, 
-0.215534) -- (1.11607, -0.201678) -- (1.10189, -0.178571) -- (1.09166, 
-0.158343) -- (1.07661, -0.133929) -- (1.07143, -0.123167) -- (1.05933, 
-0.101384) -- (1.05018, -0.0892857) -- (1.02679, -0.0460447) -- (1.02622, 
-0.0452116) -- (1.02541, -0.0445882) -- (1.02378, -0.0416355) -- (1.01536, 
-0.0251399); 

\draw [<->,thick,black] (2.2,-2.08) arc (-60:60:2.4); 
\node [right] at (3,1.5) {Valid};
\node [right] at (-4,2.5) {$\chi_2$: exp. large};
\node [right] at (-4,-2.5) {$\chi_2$: exp. large};
\draw [fill=gray!80!white] (-5.5,4.4) rectangle (-5,4.9);
\node [right] at (-5,4.65) {Exp. Large Contribution};
\draw [fill=gray!30!white] (-5.5,3.6) rectangle (-5,4.1);
\node [right] at (-5,3.85) {Exp. Small Contribution};
\end{tikzpicture}}
\subcaption{Regions of validity.}\label{fig:2b}
\end{minipage}
\caption{These figures depict the Stokes structure for parameter values  $\alpha =1$ and $\beta =1$. Figure \ref{fig:2a} shows the switching behaviour of the Stokes multiplier, $\mathcal{S}_i$, as Stokes curves are crossed. Figure \ref{fig:2b} illustrates the regions of validity for the general asymptotic solution \eqref{COMPLETE OPTIMALLY TRUNC ASYM SERIES} with $\mathcal{S}_1=0$ and a free parameter, $\mathcal{S}_2$. The dark gray regions depict where exponentially-large terms are present, whereas the light gray regions indicate the presence of exponentially-small terms present. The asymptotic solution is therefore valid in the region bounded by the anti-Stokes curves containing the positive real axis. Elsewhere, it will be exponentially dominated and will therefore no longer be a valid asymptotic approximation.}
\end{figure}
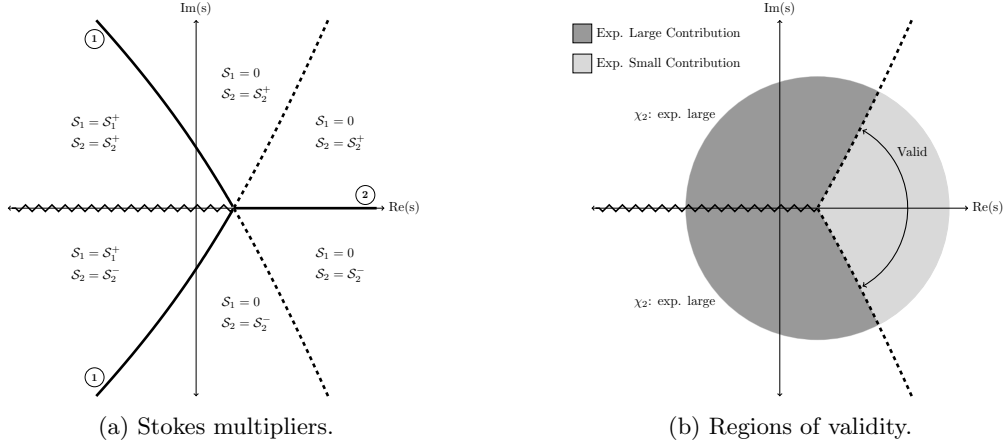
$\mathcal{S}_2$ is non-zero on either side of the positive real axis, then the asymptotic series \eqref{COMPLETE OPTIMALLY TRUNC ASYM SERIES} is valid in the region bounded by the anti-Stokes curves containing the positive real axis and contains exponentially-small contributions; this is illustrated in Figure \ref{fig:2b}.

We may repeat the process for the remaining five critical curves in order to obtain other types of asymptotic solutions with different ranges of validity. This results in the determination of two types of asymptotic solution classes. Type one solutions describe those in which the asymptotic expansion is valid within some region which contain a free parameter hidden beyond all orders. However, for special choices of the free parameter of type one solutions, we can obtain asymptotic solutions with an extended range of validity; these are referred to as type two solutions. Type two asymptotic solutions are illustrated in Figure 3.
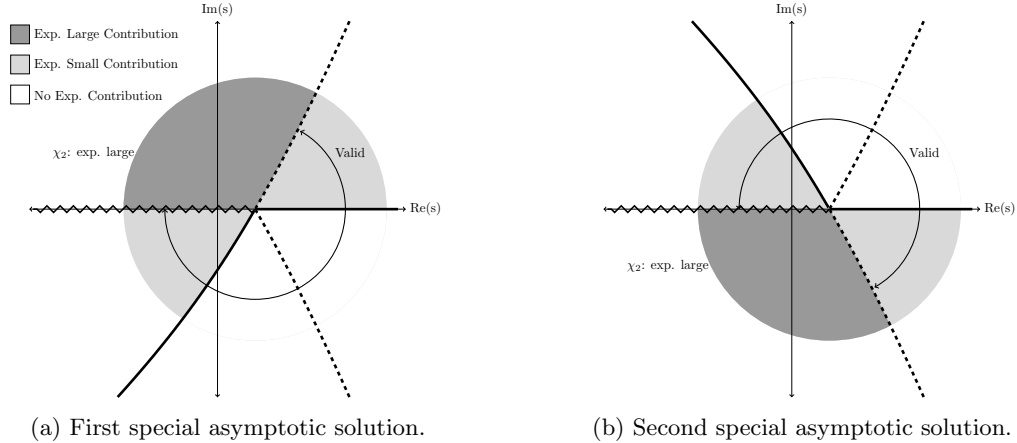
\begin{figure}[H]
\begin{minipage}{0.5\linewidth} \centering
\scalebox{0.5}{\begin{tikzpicture}{center}
\fill [gray!30!white] (1,0) circle (3.5);
\begin{scope}
\clip (1,0) circle (3.5);
\clip (-2.6549, -5.) -- (-2.58148, -4.91852) -- (-2.5, -4.82719) -- (-2.49477, 
-4.82143) -- (-2.41313, -4.72973) -- (-2.33654, -4.64286) -- (-2.32143, 
-4.62536) -- (-2.24606, -4.53966) -- (-2.18039, -4.46429) -- (-2.14286, 
-4.42024) -- (-2.08032, -4.34826) -- (-2.02645, -4.28571) -- (-1.96429, 
-4.21179) -- (-1.91595, -4.15548) -- (-1.87479, -4.10714) -- (-1.78571, 
-3.99981) -- (-1.75302, -3.96127) -- (-1.72476, -3.92857) -- (-1.59156, 
-3.76558) -- (-1.45722, -3.60008) -- (-1.43402, -3.57143) -- (-1.43162, 
-3.56838) -- (-1.42857, -3.56446) -- (-1.2912, -3.39286) -- (-1.27323, 
-3.36963) -- (-1.25, -3.33944) -- (-1.15087, -3.21429) -- (-1.11649, 
-3.16923) -- (-1.07143, -3.11001) -- (-1.01311, -3.03571) -- (-0.961437, 
-2.96713) -- (-0.892857, -2.87615) -- (-0.878049, -2.85714) -- (-0.808112, 
-2.76332) -- (-0.744704, -2.67857) -- (-0.714286, -2.63591) -- (-0.65653, 
-2.55776) -- (-0.613507, -2.5) -- (-0.535714, -2.38995) -- (-0.50669, 
-2.35045) -- (-0.365074, -2.15079) -- (-0.35951, -2.14286) -- (-0.358571, 
-2.14143) -- (-0.357143, -2.13929) -- (-0.234589, -1.96429) -- (-0.212474, 
-1.93038) -- (-0.178571, -1.87962) -- (-0.112545, -1.78571) -- (-0.0682017, 
-1.71751) -- (6.66134*10^-16, -1.61601) -- (0.00630805, -1.60714) -- 
(0.0743253, -1.5029) -- (0.125046, -1.42857) -- (0.178571, -1.34219) -- 
(0.21553, -1.28696) -- (0.241031, -1.25) -- (0.28536, -1.17822) -- 
(0.347271, -1.0813) -- (0.35353, -1.07143) -- (0.354819, -1.0691) -- 
(0.357143, -1.06529) -- (0.41007, -0.982143) -- (0.423592, -0.959306) -- 
(0.446429, -0.922488) -- (0.465711, -0.892857) -- (0.492004, -0.849147) -- 
(0.520984, -0.803571) -- (0.535714, -0.778328) -- (0.560119, -0.738691) -- 
(0.576189, -0.714286) -- (0.625, -0.632818) -- (0.628072, -0.628072) -- 
(0.630146, -0.625) -- (0.638065, -0.611935) -- (0.66173, -0.572444) -- 
(0.684784, -0.535714) -- (0.695045, -0.516474) -- (0.714286, -0.483871) -- 
(0.738456, -0.446429) -- (0.762251, -0.405108) -- (0.77289, -0.387825) -- 
(0.791133, -0.357143) -- (0.795487, -0.349058) -- (0.803571, -0.335187) -- 
(0.817658, -0.3125) -- (0.828571, -0.292856) -- (0.843867, -0.267857) -- 
(0.848214, -0.259794) -- (0.861662, -0.236662) -- (0.870496, -0.223214) -- 
(0.892857, -0.183834) -- (0.894877, -0.180592) -- (0.89636, -0.178571) -- 
(0.902062, -0.169366) -- (0.927481, -0.123909) -- (0.952603, -0.0892857) -- 
(0.95952, -0.0666624) -- (0.976384, -0.0428631) -- (1,0) -- (4.5,0) -- (4.5,-4.5) -- cycle; 
\fill [white] (-4.5,-4.5) rectangle (4.5,4.5);
\end{scope}
\begin{scope}
\clip (1,0) circle (3.5);
\clip (3.50734, 5.) -- (3.45431, 4.88288) -- (3.4165, 4.79778) -- (3.34515, 
4.64286) -- (3.3051, 4.55204) -- (3.26428, 4.46429) -- (3.21429, 4.3544) -- 
(3.19266, 4.30734) -- (3.18232, 4.28571) -- (3.15428, 4.22571) -- (3.07988, 
4.06298) -- (3.03571, 3.96857) -- (3.01677, 3.92857) -- (2.96638, 3.81933) -- 
(2.93297, 3.75) -- (2.85714, 3.58829) -- (2.85171, 3.57686) -- (2.84901, 
3.57143) -- (2.84138, 3.55567) -- (2.73701, 3.33442) -- (2.67742, 3.21429) -- 
(2.6212, 3.09309) -- (2.51146, 2.8686) -- (2.50602, 2.85714) -- (2.50413, 
2.85301) -- (2.5, 2.84446) -- (2.4182, 2.67857) -- (2.38714, 2.61286) -- 
(2.32851, 2.5) -- (2.26875, 2.37411) -- (2.16039, 2.16039) -- (2.15181, 
2.14286) -- (2.14898, 2.13673) -- (2.14286, 2.12443) -- (2.06076, 1.96429) -- 
(2.02937, 1.8992) -- (1.96998, 1.78571) -- (1.96429, 1.77408) -- (1.90815, 
1.66328) -- (1.87697, 1.60714) -- (1.78571, 1.43051) -- (1.78504, 1.42925) -- 
(1.78463, 1.42857) -- (1.7833, 1.42616) -- (1.66264, 1.19451) -- (1.60714, 
1.09245) -- (1.59498, 1.07143) -- (1.5807, 1.04498) -- (1.53727, 0.962729) -- 
(1.49747, 0.892857) -- (1.42857, 0.760484) -- (1.41154, 0.731314) -- 
(1.40127, 0.714286) -- (1.36742, 0.653139) -- (1.34807, 0.616212) -- 
(1.30078, 0.535714) -- (1.28484, 0.500871) -- (1.25336, 0.446429) -- (1.25, 
0.439713) -- (1.22097, 0.386171) -- (1.2023, 0.357143) -- (1.16071, 
0.280141) -- (1.15596, 0.272607) -- (1.15258, 0.267857) -- (1.14196, 
0.249098) -- (1.12375, 0.215534) -- (1.11607, 0.201678) -- (1.10189, 
0.178571) -- (1.09166, 0.158343) -- (1.07661, 0.133929) -- (1.07143, 
0.123167) -- (1.05933, 0.101384) -- (1.05018, 0.0892857) -- (1.02679, 
0.0460447) -- (1.02622, 0.0452116) -- (1.02378, 0.0416322) -- (1.01532, 
0.0233344) -- (1,0) -- (-4.5,0) -- (-4.5,4.5) -- cycle;
\fill [gray!80!white] (-4.5,-4.5) rectangle (4.5,4.5);
\end{scope}
\draw[<->] (-5,0) -- (5,0);
\node [right] at (5,0) {Re(s)};
\draw[<->] (0,-5) -- (0,5);
\node [above] at (0,5) {Im(s)};
\draw [-,decorate,decoration=zigzag,very thick] (1,0) -- (-4.9,0); 
\draw [dashed,line width=2] (3.50734, 5.) -- (3.45431, 4.88288) -- (3.4165, 4.79778) -- (3.34515, 
4.64286) -- (3.3051, 4.55204) -- (3.26428, 4.46429) -- (3.21429, 4.3544) -- 
(3.19266, 4.30734) -- (3.18232, 4.28571) -- (3.15428, 4.22571) -- (3.07988, 
4.06298) -- (3.03571, 3.96857) -- (3.01677, 3.92857) -- (2.96638, 3.81933) -- 
(2.93297, 3.75) -- (2.85714, 3.58829) -- (2.85171, 3.57686) -- (2.84901, 
3.57143) -- (2.84138, 3.55567) -- (2.73701, 3.33442) -- (2.67742, 3.21429) -- 
(2.6212, 3.09309) -- (2.51146, 2.8686) -- (2.50602, 2.85714) -- (2.50413, 
2.85301) -- (2.5, 2.84446) -- (2.4182, 2.67857) -- (2.38714, 2.61286) -- 
(2.32851, 2.5) -- (2.26875, 2.37411) -- (2.16039, 2.16039) -- (2.15181, 
2.14286) -- (2.14898, 2.13673) -- (2.14286, 2.12443) -- (2.06076, 1.96429) -- 
(2.02937, 1.8992) -- (1.96998, 1.78571) -- (1.96429, 1.77408) -- (1.90815, 
1.66328) -- (1.87697, 1.60714) -- (1.78571, 1.43051) -- (1.78504, 1.42925) -- 
(1.78463, 1.42857) -- (1.7833, 1.42616) -- (1.66264, 1.19451) -- (1.60714, 
1.09245) -- (1.59498, 1.07143) -- (1.5807, 1.04498) -- (1.53727, 0.962729) -- 
(1.49747, 0.892857) -- (1.42857, 0.760484) -- (1.41154, 0.731314) -- 
(1.40127, 0.714286) -- (1.36742, 0.653139) -- (1.34807, 0.616212) -- 
(1.30078, 0.535714) -- (1.28484, 0.500871) -- (1.25336, 0.446429) -- (1.25, 
0.439713) -- (1.22097, 0.386171) -- (1.2023, 0.357143) -- (1.16071, 
0.280141) -- (1.15596, 0.272607) -- (1.15258, 0.267857) -- (1.14196, 
0.249098) -- (1.12375, 0.215534) -- (1.11607, 0.201678) -- (1.10189, 
0.178571) -- (1.09166, 0.158343) -- (1.07661, 0.133929) -- (1.07143, 
0.123167) -- (1.05933, 0.101384) -- (1.05018, 0.0892857) -- (1.02679, 
0.0460447) -- (1.02622, 0.0452116) -- (1.02378, 0.0416322) -- (1.01532, 
0.0233344); 
\draw [dashed,line width=2] (3.50734, -5.) -- (3.45431, -4.88288) -- (3.4165, -4.79778) -- (3.34515, 
-4.64286) -- (3.3051, -4.55204) -- (3.26428, -4.46429) -- (3.21429, 
-4.3544) -- (3.19266, -4.30734) -- (3.18232, -4.28571) -- (3.15428, 
-4.22571) -- (3.07988, -4.06298) -- (3.03571, -3.96857) -- (3.01677, 
-3.92857) -- (2.96638, -3.81933) -- (2.93297, -3.75) -- (2.85714, 
-3.58829) -- (2.85171, -3.57686) -- (2.84901, -3.57143) -- (2.84138, 
-3.55567) -- (2.73701, -3.33442) -- (2.67742, -3.21429) -- (2.6212, 
-3.09309) -- (2.51146, -2.8686) -- (2.50602, -2.85714) -- (2.50413, 
-2.85301) -- (2.5, -2.84446) -- (2.4182, -2.67857) -- (2.38714, -2.61286) -- 
(2.32851, -2.5) -- (2.26875, -2.37411) -- (2.16039, -2.16039) -- (2.15181, 
-2.14286) -- (2.14898, -2.13673) -- (2.14286, -2.12443) -- (2.06076, 
-1.96429) -- (2.02937, -1.8992) -- (1.96998, -1.78571) -- (1.96429, 
-1.77408) -- (1.90815, -1.66328) -- (1.87697, -1.60714) -- (1.78571, 
-1.43051) -- (1.78504, -1.42925) -- (1.78463, -1.42857) -- (1.7834, 
-1.42626) -- (1.72369, -1.31203) -- (1.68935, -1.25) -- (1.66189, 
-1.19525) -- (1.60714, -1.09245) -- (1.59498, -1.07143) -- (1.5807, 
-1.04498) -- (1.53727, -0.962729) -- (1.49747, -0.892857) -- (1.42857, 
-0.760484) -- (1.41154, -0.731314) -- (1.40127, -0.714286) -- (1.36742, 
-0.653139) -- (1.34807, -0.616212) -- (1.30078, -0.535714) -- (1.28484, 
-0.500871) -- (1.25336, -0.446429) -- (1.25, -0.439713) -- (1.22097, 
-0.386171) -- (1.2023, -0.357143) -- (1.16071, -0.280141) -- (1.15596, 
-0.272607) -- (1.15258, -0.267857) -- (1.14196, -0.249098) -- (1.12375, 
-0.215534) -- (1.11607, -0.201678) -- (1.10189, -0.178571) -- (1.09166, 
-0.158343) -- (1.07661, -0.133929) -- (1.07143, -0.123167) -- (1.05933, 
-0.101384) -- (1.05018, -0.0892857) -- (1.02679, -0.0460447) -- (1.02622, 
-0.0452116) -- (1.02541, -0.0445882) -- (1.02378, -0.0416355) -- (1.01536, 
-0.0251399); 
\draw[line width=2] (-2.6549, -5.) -- (-2.58148, -4.91852) -- (-2.5, -4.82719) -- (-2.49477, 
-4.82143) -- (-2.41313, -4.72973) -- (-2.33654, -4.64286) -- (-2.32143, 
-4.62536) -- (-2.24606, -4.53966) -- (-2.18039, -4.46429) -- (-2.14286, 
-4.42024) -- (-2.08032, -4.34826) -- (-2.02645, -4.28571) -- (-1.96429, 
-4.21179) -- (-1.91595, -4.15548) -- (-1.87479, -4.10714) -- (-1.78571, 
-3.99981) -- (-1.75302, -3.96127) -- (-1.72476, -3.92857) -- (-1.59156, 
-3.76558) -- (-1.45722, -3.60008) -- (-1.43402, -3.57143) -- (-1.43162, 
-3.56838) -- (-1.42857, -3.56446) -- (-1.2912, -3.39286) -- (-1.27323, 
-3.36963) -- (-1.25, -3.33944) -- (-1.15087, -3.21429) -- (-1.11649, 
-3.16923) -- (-1.07143, -3.11001) -- (-1.01311, -3.03571) -- (-0.961437, 
-2.96713) -- (-0.892857, -2.87615) -- (-0.878049, -2.85714) -- (-0.808112, 
-2.76332) -- (-0.744704, -2.67857) -- (-0.714286, -2.63591) -- (-0.65653, 
-2.55776) -- (-0.613507, -2.5) -- (-0.535714, -2.38995) -- (-0.50669, 
-2.35045) -- (-0.365074, -2.15079) -- (-0.35951, -2.14286) -- (-0.358571, 
-2.14143) -- (-0.357143, -2.13929) -- (-0.234589, -1.96429) -- (-0.212474, 
-1.93038) -- (-0.178571, -1.87962) -- (-0.112545, -1.78571) -- (-0.0682017, 
-1.71751) -- (6.66134*10^-16, -1.61601) -- (0.00630805, -1.60714) -- 
(0.0743253, -1.5029) -- (0.125046, -1.42857) -- (0.178571, -1.34219) -- 
(0.21553, -1.28696) -- (0.241031, -1.25) -- (0.28536, -1.17822) -- 
(0.347271, -1.0813) -- (0.35353, -1.07143) -- (0.354819, -1.0691) -- 
(0.357143, -1.06529) -- (0.41007, -0.982143) -- (0.423592, -0.959306) -- 
(0.446429, -0.922488) -- (0.465711, -0.892857) -- (0.492004, -0.849147) -- 
(0.520984, -0.803571) -- (0.535714, -0.778328) -- (0.560119, -0.738691) -- 
(0.576189, -0.714286) -- (0.625, -0.632818) -- (0.628072, -0.628072) -- 
(0.630146, -0.625) -- (0.638065, -0.611935) -- (0.66173, -0.572444) -- 
(0.684784, -0.535714) -- (0.695045, -0.516474) -- (0.714286, -0.483871) -- 
(0.738456, -0.446429) -- (0.762251, -0.405108) -- (0.77289, -0.387825) -- 
(0.791133, -0.357143) -- (0.795487, -0.349058) -- (0.803571, -0.335187) -- 
(0.817658, -0.3125) -- (0.828571, -0.292856) -- (0.843867, -0.267857) -- 
(0.848214, -0.259794) -- (0.861662, -0.236662) -- (0.870496, -0.223214) -- 
(0.892857, -0.183834) -- (0.894877, -0.180592) -- (0.89636, -0.178571) -- 
(0.902062, -0.169366) -- (0.927481, -0.123909) -- (0.952603, -0.0892857) -- 
(0.95952, -0.0666624) -- (0.976384, -0.0428631); 
\draw[line width=2] (1,0) -- (4.8,0); 

\draw [<->,thick,black] (-1.4,0) arc (-180:60:2.4); 
\node [right] at (3,1.5) {Valid};
\node [right] at (-4.5,1.5) {$\chi_2$: exp. large};
\draw [fill=gray!80!white] (-5.5,4.4) rectangle (-5,4.9);
\node [right] at (-5,4.65) {Exp. Large Contribution};
\draw [fill=gray!30!white] (-5.5,3.6) rectangle (-5,4.1);
\node [right] at (-5,3.85) {Exp. Small Contribution};
\draw [fill=white] (-5.5,2.8) rectangle (-5,3.3);
\node [right] at (-5,3) {No Exp. Contribution};
\end{tikzpicture}}
\subcaption{First special asymptotic solution.}\label{fig:3a}
\end{minipage}
\begin{minipage}{0.5\linewidth} 
\centering
\scalebox{0.5}{\begin{tikzpicture}{center}
\fill [gray!30!white] (1,0) circle (3.5);
\begin{scope}
\clip (1,0) circle (3.5);
\clip (-2.6549, 5.) -- (-2.58148, 4.91852) -- (-2.5, 4.82719) -- (-2.49477, 
4.82143) -- (-2.41313, 4.72973) -- (-2.33654, 4.64286) -- (-2.32143, 
4.62536) -- (-2.24606, 4.53966) -- (-2.18039, 4.46429) -- (-2.14286, 
4.42024) -- (-2.08032, 4.34826) -- (-2.02645, 4.28571) -- (-1.96429, 
4.21179) -- (-1.91595, 4.15548) -- (-1.87479, 4.10714) -- (-1.78571, 
3.99981) -- (-1.75302, 3.96127) -- (-1.72476, 3.92857) -- (-1.59156, 
3.76558) -- (-1.45722, 3.60008) -- (-1.43402, 3.57143) -- (-1.43162, 
3.56838) -- (-1.42857, 3.56446) -- (-1.2912, 3.39286) -- (-1.27323, 
3.36963) -- (-1.25, 3.33944) -- (-1.15087, 3.21429) -- (-1.11649, 3.16923) -- 
(-1.07143, 3.11001) -- (-1.01311, 3.03571) -- (-0.961437, 2.96713) -- 
(-0.892857, 2.87615) -- (-0.878049, 2.85714) -- (-0.808112, 2.76332) -- 
(-0.744704, 2.67857) -- (-0.714286, 2.63591) -- (-0.65653, 2.55776) -- 
(-0.613507, 2.5) -- (-0.535714, 2.38995) -- (-0.50669, 2.35045) -- 
(-0.365074, 2.15079) -- (-0.35951, 2.14286) -- (-0.358571, 2.14143) -- 
(-0.357143, 2.13929) -- (-0.234589, 1.96429) -- (-0.212474, 1.93038) -- 
(-0.178571, 1.87962) -- (-0.112545, 1.78571) -- (-0.0682017, 1.71751) -- (0.00630805, 1.60714) -- (0.0743253, 
1.5029) -- (0.125046, 1.42857) -- (0.178571, 1.34219) -- (0.21553, 
1.28696) -- (0.241031, 1.25) -- (0.28536, 1.17822) -- (0.347271, 1.0813) -- 
(0.35353, 1.07143) -- (0.354819, 1.0691) -- (0.357143, 1.06529) -- 
(0.41007, 0.982143) -- (0.423592, 0.959306) -- (0.446429, 0.922488) -- 
(0.465711, 0.892857) -- (0.492004, 0.849147) -- (0.520984, 0.803571) -- 
(0.535714, 0.778328) -- (0.560119, 0.738691) -- (0.576189, 0.714286) -- 
(0.625, 0.632818) -- (0.628042, 0.628042) -- (0.63866, 0.61134) -- 
(0.684784, 0.535714) -- (0.695045, 0.516474) -- (0.714286, 0.483871) -- 
(0.738456, 0.446429) -- (0.762251, 0.405108) -- (0.77289, 0.387825) -- 
(0.791472, 0.357143) -- (0.803571, 0.333968) -- (0.827991, 0.292276) -- 
(0.845383, 0.267857) -- (0.892857, 0.183451) -- (0.894877, 0.180592) -- 
(0.89636, 0.178571) -- (0.902062, 0.169366) -- (0.927481, 0.123909) -- 
(0.952603, 0.0892857) -- (0.95952, 0.0666632) -- (0.976398, 0.0428371) -- (1,0) -- (5,0) -- (5,4.5) -- cycle;
\fill [white] (-4.5,-4.5) rectangle (4.5,4.5);
\end{scope}
\begin{scope}
\clip (1,0) circle (3.5);
\clip (3.50734, -5.) -- (3.45431, -4.88288) -- (3.4165, -4.79778) -- (3.34515, 
-4.64286) -- (3.3051, -4.55204) -- (3.26428, -4.46429) -- (3.21429, 
-4.3544) -- (3.19266, -4.30734) -- (3.18232, -4.28571) -- (3.15428, 
-4.22571) -- (3.07988, -4.06298) -- (3.03571, -3.96857) -- (3.01677, 
-3.92857) -- (2.96638, -3.81933) -- (2.93297, -3.75) -- (2.85714, 
-3.58829) -- (2.85171, -3.57686) -- (2.84901, -3.57143) -- (2.84138, 
-3.55567) -- (2.73701, -3.33442) -- (2.67742, -3.21429) -- (2.6212, 
-3.09309) -- (2.51146, -2.8686) -- (2.50602, -2.85714) -- (2.50413, 
-2.85301) -- (2.5, -2.84446) -- (2.4182, -2.67857) -- (2.38714, -2.61286) -- 
(2.32851, -2.5) -- (2.26875, -2.37411) -- (2.16039, -2.16039) -- (2.15181, 
-2.14286) -- (2.14898, -2.13673) -- (2.14286, -2.12443) -- (2.06076, 
-1.96429) -- (2.02937, -1.8992) -- (1.96998, -1.78571) -- (1.96429, 
-1.77408) -- (1.90815, -1.66328) -- (1.87697, -1.60714) -- (1.78571, 
-1.43051) -- (1.78504, -1.42925) -- (1.78463, -1.42857) -- (1.7834, 
-1.42626) -- (1.72369, -1.31203) -- (1.68935, -1.25) -- (1.66189, 
-1.19525) -- (1.60714, -1.09245) -- (1.59498, -1.07143) -- (1.5807, 
-1.04498) -- (1.53727, -0.962729) -- (1.49747, -0.892857) -- (1.42857, 
-0.760484) -- (1.41154, -0.731314) -- (1.40127, -0.714286) -- (1.36742, 
-0.653139) -- (1.34807, -0.616212) -- (1.30078, -0.535714) -- (1.28484, 
-0.500871) -- (1.25336, -0.446429) -- (1.25, -0.439713) -- (1.22097, 
-0.386171) -- (1.2023, -0.357143) -- (1.16071, -0.280141) -- (1.15596, 
-0.272607) -- (1.15258, -0.267857) -- (1.14196, -0.249098) -- (1.12375, 
-0.215534) -- (1.11607, -0.201678) -- (1.10189, -0.178571) -- (1.09166, 
-0.158343) -- (1.07661, -0.133929) -- (1.07143, -0.123167) -- (1.05933, 
-0.101384) -- (1.05018, -0.0892857) -- (1.02679, -0.0460447) -- (1.02622, 
-0.0452116) -- (1.02541, -0.0445882) -- (1.02378, -0.0416355) -- (1.01536, 
-0.0251399) -- (1,0) -- (-4.5,0) -- (-4.5,-4.5) -- cycle;
\fill [gray!80!white] (-4.5,-4.5) rectangle (4.5,4.5);
\end{scope}
\draw[<->] (-5,0) -- (5,0);
\node [right] at (5,0) {Re(s)};
\draw[<->] (0,-5) -- (0,5);
\node [above] at (0,5) {Im(s)};
\draw [-,decorate,decoration=zigzag,very thick] (1,0) -- (-4.9,0); 
\draw [dashed,line width=2] (3.50734, 5.) -- (3.45431, 4.88288) -- (3.4165, 4.79778) -- (3.34515, 
4.64286) -- (3.3051, 4.55204) -- (3.26428, 4.46429) -- (3.21429, 4.3544) -- 
(3.19266, 4.30734) -- (3.18232, 4.28571) -- (3.15428, 4.22571) -- (3.07988, 
4.06298) -- (3.03571, 3.96857) -- (3.01677, 3.92857) -- (2.96638, 3.81933) -- 
(2.93297, 3.75) -- (2.85714, 3.58829) -- (2.85171, 3.57686) -- (2.84901, 
3.57143) -- (2.84138, 3.55567) -- (2.73701, 3.33442) -- (2.67742, 3.21429) -- 
(2.6212, 3.09309) -- (2.51146, 2.8686) -- (2.50602, 2.85714) -- (2.50413, 
2.85301) -- (2.5, 2.84446) -- (2.4182, 2.67857) -- (2.38714, 2.61286) -- 
(2.32851, 2.5) -- (2.26875, 2.37411) -- (2.16039, 2.16039) -- (2.15181, 
2.14286) -- (2.14898, 2.13673) -- (2.14286, 2.12443) -- (2.06076, 1.96429) -- 
(2.02937, 1.8992) -- (1.96998, 1.78571) -- (1.96429, 1.77408) -- (1.90815, 
1.66328) -- (1.87697, 1.60714) -- (1.78571, 1.43051) -- (1.78504, 1.42925) -- 
(1.78463, 1.42857) -- (1.7833, 1.42616) -- (1.66264, 1.19451) -- (1.60714, 
1.09245) -- (1.59498, 1.07143) -- (1.5807, 1.04498) -- (1.53727, 0.962729) -- 
(1.49747, 0.892857) -- (1.42857, 0.760484) -- (1.41154, 0.731314) -- 
(1.40127, 0.714286) -- (1.36742, 0.653139) -- (1.34807, 0.616212) -- 
(1.30078, 0.535714) -- (1.28484, 0.500871) -- (1.25336, 0.446429) -- (1.25, 
0.439713) -- (1.22097, 0.386171) -- (1.2023, 0.357143) -- (1.16071, 
0.280141) -- (1.15596, 0.272607) -- (1.15258, 0.267857) -- (1.14196, 
0.249098) -- (1.12375, 0.215534) -- (1.11607, 0.201678) -- (1.10189, 
0.178571) -- (1.09166, 0.158343) -- (1.07661, 0.133929) -- (1.07143, 
0.123167) -- (1.05933, 0.101384) -- (1.05018, 0.0892857) -- (1.02679, 
0.0460447) -- (1.02622, 0.0452116) -- (1.02378, 0.0416322) -- (1.01532, 
0.0233344); 
\draw [dashed,line width=2] (3.50734, -5.) -- (3.45431, -4.88288) -- (3.4165, -4.79778) -- (3.34515, 
-4.64286) -- (3.3051, -4.55204) -- (3.26428, -4.46429) -- (3.21429, 
-4.3544) -- (3.19266, -4.30734) -- (3.18232, -4.28571) -- (3.15428, 
-4.22571) -- (3.07988, -4.06298) -- (3.03571, -3.96857) -- (3.01677, 
-3.92857) -- (2.96638, -3.81933) -- (2.93297, -3.75) -- (2.85714, 
-3.58829) -- (2.85171, -3.57686) -- (2.84901, -3.57143) -- (2.84138, 
-3.55567) -- (2.73701, -3.33442) -- (2.67742, -3.21429) -- (2.6212, 
-3.09309) -- (2.51146, -2.8686) -- (2.50602, -2.85714) -- (2.50413, 
-2.85301) -- (2.5, -2.84446) -- (2.4182, -2.67857) -- (2.38714, -2.61286) -- 
(2.32851, -2.5) -- (2.26875, -2.37411) -- (2.16039, -2.16039) -- (2.15181, 
-2.14286) -- (2.14898, -2.13673) -- (2.14286, -2.12443) -- (2.06076, 
-1.96429) -- (2.02937, -1.8992) -- (1.96998, -1.78571) -- (1.96429, 
-1.77408) -- (1.90815, -1.66328) -- (1.87697, -1.60714) -- (1.78571, 
-1.43051) -- (1.78504, -1.42925) -- (1.78463, -1.42857) -- (1.7834, 
-1.42626) -- (1.72369, -1.31203) -- (1.68935, -1.25) -- (1.66189, 
-1.19525) -- (1.60714, -1.09245) -- (1.59498, -1.07143) -- (1.5807, 
-1.04498) -- (1.53727, -0.962729) -- (1.49747, -0.892857) -- (1.42857, 
-0.760484) -- (1.41154, -0.731314) -- (1.40127, -0.714286) -- (1.36742, 
-0.653139) -- (1.34807, -0.616212) -- (1.30078, -0.535714) -- (1.28484, 
-0.500871) -- (1.25336, -0.446429) -- (1.25, -0.439713) -- (1.22097, 
-0.386171) -- (1.2023, -0.357143) -- (1.16071, -0.280141) -- (1.15596, 
-0.272607) -- (1.15258, -0.267857) -- (1.14196, -0.249098) -- (1.12375, 
-0.215534) -- (1.11607, -0.201678) -- (1.10189, -0.178571) -- (1.09166, 
-0.158343) -- (1.07661, -0.133929) -- (1.07143, -0.123167) -- (1.05933, 
-0.101384) -- (1.05018, -0.0892857) -- (1.02679, -0.0460447) -- (1.02622, 
-0.0452116) -- (1.02541, -0.0445882) -- (1.02378, -0.0416355) -- (1.01536, 
-0.0251399); 
\draw[line width=2] (-2.6549, 5.) -- (-2.58148, 4.91852) -- (-2.5, 4.82719) -- (-2.49477, 
4.82143) -- (-2.41313, 4.72973) -- (-2.33654, 4.64286) -- (-2.32143, 
4.62536) -- (-2.24606, 4.53966) -- (-2.18039, 4.46429) -- (-2.14286, 
4.42024) -- (-2.08032, 4.34826) -- (-2.02645, 4.28571) -- (-1.96429, 
4.21179) -- (-1.91595, 4.15548) -- (-1.87479, 4.10714) -- (-1.78571, 
3.99981) -- (-1.75302, 3.96127) -- (-1.72476, 3.92857) -- (-1.59156, 
3.76558) -- (-1.45722, 3.60008) -- (-1.43402, 3.57143) -- (-1.43162, 
3.56838) -- (-1.42857, 3.56446) -- (-1.2912, 3.39286) -- (-1.27323, 
3.36963) -- (-1.25, 3.33944) -- (-1.15087, 3.21429) -- (-1.11649, 3.16923) -- 
(-1.07143, 3.11001) -- (-1.01311, 3.03571) -- (-0.961437, 2.96713) -- 
(-0.892857, 2.87615) -- (-0.878049, 2.85714) -- (-0.808112, 2.76332) -- 
(-0.744704, 2.67857) -- (-0.714286, 2.63591) -- (-0.65653, 2.55776) -- 
(-0.613507, 2.5) -- (-0.535714, 2.38995) -- (-0.50669, 2.35045) -- 
(-0.365074, 2.15079) -- (-0.35951, 2.14286) -- (-0.358571, 2.14143) -- 
(-0.357143, 2.13929) -- (-0.234589, 1.96429) -- (-0.212474, 1.93038) -- 
(-0.178571, 1.87962) -- (-0.112545, 1.78571) -- (-0.0682017, 1.71751) -- (0.00630805, 1.60714) -- (0.0743253, 
1.5029) -- (0.125046, 1.42857) -- (0.178571, 1.34219) -- (0.21553, 
1.28696) -- (0.241031, 1.25) -- (0.28536, 1.17822) -- (0.347271, 1.0813) -- 
(0.35353, 1.07143) -- (0.354819, 1.0691) -- (0.357143, 1.06529) -- 
(0.41007, 0.982143) -- (0.423592, 0.959306) -- (0.446429, 0.922488) -- 
(0.465711, 0.892857) -- (0.492004, 0.849147) -- (0.520984, 0.803571) -- 
(0.535714, 0.778328) -- (0.560119, 0.738691) -- (0.576189, 0.714286) -- 
(0.625, 0.632818) -- (0.628042, 0.628042) -- (0.63866, 0.61134) -- 
(0.684784, 0.535714) -- (0.695045, 0.516474) -- (0.714286, 0.483871) -- 
(0.738456, 0.446429) -- (0.762251, 0.405108) -- (0.77289, 0.387825) -- 
(0.791472, 0.357143) -- (0.803571, 0.333968) -- (0.827991, 0.292276) -- 
(0.845383, 0.267857) -- (0.892857, 0.183451) -- (0.894877, 0.180592) -- 
(0.89636, 0.178571) -- (0.902062, 0.169366) -- (0.927481, 0.123909) -- 
(0.952603, 0.0892857) -- (0.95952, 0.0666632) -- (0.976398, 0.0428371); 
\draw[line width=2] (1,0) -- (4.8,0); 

\draw [<->,thick,black] (2.2,-2.08) arc (-60:180:2.4); 
\node [right] at (3,1.5) {Valid};
\node [right] at (-4.5,-1.5) {$\chi_2$: exp. large};
\draw [white] (-5.5,4.4) rectangle (-5,4.9);
\end{tikzpicture}}
\subcaption{Second special asymptotic solution.}\label{fig:3b}
\end{minipage}
\caption{This figure illustrates special asymptotic solutions given in \eqref{COMPLETE OPTIMALLY TRUNC ASYM SERIES}, for $\alpha =1$ and $\beta =1$. Figure \ref{fig:3a} demonstrates that if we demand that the exponential contribution due to $\chi_2$ be inactive in the region below the real positive axis ($\mathcal{S}_2^-=0$) then the range of validity can be extended. This is also equivalent to specifying that our algebraic solution be valid about the lower anti-Stokes curve. Figure \ref{fig:3b} demonstrates the extended region of validity if $\chi_2$ is inactive in the region above the real positive axis ($\mathcal{S}_2^+=0$). This is also equivalent to specifying that the algebraic solution be valid about the upper anti-Stokes curve. Unshaded regions indicate regions in which there are no exponential contributions.}
\end{figure}
\begin{figure}[H]
\begin{minipage}{0.5\linewidth} 
\centering
\scalebox{0.5}{\begin{tikzpicture}{center}
\fill[gray!80!white] (0.707,-0.707) circle (3.5);
\begin{scope}
\clip (0.707,-0.707) circle (3.5);
\clip (5., 4.49464) -- (4.98739, 4.4769) -- (4.97835, 4.46429) -- (4.92492, 
4.38921) -- (4.91313, 4.37259) -- (4.85091, 4.28571) -- (4.82143, 4.24411) -- 
(4.76425, 4.16432) -- (4.7225, 4.10714) -- (4.64286, 3.99625) -- (4.61437, 
3.95706) -- (4.59294, 3.92857) -- (4.46365, 3.75063) -- (4.46004, 3.74575) -- 
(4.33158, 3.57143) -- (4.31234, 3.54481) -- (4.28571, 3.5087) -- (4.19902, 
3.39286) -- (4.16014, 3.33986) -- (4.10714, 3.26906) -- (4.06548, 3.21429) -- 
(4.00704, 3.13581) -- (3.93089, 3.03571) -- (3.92857, 3.03259) -- (3.853, 
2.93271) -- (3.79454, 2.85714) -- (3.75, 2.79819) -- (3.69798, 2.73059) -- 
(3.65702, 2.67857) -- (3.57143, 2.56709) -- (3.54196, 2.52947) -- (3.51832, 
2.5) -- (3.39286, 2.33936) -- (3.3849, 2.32939) -- (3.32218, 2.25076) -- 
(3.23663, 2.14286) -- (3.22691, 2.13023) -- (3.21429, 2.1142) -- (3.09312, 
1.96429) -- (3.06792, 1.93208) -- (3.03571, 1.8919) -- (2.94812, 1.78571) -- 
(2.9078, 1.73506) -- (2.85714, 1.67305) -- (2.80159, 1.60714) -- (2.7465, 
1.53922) -- (2.67857, 1.45769) -- (2.65345, 1.42857) -- (2.58398, 1.34459) -- 
(2.50352, 1.25) -- (2.5, 1.24566) -- (2.42022, 1.15121) -- (2.35085, 
1.07143) -- (2.32143, 1.03581) -- (2.25518, 0.959105) -- (2.19619, 
0.892857) -- (2.14286, 0.8294) -- (2.08882, 0.768322) -- (2.03946, 
0.714286) -- (1.96429, 0.626423) -- (1.92111, 0.578887) -- (1.88055, 
0.535714) -- (1.78571, 0.426858) -- (1.75202, 0.390833) -- (1.71935, 
0.357143) -- (1.60714, 0.230671) -- (1.5814, 0.204313) -- (1.55575, 
0.178571) -- (1.42857, 0.037803) -- (1.4095, 0.0190751) -- (1.38962, 
 6.66134*10^-16) -- (1.25, -0.151838) -- (1.23613, -0.164706) -- (1.22076, 
-0.178571) -- (1.07143, -0.338429) -- (1.0613, -0.347017) -- (1.05037, 
-0.357143) -- (0.982143, -0.429559) -- (0.973337, -0.437622) -- (0.963337, 
-0.446429) -- (0.892857, -0.521152) -- (0.885016, -0.527873) -- (0.875234, 
-0.535714) -- (0.803571, -0.612579) -- (0.796259, -0.617687) -- (0.784543, 
-0.625) -- (0.751864, -0.662578) -- (0.738044, -0.674735) -- (0.729396, 
-0.685352) -- (0.707,-0.707) -- (5,-0.707) -- cycle;
\fill [gray!30!white] (-5,-5) rectangle (5,5);
\end{scope}
\begin{scope}
\clip (0.707,-0.707) circle (3.5); 
\clip (1.73936, -5.) -- (1.72552, -4.9398) -- (1.68361, -4.74496) -- (1.65957,  
-4.64286) -- (1.62352, -4.48066) -- (1.61787, -4.45356) -- (1.60714,  
-4.40603) -- (1.57977, -4.28571) -- (1.55253, -4.16176) -- (1.5182,  
-4.0182) -- (1.49845, -3.92857) -- (1.48627, -3.87087) -- (1.45753, -3.75) --  
(1.42857, -3.62324) -- (1.41886, -3.58114) -- (1.41633, -3.57143) --  
(1.41223, -3.55509) -- (1.35215, -3.29071) -- (1.3322, -3.21429) --  
(1.30215, -3.08787) -- (1.28375, -3.00197) -- (1.25, -2.8634) -- (1.24845,  
-2.85714) -- (1.21509, -2.71348) -- (1.18966, -2.61823) -- (1.16131, -2.5) --  
(1.14626, -2.42517) -- (1.07629, -2.14772) -- (1.07527, -2.14286) --  
(1.07468, -2.1396) -- (1.07143, -2.12629) -- (1.03067, -1.96429) --  
(1.01688, -1.90973) -- (1.00394, -1.85321) -- (0.985733, -1.78571) --  
(0.968027, -1.71054) -- (0.956966, -1.67125) -- (0.940694, -1.60714) --  
(0.93238, -1.56762) -- (0.896779, -1.43249) -- (0.895922, -1.42857) --  
(0.895434, -1.42599) -- (0.892857, -1.41584) -- (0.872805, -1.33929) --  
(0.859888, -1.28297) -- (0.84959, -1.25) -- (0.832949, -1.19009) --  
(0.823088, -1.1412) -- (0.803571, -1.07151) -- (0.803553, -1.07145) --  
(0.803546, -1.07143) -- (0.803535, -1.07139) -- (0.785559, -1.00016) --  
(0.779475, -0.982143) -- (0.770823, -0.949395) -- (0.766833, -0.929595) --  
(0.755281, -0.892857) -- (0.748507, -0.858636) -- (0.737117, -0.826402) --  
(0.732253, -0.803571) -- (0.730726, -0.797217) -- (0.719664, -0.760578) --  
(0.714286, -0.729005) -- (0.711413, -0.721922) -- (0.707,-0.707) -- (5,1) -- (5,-4) -- cycle;
\fill [gray!30!white] (-5,-5) rectangle (5,5);
\end{scope}
\draw[<->] (-5,0) -- (5,0);
\node [right] at (5,0) {Re(s)};
\draw[<->] (0,-5) -- (0,5);
\node [above] at (0,5) {Im(s)};
\draw [-,decorate,decoration=zigzag,very thick] (0.707,-0.707) -- (-4.9,4.7); 
\draw [dashed,line width=2] (5., 4.49464) -- (4.98739, 4.4769) -- (4.97835, 4.46429) -- (4.92492, 
4.38921) -- (4.91313, 4.37259) -- (4.85091, 4.28571) -- (4.82143, 4.24411) -- 
(4.76425, 4.16432) -- (4.7225, 4.10714) -- (4.64286, 3.99625) -- (4.61437, 
3.95706) -- (4.59294, 3.92857) -- (4.46365, 3.75063) -- (4.46004, 3.74575) -- 
(4.33158, 3.57143) -- (4.31234, 3.54481) -- (4.28571, 3.5087) -- (4.19902, 
3.39286) -- (4.16014, 3.33986) -- (4.10714, 3.26906) -- (4.06548, 3.21429) -- 
(4.00704, 3.13581) -- (3.93089, 3.03571) -- (3.92857, 3.03259) -- (3.853, 
2.93271) -- (3.79454, 2.85714) -- (3.75, 2.79819) -- (3.69798, 2.73059) -- 
(3.65702, 2.67857) -- (3.57143, 2.56709) -- (3.54196, 2.52947) -- (3.51832, 
2.5) -- (3.39286, 2.33936) -- (3.3849, 2.32939) -- (3.32218, 2.25076) -- 
(3.23663, 2.14286) -- (3.22691, 2.13023) -- (3.21429, 2.1142) -- (3.09312, 
1.96429) -- (3.06792, 1.93208) -- (3.03571, 1.8919) -- (2.94812, 1.78571) -- 
(2.9078, 1.73506) -- (2.85714, 1.67305) -- (2.80159, 1.60714) -- (2.7465, 
1.53922) -- (2.67857, 1.45769) -- (2.65345, 1.42857) -- (2.58398, 1.34459) -- 
(2.50352, 1.25) -- (2.5, 1.24566) -- (2.42022, 1.15121) -- (2.35085, 
1.07143) -- (2.32143, 1.03581) -- (2.25518, 0.959105) -- (2.19619, 
0.892857) -- (2.14286, 0.8294) -- (2.08882, 0.768322) -- (2.03946, 
0.714286) -- (1.96429, 0.626423) -- (1.92111, 0.578887) -- (1.88055, 
0.535714) -- (1.78571, 0.426858) -- (1.75202, 0.390833) -- (1.71935, 
0.357143) -- (1.60714, 0.230671) -- (1.5814, 0.204313) -- (1.55575, 
0.178571) -- (1.42857, 0.037803) -- (1.4095, 0.0190751) -- (1.38962, 
 6.66134*10^-16) -- (1.25, -0.151838) -- (1.23613, -0.164706) -- (1.22076, 
-0.178571) -- (1.07143, -0.338429) -- (1.0613, -0.347017) -- (1.05037, 
-0.357143) -- (0.982143, -0.429559) -- (0.973337, -0.437622) -- (0.963337, 
-0.446429) -- (0.892857, -0.521152) -- (0.885016, -0.527873) -- (0.875234, 
-0.535714) -- (0.803571, -0.612579) -- (0.796259, -0.617687) -- (0.784543, 
-0.625) -- (0.751864, -0.662578) -- (0.738044, -0.674735) -- (0.729396, 
-0.685352)
; 
\draw [dashed,line width=2] (-5., 1.52244) -- (-4.9405, 1.48807) -- (-4.83713, 1.42857) -- (-4.82143, 
1.41938) -- (-4.7134, 1.35802) -- (-4.64286, 1.31803) -- (-4.51878, 1.25) -- 
(-4.48382, 1.23047) -- (-4.39816, 1.18387) -- (-4.28571, 1.12324) -- 
(-4.25159, 1.10556) -- (-4.18587, 1.07143) -- (-4.10714, 1.02963) -- 
(-4.01648, 0.98352) -- (-3.92857, 0.936963) -- (-3.77842, 0.864442) -- 
(-3.6673, 0.810161) -- (-3.57143, 0.764153) -- (-3.53712, 0.748593) -- 
(-3.46285, 0.714286) -- (-3.39286, 0.681072) -- (-3.29241, 0.63616) -- 
(-3.21429, 0.600292) -- (-3.06465, 0.535714) -- (-3.04461, 0.52682) -- 
(-3.01288, 0.512876) -- (-2.85714, 0.446789) -- (-2.79281, 0.421477) -- 
(-2.67857, 0.374211) -- (-2.63406, 0.357143) -- (-2.53799, 0.319151) -- 
(-2.5, 0.302353) -- (-2.41158, 0.268719) -- (-2.27969, 0.220306) -- 
(-2.16825, 0.178571) -- (-2.14286, 0.168746) -- (-2.12735, 0.163067) -- 
(-2.01871, 0.124147) -- (-1.96429, 0.103697) -- (-1.88705, 0.0772353) -- 
(-1.84931, 0.0635929) -- (-1.78571, 0.041185) -- (-1.75411, 0.0316032) -- 
(-1.66222, 
 6.66134*10^-16) -- (-1.60714, -0.019611) -- (-1.48679, -0.0582175) -- 
(-1.42857, -0.0796137) -- (-1.35328, -0.103277) -- (-1.31093, -0.117643) -- 
(-1.25, -0.137533) -- (-1.21817, -0.146745) -- (-1.16071, -0.165617) -- 
(-1.1189, -0.178571) -- (-1.08282, -0.189962) -- (-1.07143, -0.19389) -- 
(-1.0487, -0.201297) -- (-0.94656, -0.232275) -- (-0.892857, -0.250272) -- 
(-0.810297, -0.274583) -- (-0.803571, -0.276885) -- (-0.790523, 
-0.280905) -- (-0.741864, -0.295436) -- (-0.714286, -0.30417) -- 
(-0.673376, -0.316233) -- (-0.662498, -0.319645) -- (-0.625, -0.331027) -- 
(-0.536716, -0.357143) -- (-0.535947, -0.357376) -- (-0.535714, 
-0.357456) -- (-0.535253, -0.357604) -- (-0.397443, -0.397443) -- 
(-0.357143, -0.410314) -- (-0.328804, -0.41809) -- (-0.282287, 
-0.431998) -- (-0.267857, -0.436114) -- (-0.259649, -0.43822) -- 
(-0.230615, -0.446429) -- (-0.178571, -0.461921) -- (-0.156757, 
-0.468243) -- (-0.120887, -0.47803) -- (-0.0892857, -0.487637) -- 
(-0.0514527, -0.497881) -- (-0.0318833, -0.503831) -- (6.66134*10^-16, 
-0.512928) -- (0.0182527, -0.517462) -- (0.0824211, -0.535714) -- 
(0.0892857, -0.537809) -- (0.0922674, -0.538696) -- (0.133929, 
-0.550384) -- (0.157751, -0.556535) -- (0.178571, -0.562866) -- (0.216335, 
-0.573477) -- (0.227405, -0.576166) -- (0.267857, -0.588139) -- (0.297583, 
-0.595275) -- (0.340196, -0.608053) -- (0.357143, -0.612535) -- (0.407129, 
-0.625) -- (0.438155, -0.633274) -- (0.446429, -0.637053) -- (0.466022, 
-0.644593) -- (0.535714, -0.662326) -- (0.577983, -0.672017) -- (0.586154, 
-0.675439) -- (0.624668, -0.684989)
; 
\draw [dashed,line width=2] (1.73936, -5.) -- (1.72552, -4.9398) -- (1.68361, -4.74496) -- (1.65957,  
-4.64286) -- (1.62352, -4.48066) -- (1.61787, -4.45356) -- (1.60714,  
-4.40603) -- (1.57977, -4.28571) -- (1.55253, -4.16176) -- (1.5182,  
-4.0182) -- (1.49845, -3.92857) -- (1.48627, -3.87087) -- (1.45753, -3.75) --  
(1.42857, -3.62324) -- (1.41886, -3.58114) -- (1.41633, -3.57143) --  
(1.41223, -3.55509) -- (1.35215, -3.29071) -- (1.3322, -3.21429) --  
(1.30215, -3.08787) -- (1.28375, -3.00197) -- (1.25, -2.8634) -- (1.24845,  
-2.85714) -- (1.21509, -2.71348) -- (1.18966, -2.61823) -- (1.16131, -2.5) --  
(1.14626, -2.42517) -- (1.07629, -2.14772) -- (1.07527, -2.14286) --  
(1.07468, -2.1396) -- (1.07143, -2.12629) -- (1.03067, -1.96429) --  
(1.01688, -1.90973) -- (1.00394, -1.85321) -- (0.985733, -1.78571) --  
(0.968027, -1.71054) -- (0.956966, -1.67125) -- (0.940694, -1.60714) --  
(0.93238, -1.56762) -- (0.896779, -1.43249) -- (0.895922, -1.42857) --  
(0.895434, -1.42599) -- (0.892857, -1.41584) -- (0.872805, -1.33929) --  
(0.859888, -1.28297) -- (0.84959, -1.25) -- (0.832949, -1.19009) --  
(0.823088, -1.1412) -- (0.803571, -1.07151) -- (0.803553, -1.07145) --  
(0.803546, -1.07143) -- (0.803535, -1.07139) -- (0.785559, -1.00016) --  
(0.779475, -0.982143) -- (0.770823, -0.949395) -- (0.766833, -0.929595) --  
(0.755281, -0.892857) -- (0.748507, -0.858636) -- (0.737117, -0.826402) --  
(0.732253, -0.803571) -- (0.730726, -0.797217) -- (0.719664, -0.760578) --  
(0.714286, -0.729005) -- (0.711413, -0.721922); 
\draw[line width=2] (-1.52244, 5.) -- (-1.48807, 4.9405) -- (-1.42857, 4.83713) -- (-1.41938,  
4.82143) -- (-1.35802, 4.7134) -- (-1.31803, 4.64286) -- (-1.25, 4.51878) --  
(-1.23047, 4.48382) -- (-1.18387, 4.39816) -- (-1.12324, 4.28571) --  
(-1.10556, 4.25159) -- (-1.07143, 4.18587) -- (-1.02963, 4.10714) --  
(-0.98352, 4.01648) -- (-0.936963, 3.92857) -- (-0.864442, 3.77842) --  
(-0.810161, 3.6673) -- (-0.764153, 3.57143) -- (-0.748593, 3.53712) --  
(-0.714286, 3.46285) -- (-0.681072, 3.39286) -- (-0.63616, 3.29241) --  
(-0.600292, 3.21429) -- (-0.535714, 3.06465) -- (-0.52682, 3.04461) --  
(-0.512876, 3.01288) -- (-0.446789, 2.85714) -- (-0.421477, 2.79281) --  
(-0.374211, 2.67857) -- (-0.357143, 2.63406) -- (-0.319151, 2.53799) --  
(-0.302353, 2.5) -- (-0.268719, 2.41158) -- (-0.220306, 2.27969) --  
(-0.178571, 2.16825) -- (-0.168746, 2.14286) -- (-0.124653, 2.0182) --  
(-0.0622257, 1.84794) -- (-0.041185, 1.78571) -- (-0.0316032, 1.75411) --  
(6.66134*10^-16, 1.66222) -- (0.019611, 1.60714) -- (0.0582175, 1.48679) --  
(0.0796137, 1.42857) -- (0.103277, 1.35328) -- (0.117643, 1.31093) --  
(0.137533, 1.25) -- (0.146745, 1.21817) -- (0.165617, 1.16071) --  
(0.178571, 1.1189) -- (0.189962, 1.08282) -- (0.19389, 1.07143) --  
(0.201297, 1.0487) -- (0.232275, 0.94656) -- (0.24955, 0.892857) --  
(0.267857, 0.832151) -- (0.274583, 0.810297) -- (0.281166, 0.790263) --  
(0.30417, 0.714286) -- (0.315777, 0.67292) -- (0.331027, 0.625) --  
(0.357143, 0.536716) -- (0.357376, 0.535947) -- (0.357456, 0.535714) --  
(0.357604, 0.535253) -- (0.397443, 0.397443) -- (0.411015, 0.357143) --  
(0.432385, 0.2819) -- (0.438085, 0.259514) -- (0.463052, 0.178571) --  
(0.477449, 0.120306) -- (0.504987, 0.0307275) -- (0.512928, 
 6.66134*10^-16) -- (0.517042, -0.0186727) -- (0.535714, -0.0824211) --  
(0.537809, -0.0892857) -- (0.55597, -0.158316) -- (0.563182, -0.178571) --  
(0.575637, -0.218494) -- (0.588139, -0.267857) -- (0.594132, -0.298725) --  
(0.612535, -0.357143) -- (0.625, -0.407129) -- (0.633722, -0.437706) --  
(0.637053, -0.446429) -- (0.642146, -0.463574) -- (0.652428, -0.508286) --  
(0.662326, -0.535714) -- (0.671833, -0.584502) -- (0.685476, -0.626175); 
\draw[line width=2]  (5., -1.73936) --  (4.9398, -1.72552) --  (4.74496, -1.68361) --  (4.64286,  
-1.65957) --  (4.48066, -1.62352) --  (4.45356, -1.61787) --  (4.40603,  
-1.60714) --  (4.28571, -1.57977) --  (4.16176, -1.55253) --  (4.0182,  
-1.5182) --  (3.92857, -1.49845) --  (3.87087, -1.48627) --  (3.75, -1.45753) --  
 (3.62324, -1.42857) --  (3.58114, -1.41886) --  (3.57143, -1.41633) --  
 (3.55509, -1.41223) --  (3.29071, -1.35215) --  (3.21429, -1.3322) --  
 (3.08787, -1.30215) --  (3.00197, -1.28375) --  (2.8634, -1.25) --  (2.85714,  
-1.24845) --  (2.71348, -1.21509) --  (2.61823, -1.18966) --  (2.5, -1.16131) --  
 (2.42517, -1.14626) --  (2.14772, -1.07629) --  (2.14286, -1.07527) --  
 (2.1396, -1.07468) --  (2.12629, -1.07143) --  (1.96429, -1.03067) --  
 (1.90973, -1.01688) --  (1.85321, -1.00394) --  (1.78571, -0.985733) --  
 (1.71054, -0.968027) --  (1.67125, -0.956966) --  (1.60714, -0.940694) --  
 (1.56762, -0.93238) --  (1.43249, -0.896779) --  (1.42857, -0.895922) --  
 (1.42599, -0.895434) --  (1.41584, -0.892857) --  (1.33929, -0.872805) --  
 (1.28297, -0.859888) --  (1.25, -0.84959) --  (1.19009, -0.832949) --  
 (1.1412, -0.823088) --  (1.07151, -0.803571) --  (1.07143, -0.803546) --  
 (1.07139, -0.803535) --  (1.00016, -0.785559) --  (0.982143, -0.779475) --  
 (0.949395, -0.770823) --  (0.929595, -0.766833) --  (0.892857, -0.755281) --  
 (0.85511, -0.752033) --  (0.803571, -0.731186) --  (0.721449, -0.714286) --  
 (0.716252, -0.71232) --  (0.714286, -0.709162) --  (0.707105, -0.707108) --  
 (0.709517, -0.714096) --  (0.669643, -0.740858) --  (0.662578, -0.751864) --  
 (0.625, -0.784543) --  (0.617687, -0.796259) --  (0.612579, -0.803571) --  
 (0.535714, -0.875234) --  (0.527873, -0.885016) --  (0.521152, -0.892857) --  
 (0.446429, -0.963337) --  (0.437622, -0.973337) --  (0.429559, -0.982143) --  
 (0.357143, -1.05037) --  (0.347017, -1.0613) --  (0.338429, -1.07143) --  
 (0.178571, -1.22076) --  (0.164706, -1.23613) --  (0.151838, -1.25) --  
 (6.66134*10^-16, -1.38962) --  (-0.0190751, -1.4095) --  (-0.037803,  
-1.42857) --  (-0.178571, -1.55575) --  (-0.204185, -1.58153) --  (-0.230671,  
-1.60714) --  (-0.357143, -1.71935) --  (-0.390833, -1.75202) --  (-0.426858,  
-1.78571) --  (-0.535714, -1.88055) --  (-0.578887, -1.92111) --  (-0.626423,  
-1.96429) --  (-0.714286, -2.03946) --  (-0.768322, -2.08882) --  (-0.8294,  
-2.14286) --  (-0.892857, -2.19619) --  (-0.959105, -2.25518) --  (-1.03581,  
-2.32143) --  (-1.07143, -2.35085) --  (-1.15121, -2.42022) --  (-1.24566,  
-2.5) --  (-1.25, -2.50352) --  (-1.34459, -2.58398) --  (-1.42857, -2.65345) --  
 (-1.45769, -2.67857) --  (-1.53922, -2.7465) --  (-1.60714, -2.80159) --  
 (-1.67305, -2.85714) --  (-1.73506, -2.9078) --  (-1.78571, -2.94812) --  
 (-1.8919, -3.03571) --  (-1.93208, -3.06792) --  (-1.96429, -3.09312) --  
 (-2.1142, -3.21429) --  (-2.13023, -3.22691) --  (-2.14286, -3.23663) --  
 (-2.25076, -3.32218) --  (-2.32939, -3.3849) --  (-2.33936, -3.39286) --  
 (-2.5, -3.51832) --  (-2.52947, -3.54196) --  (-2.56709, -3.57143) --  
 (-2.67857, -3.65702) --  (-2.73059, -3.69798) --  (-2.79819, -3.75) --  
 (-2.85714, -3.79454) --  (-2.93271, -3.853) --  (-3.03259, -3.92857) --  
 (-3.03571, -3.93089) --  (-3.13581, -4.00704) --  (-3.21429, -4.06548) --  
 (-3.26906, -4.10714) --  (-3.33986, -4.16014) --  (-3.39286, -4.19902) --  
 (-3.5087, -4.28571) --  (-3.54481, -4.31234) --  (-3.57143, -4.33158) --  
 (-3.74575, -4.46004) --  (-3.75063, -4.46365) --  (-3.92857, -4.59294) --  
 (-3.95706, -4.61437) --  (-3.99625, -4.64286) --  (-4.10714, -4.7225) --  
 (-4.16432, -4.76425) --  (-4.24411, -4.82143) --  (-4.28571, -4.85091) --  
 (-4.37259, -4.91313) --  (-4.38921, -4.92492) --  (-4.46429, -4.97835) --  
 (-4.4769, -4.98739) --  (-4.49464, -5.); 
\draw [<->,thick,black] (2.47,1.06) arc (45:-74:2.5); 
\node [right] at (2.7,-2.5) {Valid};
\end{tikzpicture}}
\subcaption{Stokes structure with complex parameters.}\label{fig:4a}
\end{minipage}
\begin{minipage}{0.5\linewidth} \centering
\scalebox{0.5}{\begin{tikzpicture}{center}
\draw[<->] (-5,0) -- (5,0);
\node [right] at (5,0) {Re(s)};
\draw[<->] (0,-5) -- (0,5);
\node [above] at (0,5) {Im(s)};
\draw [-,decorate,decoration=zigzag,very thick] (-1,0) -- (4.8,0); 
\draw [line width=2] (2.6549, -5.) -- (2.58148, -4.91852) -- (2.5, -4.82719) -- (2.49477, 
-4.82143) -- (2.41313, -4.72973) -- (2.33654, -4.64286) -- (2.32143, 
-4.62536) -- (2.24606, -4.53966) -- (2.18039, -4.46429) -- (2.14286, 
-4.42024) -- (2.08032, -4.34826) -- (2.02645, -4.28571) -- (1.96429, 
-4.21179) -- (1.91595, -4.15548) -- (1.87479, -4.10714) -- (1.78571, 
-3.99981) -- (1.75302, -3.96127) -- (1.72476, -3.92857) -- (1.59156, 
-3.76558) -- (1.45722, -3.60008) -- (1.43402, -3.57143) -- (1.43162, 
-3.56838) -- (1.42857, -3.56446) -- (1.2912, -3.39286) -- (1.27323, 
-3.36963) -- (1.25, -3.33944) -- (1.15087, -3.21429) -- (1.11649, 
-3.16923) -- (1.07143, -3.11001) -- (1.01311, -3.03571) -- (0.961437, 
-2.96713) -- (0.892857, -2.87615) -- (0.878049, -2.85714) -- (0.808112, 
-2.76332) -- (0.744704, -2.67857) -- (0.714286, -2.63591) -- (0.65653, 
-2.55776) -- (0.613507, -2.5) -- (0.535714, -2.38995) -- (0.50669, 
-2.35045) -- (0.365074, -2.15079) -- (0.35951, -2.14286) -- (0.358571, 
-2.14143) -- (0.357143, -2.13929) -- (0.234589, -1.96429) -- (0.212474, 
-1.93038) -- (0.178571, -1.87962) -- (0.112545, -1.78571) -- (0.0682017, 
-1.71751) -- (6.66134*10^-16, -1.61601) -- (-0.00630805, -1.60714) -- 
(-0.0743253, -1.5029) -- (-0.125046, -1.42857) -- (-0.178571, -1.34219) -- 
(-0.21553, -1.28696) -- (-0.241031, -1.25) -- (-0.28536, -1.17822) -- 
(-0.347271, -1.0813) -- (-0.35353, -1.07143) -- (-0.354819, -1.0691) -- 
(-0.357143, -1.06529) -- (-0.41007, -0.982143) -- (-0.423592, -0.959306) -- 
(-0.446429, -0.922488) -- (-0.465711, -0.892857) -- (-0.492004, 
-0.849147) -- (-0.520984, -0.803571) -- (-0.535714, -0.778328) -- 
(-0.560119, -0.738691) -- (-0.576189, -0.714286) -- (-0.625, -0.632818) -- 
(-0.628072, -0.628072) -- (-0.630146, -0.625) -- (-0.638065, -0.611935) -- 
(-0.66173, -0.572444) -- (-0.684784, -0.535714) -- (-0.695267, 
-0.516695) -- (-0.714286, -0.483871) -- (-0.728763, -0.460906) -- 
(-0.737903, -0.446429) -- (-0.758929, -0.410523) -- (-0.762251, 
-0.405108) -- (-0.77289, -0.387825) -- (-0.791133, -0.357143) -- 
(-0.795487, -0.349058) -- (-0.803571, -0.335187) -- (-0.817658, -0.3125) -- 
(-0.828571, -0.292856) -- (-0.843867, -0.267857) -- (-0.848214, 
-0.259794) -- (-0.861662, -0.236662) -- (-0.870496, -0.223214) -- 
(-0.892857, -0.183834) -- (-0.894877, -0.180592) -- (-0.89636, 
-0.178571) -- (-0.902062, -0.169366) -- (-0.927481, -0.123909) -- 
(-0.952603, -0.0892857) -- (-0.95952, -0.0666632) -- (-0.976398, 
-0.0428371); 
\draw [line width=2] (2.6549, 5.) -- (2.58148, 4.91852) -- (2.5, 4.82719) -- (2.49477, 4.82143) -- 
(2.41313, 4.72973) -- (2.33654, 4.64286) -- (2.32143, 4.62536) -- (2.24606, 
4.53966) -- (2.18039, 4.46429) -- (2.14286, 4.42024) -- (2.08032, 4.34826) -- 
(2.02645, 4.28571) -- (1.96429, 4.21179) -- (1.91595, 4.15548) -- (1.87479, 
4.10714) -- (1.78571, 3.99981) -- (1.75302, 3.96127) -- (1.72476, 3.92857) -- 
(1.59156, 3.76558) -- (1.45722, 3.60008) -- (1.43402, 3.57143) -- (1.43162, 
3.56838) -- (1.42857, 3.56446) -- (1.2912, 3.39286) -- (1.27323, 3.36963) -- 
(1.25, 3.33944) -- (1.15087, 3.21429) -- (1.11649, 3.16923) -- (1.07143, 
3.11001) -- (1.01311, 3.03571) -- (0.961437, 2.96713) -- (0.892857, 
2.87615) -- (0.878049, 2.85714) -- (0.808112, 2.76332) -- (0.744704, 
2.67857) -- (0.714286, 2.63591) -- (0.65653, 2.55776) -- (0.613507, 2.5) -- 
(0.535714, 2.38995) -- (0.50669, 2.35045) -- (0.365074, 2.15079) -- 
(0.35951, 2.14286) -- (0.358571, 2.14143) -- (0.357143, 2.13929) -- 
(0.234589, 1.96429) -- (0.212474, 1.93038) -- (0.178571, 1.87962) -- 
(0.112545, 1.78571) -- (0.0682017, 1.71751) -- (6.66134*10^-16, 1.61601) -- 
(-0.00630805, 1.60714) -- (-0.0186277, 1.58852) -- (-0.0747275, 1.5033) -- 
(-0.125046, 1.42857) -- (-0.178571, 1.34219) -- (-0.21553, 1.28696) -- 
(-0.241031, 1.25) -- (-0.28536, 1.17822) -- (-0.347271, 1.0813) -- 
(-0.35353, 1.07143) -- (-0.354819, 1.0691) -- (-0.357143, 1.06507) -- 
(-0.467391, 0.892857) -- (-0.492004, 0.849147) -- (-0.520984, 0.803571) -- 
(-0.535714, 0.778328) -- (-0.560315, 0.738887) -- (-0.576189, 0.714286) -- 
(-0.625, 0.632818) -- (-0.628042, 0.628042) -- (-0.63866, 0.61134) -- 
(-0.684784, 0.535714) -- (-0.695045, 0.516474) -- (-0.714286, 0.483871) -- 
(-0.738456, 0.446429) -- (-0.762251, 0.405108) -- (-0.77289, 0.387825) -- 
(-0.791472, 0.357143) -- (-0.803571, 0.333968) -- (-0.827991, 0.292276) -- 
(-0.845383, 0.267857) -- (-0.861662, 0.236662) -- (-0.892857, 0.183451) -- 
(-0.894877, 0.180592) -- (-0.89636, 0.178571) -- (-0.902062, 0.169366) -- 
(-0.927481, 0.123909) -- (-0.952603, 0.0892857) -- (-0.95952, 0.0666624) -- 
(-0.976384, 0.0428631); 
\draw[line width=2] (-1,0) -- (-4.8,0); 
\draw [dashed, line width=2] (-3.50734, -5.) -- (-3.45431, -4.88288) -- (-3.4165, -4.79778) -- 
(-3.34515, -4.64286) -- (-3.3051, -4.55204) -- (-3.26428, -4.46429) -- 
(-3.21429, -4.3544) -- (-3.19266, -4.30734) -- (-3.18232, -4.28571) -- 
(-3.15428, -4.22571) -- (-3.07988, -4.06298) -- (-3.03571, -3.96857) -- 
(-3.01677, -3.92857) -- (-2.96638, -3.81933) -- (-2.93297, -3.75) -- 
(-2.85714, -3.58829) -- (-2.85171, -3.57686) -- (-2.84901, -3.57143) -- 
(-2.84138, -3.55567) -- (-2.73701, -3.33442) -- (-2.67742, -3.21429) -- 
(-2.6212, -3.09309) -- (-2.51146, -2.8686) -- (-2.50602, -2.85714) -- 
(-2.50413, -2.85301) -- (-2.5, -2.84446) -- (-2.4182, -2.67857) -- 
(-2.38714, -2.61286) -- (-2.32851, -2.5) -- (-2.26875, -2.37411) -- 
(-2.16039, -2.16039) -- (-2.15181, -2.14286) -- (-2.14898, -2.13673) -- 
(-2.14286, -2.12443) -- (-2.06076, -1.96429) -- (-2.02937, -1.8992) -- 
(-1.96998, -1.78571) -- (-1.96429, -1.77408) -- (-1.90815, -1.66328) -- 
(-1.87697, -1.60714) -- (-1.78571, -1.43051) -- (-1.78504, -1.42925) -- 
(-1.78463, -1.42857) -- (-1.7833, -1.42616) -- (-1.66264, -1.19451) -- 
(-1.60714, -1.09245) -- (-1.59498, -1.07143) -- (-1.53846, -0.96154) -- 
(-1.49747, -0.892857) -- (-1.42857, -0.760484) -- (-1.41154, -0.731314) -- 
(-1.40127, -0.714286) -- (-1.36742, -0.653139) -- (-1.34807, -0.616212) -- 
(-1.30078, -0.535714) -- (-1.28484, -0.500871) -- (-1.25336, -0.446429) -- 
(-1.25, -0.439713) -- (-1.22097, -0.386171) -- (-1.2023, -0.357143) -- 
(-1.16071, -0.280141) -- (-1.15596, -0.272607) -- (-1.15258, -0.267857) -- 
(-1.14196, -0.249098) -- (-1.12375, -0.215534) -- (-1.11607, -0.201678) -- 
(-1.10189, -0.178571) -- (-1.09166, -0.158343) -- (-1.07661, -0.133929) -- 
(-1.07143, -0.123167) -- (-1.05933, -0.101384) -- (-1.05018, -0.0892857) -- 
(-1.02679, -0.0460447) -- (-1.02622, -0.0452116) -- (-1.02541, 
-0.0445837) -- (-1.02419, -0.0420515) -- (-1.0223, -0.0386928); 
\draw [dashed, line width=2] (-3.50734, 5.) -- (-3.45431, 4.88288) -- (-3.4165, 4.79778) -- (-3.34515, 
4.64286) -- (-3.3051, 4.55204) -- (-3.26428, 4.46429) -- (-3.21429, 
4.3544) -- (-3.19266, 4.30734) -- (-3.18232, 4.28571) -- (-3.15428, 
4.22571) -- (-3.07988, 4.06298) -- (-3.03571, 3.96857) -- (-3.01677, 
3.92857) -- (-2.96638, 3.81933) -- (-2.93297, 3.75) -- (-2.85714, 3.58829) -- 
(-2.85171, 3.57686) -- (-2.84901, 3.57143) -- (-2.84138, 3.55567) -- 
(-2.73701, 3.33442) -- (-2.67742, 3.21429) -- (-2.6212, 3.09309) -- 
(-2.51146, 2.8686) -- (-2.50602, 2.85714) -- (-2.50413, 2.85301) -- (-2.5, 
2.84446) -- (-2.4182, 2.67857) -- (-2.38714, 2.61286) -- (-2.32851, 2.5) -- 
(-2.26875, 2.37411) -- (-2.16039, 2.16039) -- (-2.15181, 2.14286) -- 
(-2.14898, 2.13673) -- (-2.14286, 2.12443) -- (-2.06076, 1.96429) -- 
(-2.02937, 1.8992) -- (-1.96998, 1.78571) -- (-1.96429, 1.77408) -- 
(-1.90815, 1.66328) -- (-1.87697, 1.60714) -- (-1.78571, 1.43051) -- 
(-1.78504, 1.42925) -- (-1.78463, 1.42857) -- (-1.7834, 1.42626) -- 
(-1.72369, 1.31203) -- (-1.68935, 1.25) -- (-1.66189, 1.19525) -- 
(-1.60714, 1.09245) -- (-1.59498, 1.07143) -- (-1.53846, 0.96154) -- 
(-1.49747, 0.892857) -- (-1.42857, 0.760484) -- (-1.41154, 0.731314) -- 
(-1.40127, 0.714286) -- (-1.36742, 0.653139) -- (-1.34807, 0.616212) -- 
(-1.30078, 0.535714) -- (-1.28484, 0.500871) -- (-1.25336, 0.446429) -- 
(-1.25, 0.439713) -- (-1.22097, 0.386171) -- (-1.2023, 0.357143) -- 
(-1.16071, 0.280141) -- (-1.15596, 0.272607) -- (-1.15286, 0.267857) -- 
(-1.14196, 0.249098) -- (-1.12375, 0.215534) -- (-1.11607, 0.201678) -- 
(-1.10189, 0.178571) -- (-1.09166, 0.158343) -- (-1.07661, 0.133929) -- 
(-1.07143, 0.123167) -- (-1.05933, 0.101384) -- (-1.05018, 0.0892857) -- 
(-1.02679, 0.0460447) -- (-1.02622, 0.0452116) -- (-1.0219, 0.0383254); 
\node [right] at (1,2) {Re$(\chi_1)<0$, Im$(\chi_1)>0$};
\node [right] at (1,1.5) {Re$(\chi_2)>0$, Im$(\chi_2)<0$};
\node [right] at (1,-1.5) {Re$(\chi_2)>0$, Im$(\chi_2)>0$};
\node [right] at (1,-1) {Re$(\chi_1)<0$, Im$(\chi_1)<0$};
\node [below] at (-0.5,-4.3) {Re$(\chi_2)>0$, Im$(\chi_2)<0$};
\node [below] at (-0.5,-3.8) {Re$(\chi_1)<0$, Im$(\chi_1)>0$};
\node [above] at (-0.5,4.3) {Re$(\chi_1)<0$, Im$(\chi_1)<0$};
\node [above] at (-0.5,3.8) {Re$(\chi_2)>0$, Im$(\chi_2)>0$};
\node at (-4,1.5) {Re$(\chi_1)>0$, Im$(\chi_1)<0$};
\node at (-4,1) {Re$(\chi_2)<0$, Im$(\chi_2)>0$};
\node at (-4,-1.5) {Re$(\chi_2)<0$, Im$(\chi_2)<0$};
\node at (-4,-1) {Re$(\chi_1)>0$, Im$(\chi_1)>0$};
\node [draw,circle,scale=0.8] at (-4.5,0.35) {\textbf{1}};
\node [draw,circle,scale=0.8] at (2.7,4.5) {\textbf{2}};
\node [draw,circle,scale=0.8] at (2.7,-4.5) {\textbf{2}};
\end{tikzpicture}}
\subcaption{Stokes structure with parameters $\alpha =-1$ and $\beta=1$.}\label{fig:4b}
\end{minipage}
\caption{These figures depict the Stokes structure for complex parameters. Figure \ref{fig:4a} illustrates the Stokes structure for parameters $\alpha =\exp(i\pi/4)$ and $\beta =1$ with the region of validity for a general asymptotic solution. We see that the Stokes structure has been rotated clockwise by $\pi/4$ as a result of $\alpha$ being complex. We also note that the branch cut has been chosen arbitrarily. Figure \ref{fig:4b} illustrates the Stokes structure for $\alpha=-1$ and $\beta=1$. The structure is a rotation by $\pi$, as expected.}
\end{figure}
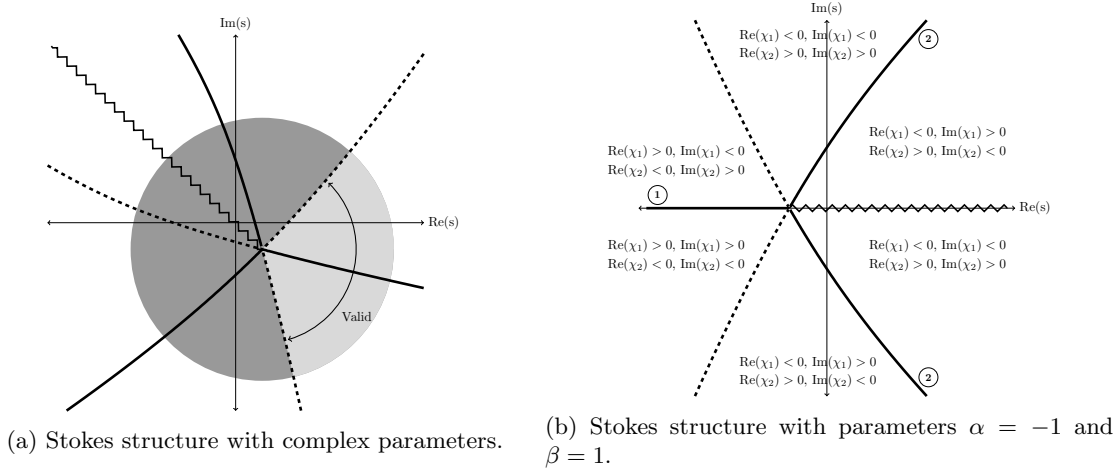

\section{Non-vanishing Asymptotics}\label{S:Non-vanishing Asymptotics}
We have completed the analysis for solutions with the behaviour $x_n\ll 1$ as $n\rightarrow\infty$ of equation \eqref{dPII}. In addition to these solutions with this behaviour, there are solutions which grow in the asymptotic limit, that is, $x_n\gg 1$ as $n\rightarrow\infty$. The analysis involved in the subsequent sections is nearly identical to Sections \ref{S:Asymptotic series expansions} and \ref{S:Exponential Asymptotics}. Hence, we will omit the details and only provide the key results. As before, we scale into the far field by setting $s=\epsilon n$. Then the scaling for non-vanishing $x_n$ behaviour, and the appropriate choice of parameter sizes, is given by
\begin{equation}\label{largescalings}
x_n=\frac{g(s)}{\epsilon}, \qquad \alpha=\frac{\hat{\alpha}}{\epsilon}, \qquad \beta=\frac{\hat{\beta}}{\epsilon^2}, \qquad \gamma=\frac{\hat{\gamma}}{\epsilon^3}.
\end{equation}
As before, we drop the hat notation for simplicity. The rescaled equation is then given by 
\begin{equation}\label{rescaledLARGElimiteqn}
\left(g\left(s+\epsilon\right)+g\left(s-\epsilon\right)\right) \left(\epsilon^{2}-g\left(s\right)^2\right)=\left(\alpha s +\beta\right)g\left(s\right)+\gamma, 
\end{equation}
as $\epsilon\rightarrow0$. We then expand $g(s)$ as an asymptotic power series in $\epsilon$, 
\begin{equation}\label{asymptoticseriesforLARGE}
g(s)\sim \sum_{r=0}^{\infty}\epsilon^rg_r(s),
\end{equation}
as $\epsilon\rightarrow \infty$. Substituting \eqref{asymptoticseriesforLARGE} into \eqref{rescaledLARGElimiteqn} and by matching coefficients of $\epsilon$ we can show that the leading order solution satisfies the equations 
\begin{equation}\label{leadingorderLARGE}
g_0 = \frac{6\alpha s+6\beta}{\left(2\sqrt{27}\right)^{2/3}(\sqrt{4\Psi^3+11664\gamma^2}+108\gamma)^{1/3}}-\frac{(\sqrt{4\Psi^3+11664\gamma^2}+108\gamma)^{1/3}}{432^{1/3}},
\end{equation}
or
\begin{equation}\label{leadingorderLARGE2}
g_0 =\frac{\left(1\pm i\sqrt{3}\right)(6\alpha s+6\beta)}{\left(2\sqrt{216}\right)^{2/3}(\sqrt{4\Psi^3+11664\gamma^2}+108\gamma)^{1/3}}+
\frac{\left(1\mp i\sqrt{3}\right)(\sqrt{4\Psi^3+11664\gamma^2}+108\gamma)^{1/3}}{\left(3456\right)^{1/3}},
\end{equation}
where $\Psi=6\alpha s+6\beta$. In general, we have 
\begin{equation}\label{generallateordertersLARGE}
\mathcal{O}(\epsilon^r): \qquad \left( \alpha s +\beta \right)g_r=\sum_{j=0}^{\lfloor\left(r-2\right)/2\rfloor}\frac{2g_{r-2j-2}^{(2j)}}{(2j)!}\sum_{m=0}^{r}g_m\sum_{l=0}^{r-m}g_l\sum_{j=0}^{\lfloor\left(r-m-l\right)/2\rfloor}\frac{2g_{r-m-l-2j}^{(2j)}}{(2j)!},
\end{equation}
for $n\geq 2$. Using similar reasoning as in Section \ref{S:Exponential Asymptotics}, our late-order terms ansatz is 
\begin{equation}\label{LARGElateorderterms}
g_r(s)\sim \frac{G\left(s\right)\Gamma\left(r+\kappa\right)}{\eta\left(s\right)^{r+\kappa}},
\end{equation}
as $r\rightarrow \infty$. Applying \eqref{LARGElateorderterms} into \eqref{generallateordertersLARGE}, it can then be shown that the singulant, $\eta(s)$, solves the equation
\begin{equation}\label{LARGEsingulanteqn}
\cosh\left(\eta'\right)=\frac{-\left(\alpha s+\beta+4g_0^2\right)}{2g_0^2},
\end{equation}
and the prefactor, $G(s)$, solves the equation
\begin{equation}\label{LARGEprefactorequation}
-2g_0^2G'\sinh\left(\eta'\right)-2g_0^2\eta''G\frac{\eta'\sinh\left(\eta'\right)-\cosh\left(\eta'\right)}{\left(\eta'\right)^2}+4g_0g_1G\cosh\left(\eta'\right)+8g_0g_1G=0.
\end{equation}
We observe that the right hand side of \eqref{LARGEsingulanteqn} has many more zeroes compared to \eqref{singulant dash equation}. This will mean that the Stokes and anti-Stokes curves will emerge from more than one singular point and is illustrated in Figure 7. As demonstrated in Section \ref{S:Stokes structure}, we may use the solution to (\ref{LARGEsingulanteqn}) to determine the Stokes structure of the asymptotic solution (\ref{asymptoticseriesforLARGE}). Hence, we have fully determined the late-order terms \eqref{LARGElateorderterms} where the singulant and prefactor are solutions to \eqref{LARGEsingulanteqn} and \eqref{LARGEprefactorequation} respectively. The constants associated with the prefactor can be determined in a similar fashion as demonstrated in Appendix \ref{S:Appendix Calculate late order}.

\subsection{Stokes Structure}\label{S:Stokes structure LARGE}
Once the singulant is determined we may determine the Stokes structure of asymptotic solution. As discussed in Section \ref{S:Stokes structure}, the exponentially-small contributions present are generally proportional to $\exp(-\eta/\epsilon)$, and we may therefore obtain the Stokes structure to \eqref{asymptoticseriesforLARGE}. We note that we have three distinct leading order solutions. We will consider \eqref{asymptoticseriesforLARGE} with leading order behaviour \eqref{leadingorderLARGE} with $\alpha=2, \beta=-1$ and $\gamma=2$. However, similar results may be obtained by considering the leading order behaviour described in \eqref{leadingorderLARGE2}. We will observe that the Stokes and anti-Stokes curves emerge from two singularities as opposed to one singularity when we compare the Stokes structure to that found in Section \ref{S:Stokes structure}. This is due to the leading order solution \eqref{leadingorderLARGE} having two singularities.
\begin{figure}[h!]
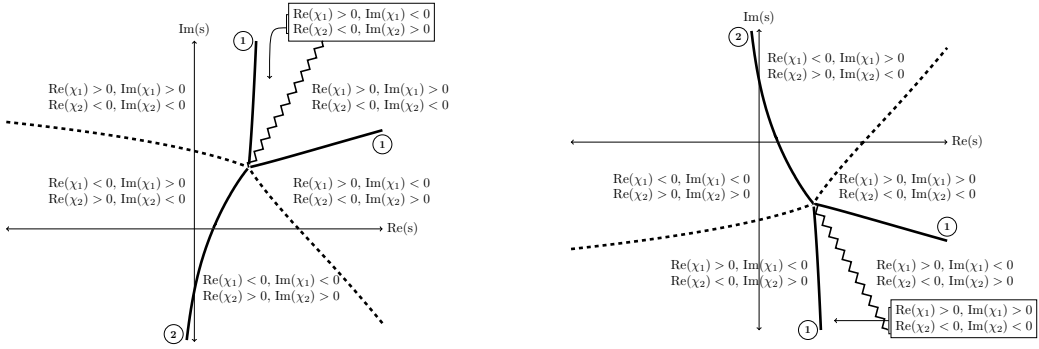

\begin{minipage}{0.5\linewidth} 
\centering
\scalebox{0.5}{
}
\subcaption{Stokes structure emerging from upper singularity.}\label{fig:5a}
\end{minipage}
\begin{minipage}{0.5\linewidth} \centering
\scalebox{0.5}{%
}
\subcaption{Stokes structure emerging from lower singularity.}\label{fig:5b}
\end{minipage}
\caption{This figure illustrates the Stokes structure for non-vanishing asymptotic solutions. In this case, the Stokes and anti-Stokes curves emerge from two singularities rather than one. We also note the Stokes structure in the upper half plane is symmetric to the Stokes structure in the lower half plane.}
\end{figure}
\begin{figure}[h!]
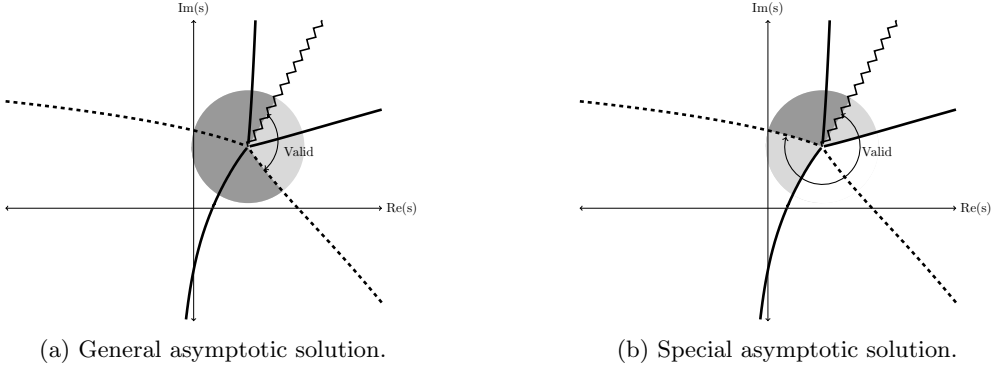

\begin{minipage}{0.5\linewidth} 
\centering
\scalebox{0.5}{%
}
\subcaption{General asymptotic solution.}\label{fig:6a}
\end{minipage}
\begin{minipage}{0.5\linewidth} 
\centering
\scalebox{0.5}{%
}
\subcaption{Special asymptotic solution.}\label{fig:6b}
\end{minipage}
\caption{This figure illustrates the asymptotic solutions valid about the Stokes curve extending to the real positive direction. The light gray shaded regions show the presence of exponentially-small contributions, while the contributions are exponentially large in the dark gray regions. Unshaded regions illustrate no exponential contributions and therefore the asymptotic behaviour is described by the leading order solution (\ref{leadingorderLARGE}). Figures \ref{fig:6a} illustrate the regions of validity of a general asymptotic solution about the upper singularity. This asymptotic solution contain one free parameter hidden beyond all orders. The regions of validity to these asymptotic solutions may be extended as shown in figure \ref{fig:6b}. This is possible if we demand that the exponential term be absent in the appropriate region. Due to the symmetry of the Stokes structure, the region of validity to the contribution due to the lower singularity is symmetric with respect to the real axis.}
\end{figure}

\begin{figure}[h!]
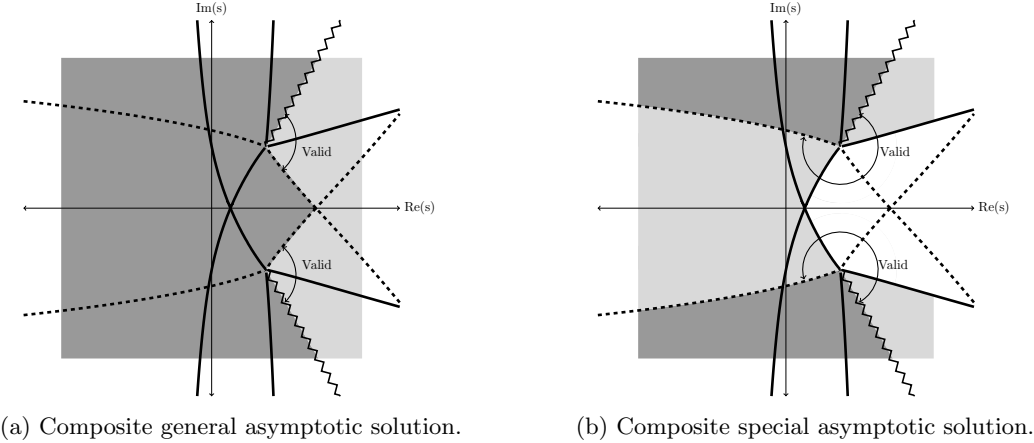

\begin{minipage}{0.5\linewidth} 
\centering
\scalebox{0.5}{
}
\subcaption{Composite general asymptotic solution.}\label{fig:7a}
\end{minipage}
\begin{minipage}{0.5\linewidth} 
\centering
\scalebox{0.5}{%
}
\subcaption{Composite special asymptotic solution.}\label{fig:7b}
\end{minipage}
\caption{This figure illustrates the regions of validity of the composite general and special asymptotic solutions. Here the shading has the meaning described in Figure 6. Due to the symmetry of the Stokes structure illustrated in Figures 5 and 6, the composite behavior for the general and special asymptotic solutions can be obtained. No interaction effects occur at the intersection between Stokes curves on the real axis, as $\textrm{Re}(\chi)$ takes the same value for both contributions at this point, and therefore both contributions are the same size as $\epsilon\rightarrow 0$.}
\end{figure}

After determining the Stokes structure of these asymptotic solutions, we may deduce their regions of validity and the switching behaviour to the exponentially-small terms present in these solutions. We obtain asymptotic solutions which exhibit similar features to those described in Section \ref{S:Stokes structure}. That is, we can obtain asymptotic solutions which contain one free parameter hidden beyond all orders of the asymptotic power series. These asymptotic solutions are valid within two adjacent regions of the complex $s$ plane. Furthermore, for special choices of the free parameter, the range of validity can be extended by two additional adjacent regions in the complex $s$ plane, as seen in Figure \ref{fig:7b}. These are special asymptotic solutions which contain no free parameters and are therefore uniquely defined.

Similarly in Section \ref{S:Stokes structure}, we have the freedom to choose any of the other Stokes or anti-Stokes curve for which the asymptotic solution is valid. As a result, other asymptotic solutions can be obtained by rotating a known asymptotic solution through two adjacent regions. Thus, we have determined the regions of validity for the asymptotic solutions of $\text{dP}_{\text{II}}$ which grow in the limit $\epsilon\rightarrow 0$ and qualitatively determined the Stokes phenomena present within these solutions.

\section{Conclusions}\label{S:Conclusions}
In this paper, we used exponential asymptotics methods to compute and investigate the asymptotic solutions to the second discrete Painlev\'{e} equation whose leading order behaviour can be described by rational expressions such as \eqref{leading order solutions}, \eqref{leadingorderLARGE}, or \eqref{leadingorderLARGE2}. We then determined the Stokes structure and used this information to deduce the regions of validity to these asymptotic solutions. The asymptotic solutions obtained are given as the sum of a truncated asymptotic power series and an exponentially-subdominant correction term given by \eqref{COMPLETE OPTIMALLY TRUNC ASYM SERIES}.

In Sections \ref{S:Asymptotic series expansions} and \ref{S:Exponential Asymptotics}, we considered asymptotic solutions which vanish as $n \rightarrow \infty$. Using exponential asymptotics, we determined the form of subdominant exponential contributions present in the asymptotic solutions, which are defined up to two free Stokes-switching parameters. From this behaviour, we deduced the associated Stokes structure, illustrated in Figure 1. By considering the Stokes switching, we found that the asymptotic series is a valid approximation in a region of the complex plane centered around the positive real axis. Furthermore, we found that it is possible to select the Stokes parameters so that the exponential contribution is absent in the region where it would normally become large. Consequently, the associated special asymptotic solutions are valid within a significantly larger region of the complex plane, shown in Figure 3. 

In Section \ref{S:Non-vanishing Asymptotics}, we considered the equivalent analysis for asymptotic solutions to the second discrete Painlev\'{e} equation which grow as $n \rightarrow \infty$, rather than vanishing. By applying exponential asymptotic methods, we again determined the Stokes structure present in these asymptotic solutions. We note that the structure of Stokes and anti-Stokes curves for this problem, illustrated in Figure 7, is significantly more complicated than in the vanishing case. Despite this, careful analysis of the exponentially-small asymptotic contributions in the problem is sufficient for us to determine the regions of validity for the asymptotic series. We again find that the asymptotic behaviour contains free Stokes-switching parameters, and that these parameters may again be chosen such that the exponential contributions disappear in regions where they would otherwise become exponentially-large. This causes the associated asymptotic series expression to have a larger region of validity, illustrated in Figure \ref{fig:7b}, including the entire real axis.

We note that, when the scalings for the vanishing case, \eqref{smallscalings}, and the non-vanishing case, \eqref{largescalings}, are undone, we find that the leading order solution to $\text{dP}_{\text{II}}$ is given by $x_n \sim -\gamma/\alpha  n $ and $x_n \sim \pm i\sqrt{\alpha n/2}$ as $n\rightarrow\infty$, respectively. From this analysis, we determine two types of asymptotic behaviours; type one solutions contain a free parameter hidden beyond all orders and type two solutions are uniquely determined with an extended region of validity. We note that the scalings for which the solutions behave as $x_n\sim \mathcal{O}(1)$ in the limit $n\rightarrow\infty$ can also be considered. However, this just reduces to either the vanishing or non-vanishing case when the scalings are undone.
  
Similar features of these asymptotic solutions are shared with the classical tronqu\'{e}e and tri-tronqu\'{e}e solutions of $\text{P}_{\text{II}}$ \eqref{PII}. The tronqu\'{e}e solutions contain free parameters hidden beyond all orders while the tri-tronqu\'{e}e are uniquely defined, both of which are valid in certain sectors in the complex plane separated by Stokes and anti-Stokes curves. In particular, as stated in Section \ref{S:Background}, the tronqu\'{e}e and tri-tronqu\'{e}e solutions are described by $w\sim \sqrt{-t/2}$ or $w\sim -\mu/t$ as $|t|\rightarrow\infty$. These similarities are shared with the asymptotic behaviours we found for $\text{dP}_{\text{II}}$.

The asymptotic study considered in \cite{Chris2015} used the same ideas to investigate asymptotic solutions for the first discrete Painlev\'{e} equation ($\text{dP}_{\text{I}}$). The qualitative features of the asymptotic solutions obtained in this study are very similar to those in \cite{Chris2015}. Using these ideas, both \cite{Chris2015} and the current study were able to determine solutions which are asymptotically free of poles to nonlinear discrete equations. An important distinction between both the Stokes structure of classic (tri-)tronqu\'{e}e solutions of the Painlev\'{e} equations and the Stokes structure found in \cite{Chris2015} is that the regions of validity for the asymptotic behaviours found in this study are bounded by curves rather than rays.  

\section{Data Accessibility}
We have no supporting data aside from that contained within the text.

\section{Authors’ Contributions}
N.J., C.J.L., and S.L. collaborated on the mathematical analysis. S.L.
drafted the manuscript. All authors gave final approval for publication.

\section{Competing Interests}
We have no competing interests.

\section{Funding}
N.J., C.J.L. and S.L. were supported by Australian Laureate Fellowship grant no. FL120100094
from the Australian Research Council.

\section{Acknowledgements}
The authors would like to thank Prof. Y. Takei for the discussions and suggestions
regarding this study.

\begin{appendices}

\section{Calculating the late-order terms near the singularity}\label{S:Appendix Calculate late order}
For $|\alpha |\neq 0$, the behaviour of the singulant can be shown to be 
\begin{align}\label{localsingulantequation}
\chi_1 \sim& -\frac{2\sqrt{\alpha }}{3}(s-s_0)^{3/2}, \\
\chi_2 \sim& \frac{2\sqrt{\alpha }}{3}(s-s_0)^{3/2}, \nonumber
\end{align}
about the singularity $s_0=(2-\beta )/\alpha $. Using \eqref{localsingulantequation} in the ordinary differential equation for the prefactor, we obtain 
\begin{equation}\label{local prefactor equation}
-\sqrt{\alpha (s-s_0)}F'-\frac{\sqrt{\alpha }}{4\sqrt{s-s_0}}F=0.
\end{equation}
We note that both $\chi_1$ and $\chi_2$ produce the same governing equation for the prefactor. Solving equation \eqref{local prefactor equation}, we find that the local behaviour of the prefactor about the singularity is given by 
\begin{equation*}\label{local prefactor form}
F\sim \frac{\Lambda}{(s-s_0)^{1/4}}
\end{equation*}
where $s_0=(2-\beta)/\alpha$. Recalling that there are two distinct singulant contributions, we therefore have two distinct constants associated with each singulant denoted by $\Lambda_1$ and $\Lambda_2$.

Finally, we require that the strength of the singularity in the late-order ansatz, \eqref{late-order terms}, must be consistent with the strength of the singularity of \eqref{leading order solutions}. In order to determine the correct value of $k$ in \eqref{late-order terms} we recall that $f_0$ has a singularity of strength one. Thus, in order to be consistent, the late-order terms must have the same strength as $n\rightarrow 0$. In the limit as $n \rightarrow 0$, the late-order term expression near the singularity becomes
\begin{equation}\label{determine k}
\frac{\Lambda_1\Gamma(k)}{(s-s_0)^{1/4}\big(-\tfrac{2}{3}\sqrt{\alpha }(s-s_0)^{3/2}\big)^{k}}+\frac{\Lambda_2\Gamma(k)}{(s-s_0)^{1/4}\big(\tfrac{2}{3}\sqrt{\alpha }(s-s_0)^{3/2}\big)^{k}},
\end{equation} 
which has singularity strength $1/4 + 3k/2$. The singularity in $f_0$ has strength one, and therefore, the strength of the singularity in \eqref{determine k} matches the strength of the singularity in $f_0$ only if $k=1/2$.

\section{Calculating the prefactor constants}\label{S:Appendix calculate prefactors}
We are yet to determine the values of the constants, $\Lambda_i$, appearing in the late-order terms \eqref{late-order terms}. In Appendix \ref{S:Appendix Calculate late order}, we showed that $\chi_1=-\chi_2$ and $F_1=\left(\Lambda_2/\Lambda_1\right)F_2$. Using these facts, we may rewrite the expression for the late-order terms as
\begin{equation}\label{prefactorEXPRESSION}
\frac{f_r(s)\,\chi_1^{r+1/2}}{F_1(s)\, \Gamma\left(r+1/2\right)} \sim \Lambda_1+(-1)^{r+1/2}\Lambda_2,
\end{equation} 
as $r\rightarrow\infty$. By appropriately adding (or subtracting) successive terms of \eqref{prefactorEXPRESSION}, we can obtain formulas for the constants, $\Lambda_i$, in the limit $r\rightarrow\infty$. Doing this, we obtain
\begin{align}\label{prefactor coefficients LIMITS}
2\Lambda_1 &= \lim\limits_{r\rightarrow\infty}\left[
\frac{f_{2r}\chi_1^{2r+1/2}}{F_1\,\Gamma\left(2r+1/2\right)}+\frac{f_{2r-1}\chi_1^{2r-1/2}}{F_1\,\Gamma\left(2r-1/2\right)}
\right], \\
-2i\Lambda_2 &= \lim\limits_{r\rightarrow\infty}\left[
\frac{f_{2r}\chi_1^{2r+1/2}}{F_1\,\Gamma\left(2r+1/2\right)}-\frac{f_{2r-1}\chi_1^{2r-1/2}}{F_1\,\Gamma\left(2r-1/2\right)}
\right]. \label{other limiting equation}
\end{align}
In Section \ref{S:Asymptotic series expansions}, we showed that all the odd terms of the asymptotic series vanish. Thus, we observe that the second term of the expressions \eqref{prefactor coefficients LIMITS}-\eqref{other limiting equation} is equal to zero. As a consequence, we find that $\Lambda_1=-i\Lambda_2$. In order to determine the values of these constants we use the local behaviours of $\chi_1$ and $F_1(s)$ near the singularity provided in Appendix \ref{S:Appendix Calculate late order}. We consider a numerical example where we choose $\alpha=-2, \beta=1$ and $\gamma=1$. Using the leading order solution, $1/(1-2s)$, and computing the behaviour of $f_r$ using \eqref{reccurence for coeffs}, we can calculate the values of $\Lambda_i$ using equations \eqref{prefactor coefficients LIMITS}-\eqref{other limiting equation} numerically using the Mathematica 10 package. For sufficiently large values of $f_r$ computed, we find that
\begin{align*}
\Lambda_1 &\approx 0.0757 - 0.0757\,i, \\
\Lambda_2 &\approx 0.0757 + 0.0757\,i. 
\end{align*}  
We have therefore determined the explicit form of the late order terms, $f_r$, of the asymptotic series.
\begin{figure}[h!]
\begin{minipage}{0.5\linewidth} 
\centering
\scalebox{0.5}{ 
\includegraphics[scale=0.5]{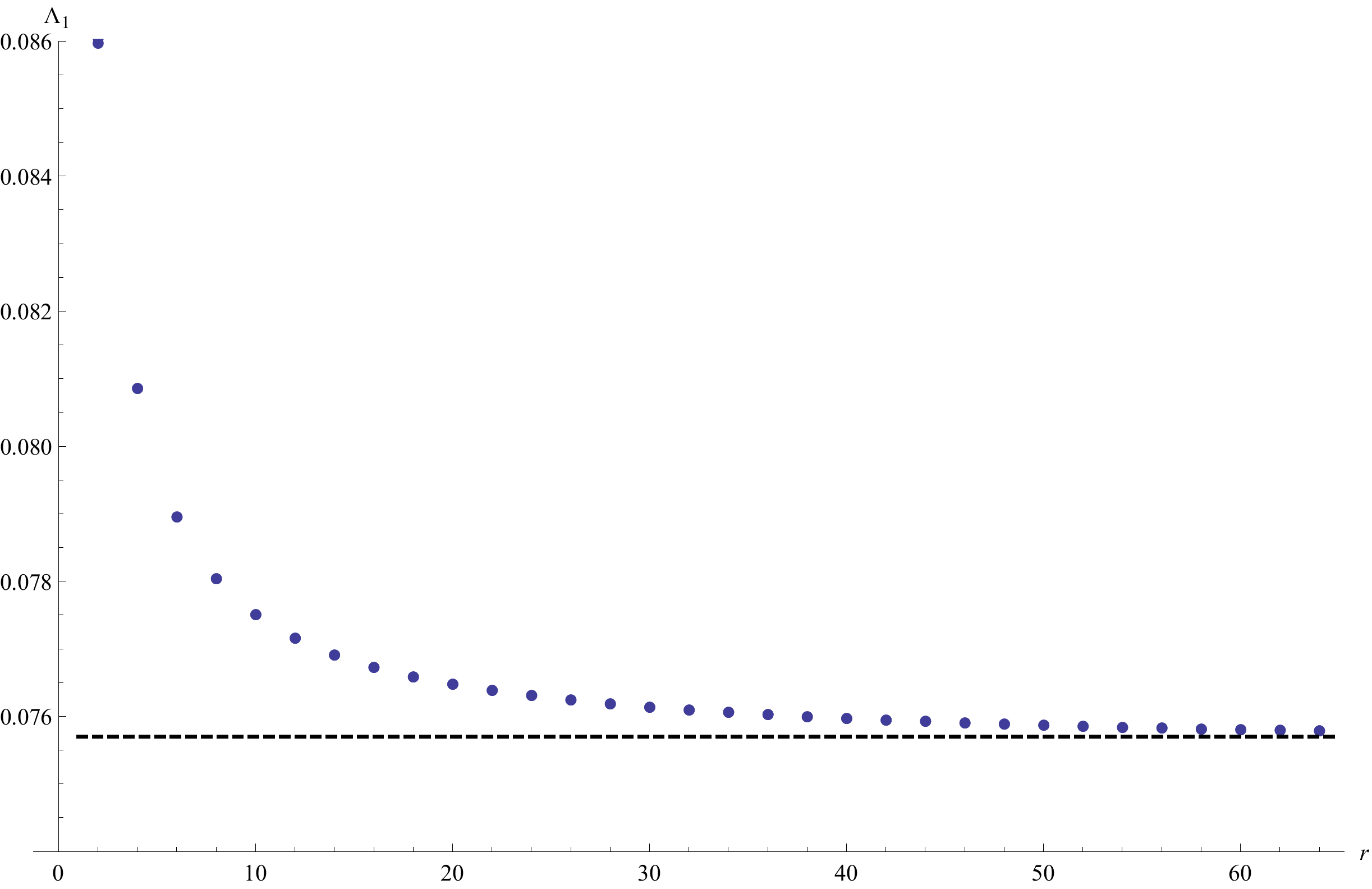}
}
\subcaption{Real part of $\Lambda_1$.}\label{fig:8a}

\end{minipage}
\begin{minipage}{0.5\linewidth} 
\centering
\scalebox{0.5}{ 
\includegraphics[scale=0.5]{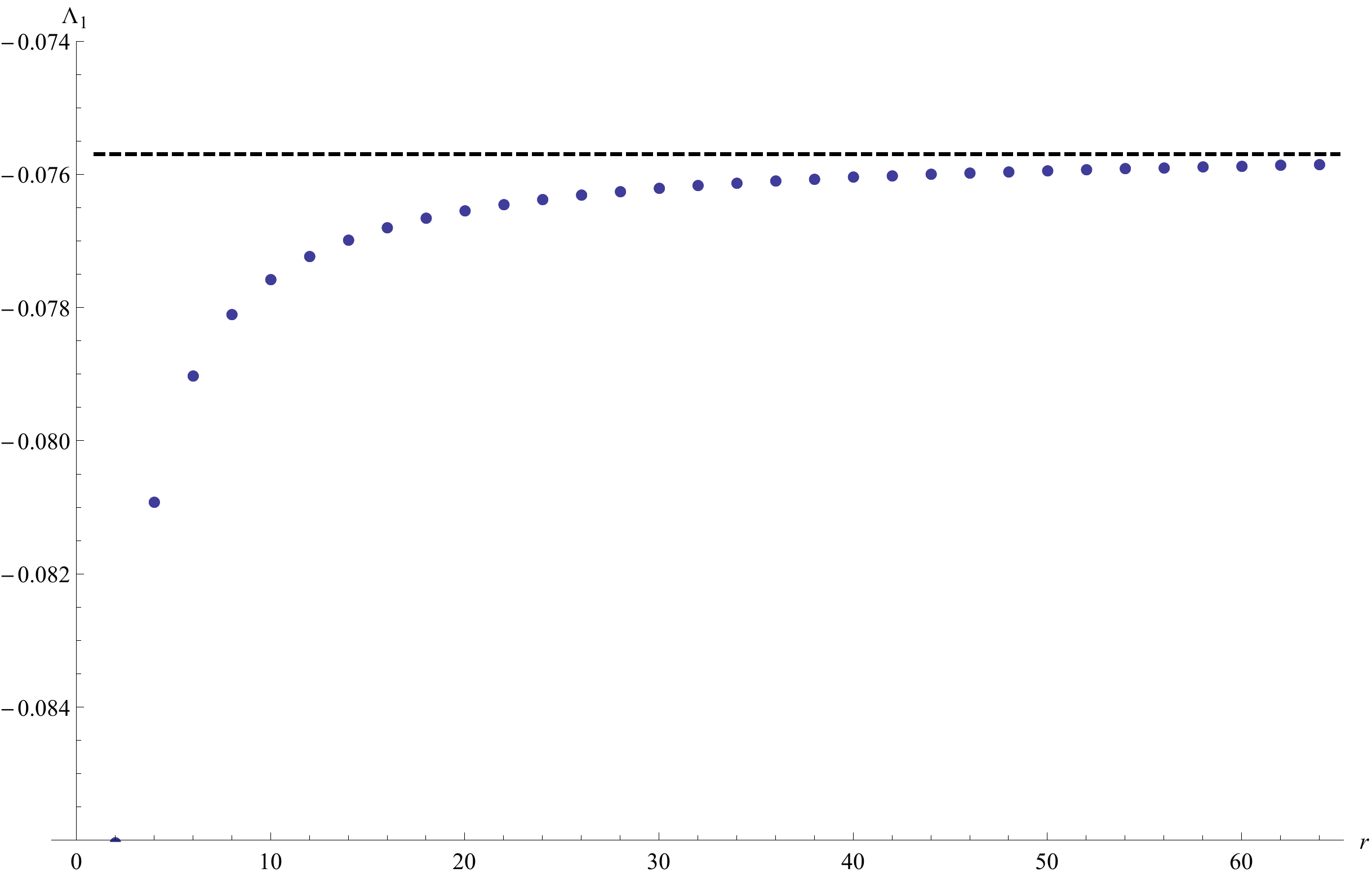}
}
\subcaption{Imaginary part of $\Lambda_1$.}\label{fig:8b}
\end{minipage}
\caption{This figure illustrates the approximation for $\Lambda_1$ with $\alpha=-2,\beta=1$ and $\gamma=1$. We see that as $r$ increases, the approximation for $\Lambda_1$ tends to the limiting value described by the black, dashed curve. The approximation for $\Lambda_2$ may be obtained from this information since $\Lambda_1=-i\,\Lambda_2$.}
\end{figure} 

\section{Stokes smoothing}\label{S:Appendix Stokes smoothing}
In order to apply the exponential asymptotic methods, we need to optimally truncate the asymptotic series \eqref{asymptotic series}. One particular way to calculate the optimal truncation point is to consider where the terms in the asymptotic series is at its smallest \cite{Boyd1999}. This heuristic is equivalent to the finding $N$ such that   
\begin{equation*}
\bigg|\frac{\epsilon^{N+1}f_{N+1}}{\epsilon^N f_N}\bigg|\sim 1,
\end{equation*}
in the limit $\epsilon\rightarrow 0$ and $N\rightarrow\infty$ (we will see that the limit $\epsilon\rightarrow 0$ is equivalent to the limit $N\rightarrow\infty$). By using the late-order form ansatz described by \eqref{late-order terms} we find that $N\sim |\chi|/\epsilon$. As this quantity may not necessarily be integer valued, we therefore choose $\omega\in[0,1)$ such that 
\begin{equation}\label{optimal truncation point}
N=\frac{|\chi|}{\epsilon}+\omega
\end{equation}
is integer valued.

We substitute the optimally-truncated series with \eqref{optimal truncation point} to the governing equation \eqref{taylorexpanded rescaled dPII}, and use the recurrence relations \eqref{reccurence for coeffs} to eliminate terms. Doing so, we obtain the equation
\begin{align}\label{governing equation for opt error}
\sum_{j=1}^{\infty}\frac{2\epsilon^{2j}R_N^{(2j)}}{(2j)!}&-\epsilon^2\sum_{r=0}^{N-1}\epsilon^r\sum_{j=0}^{\infty}\frac{2\epsilon^{2j}f_r^{(2j)}}{(2j)!}\bigg(2R_N\sum_{k=0}^{N-1}\epsilon^kf_k+R_N^2\bigg) \nonumber \\ 
&-\epsilon^2\sum_{j=0}^{\infty}\frac{2\epsilon^{2j}R_N^{(2j)}}{(2j)!}\Bigg(\bigg(\sum_{r=0}^{N-1}\epsilon^rf_r\bigg)^2-2R_N\sum_{r=0}^{N-1}\epsilon^rf_r+R_N^2\Bigg) \nonumber \\
&+\epsilon^Nf_N +\ldots \sim (\alpha s+\beta -2)R_N,
\end{align}
where the omitted terms are smaller than those which have been retained in the limit $\epsilon\rightarrow 0$.

Away from the Stokes curve, the inhomogeneous terms of equation \eqref{governing equation for opt error} is negligible, and we therefore apply a WKB analysis to the homogeneous version of \eqref{governing equation for opt error}. We therefore apply the ansatz $R_N=a(s)e^{b(s)/\epsilon}$ and match orders of $\epsilon$ as $\epsilon\rightarrow 0$. The leading order equations as $\epsilon\rightarrow 0$ can be shown to be
\begin{equation*}
\sum_{j=0}^{\infty}\frac{2\epsilon^{2j}}{(2j)!}\bigg(\frac{b'(s)}{\epsilon}\bigg)^{2j}a(s)e^{b(s)/\epsilon}=\big(\alpha s+\beta \big)a(s)e^{b(s)/\epsilon}.
\end{equation*}
Comparing this to equation \eqref{singulant equation}, we see that they coincide provided that $b(s)=\chi(s)$, where $\chi$ is the particular singulant being considered, namely $\chi_1$ or $\chi_2$. For now, we will work with general $\chi$ and specify the choice of $\chi$ in the subsequent analysis. Continuing to the next order in $\epsilon$ we find that $a(s)$ satisfies equation \eqref{prefactor equation} exactly, and hence $a(s)=F(s)$. Hence, away from the Stokes curve, the optimally-truncated error takes the form $R_N(s)\sim F(s)e^{-\chi/\epsilon}$ as $\epsilon\rightarrow 0$.

As the exponentially-small error term will experience Stokes switching, we therefore set 
\begin{equation*}
R_N(s)=\mathcal{S}(s)F(s)e^{-\chi(s)/\epsilon},
\end{equation*}
where $\mathcal{S}(s)$ is the Stokes multiplier that switches rapidly in the neighbourhood of the Stokes curve. We apply this form to equation \eqref{governing equation for opt error} and after some cancellation we find that 
\begin{equation*}
2\epsilon \mathcal{S}'Fe^{-\chi/\epsilon}\sum_{j=1}^{\infty}\frac{(2j)(-\chi')^{2j-1}}{(2j)!}=\epsilon^Nf_N
\end{equation*}
as $N\rightarrow \infty$. Rearranging this equation and applying the form of the late-order for $f_N$ as given by \eqref{late-order terms} we find that
\begin{equation}\label{Stokes multiplier differential equation}
\frac{d\mathcal{S}}{ds}\sim \epsilon^{N-1}e^{\chi/\epsilon}\frac{\Gamma(N+k)}{2\sinh(\chi')\chi^{N+k}}.
\end{equation}
Noting the form of $N$, we introduce polar coordinates by setting $\chi=\rho e^{i\theta}$ where the fast variable is $\theta$ and the slow variable, $\rho$. This transformation tells us that
\begin{equation*}
\frac{d}{ds}=\frac{-i\chi'e^{-i\theta}}{\rho}\frac{d}{d\theta}
\end{equation*}
and \eqref{optimal truncation point} becomes $N=\rho/\epsilon+\omega$. Hence \eqref{Stokes multiplier differential equation} becomes
\begin{equation}\label{Stokes multiplier differential equation transformed}
\frac{d\mathcal{S}}{d\theta}\sim \rho e^{i\theta}\epsilon^{\rho /\epsilon+\omega-1}\frac{-\Gamma(\rho /\epsilon+\omega+k)}{2i\chi'\sinh(\chi')(\rho e^{i\theta})^{\rho /\epsilon+\omega+k}}\exp\bigg(\frac{\rho e^{i\theta}}{\epsilon}\bigg).
\end{equation}
The expression $\chi'\sinh(\chi')$ is a function of $s$ which we will denote by $H(s)$. Furthermore, as we have applied the transformation $\chi=\rho e^{i\theta}$, $H(s)$ is effectively a function of $\theta$, $H(s(\theta);\rho )$, where $\rho $ is a fixed parameter. Recalling that $k=1/2$, we apply Stirling's formula \cite{Abram2012} to \eqref{Stokes multiplier differential equation transformed} and after simplification we obtain
\begin{equation}\label{Stokes multiplier differential equation transformed STIRLINGS applied}
\frac{d\mathcal{S}}{d\theta}\sim \frac{i\sqrt{2\pi}}{2H(s(\theta))}\frac{\sqrt{\rho }}{\epsilon^{k+1/2}}\exp\bigg(\frac{\rho }{\epsilon}(e^{i\theta}-1-i\theta)-i\theta(\omega+k-1)\bigg).
\end{equation}
The right hand side is exponentially-small except in the neighbourhood of $\theta=0$, which is exactly where the Stokes curve lies (where $\chi$ is purely real and positive). We now rescale about the neighbourhood of the Stokes curve in order to study the behaviour of $\mathcal{S}$. Applying the scaling $\theta=\sqrt{\epsilon}\hat{\theta}$ to \eqref{Stokes multiplier differential equation transformed STIRLINGS applied} gives us
\begin{equation}\label{rescaling stokes multipler DE}
\frac{1}{\sqrt{\epsilon}}\frac{d\mathcal{S}}{d\hat{\theta}}\sim \frac{i\sqrt{2\pi}}{2H(|\chi| )}\frac{\sqrt{|\chi| }}{\epsilon^{k+1/2}}\exp\bigg(-\frac{|\chi| \hat{\theta}^2}{2}\bigg).
\end{equation}
We note that to leading order in $\epsilon$, $H(s(\theta);\rho )$ will only depend on $\rho=|\chi|$ near the Stokes curve. Integrating \eqref{rescaling stokes multipler DE} we find that
\begin{align}\label{Stokes multiplier calculated}
\mathcal{S}\sim& \frac{i\sqrt{2\pi}}{2H(|\chi| )}\frac{\sqrt{|\chi| }}{\epsilon^{k}}\bigg[\frac{1}{\sqrt{|\chi| }}\int_{-\infty}^{\hat{\theta}/\sqrt{|\chi| }}e^{-s^2/2}ds+C\bigg], \nonumber \\ 
=&\frac{i\pi}{2\sqrt{\epsilon}H(|\chi| )}\bigg[C+\text{erf}\bigg(\sqrt{\frac{\theta}{2\epsilon |\chi|}}\bigg)\bigg], \nonumber
\end{align}
where $C$ is an arbitrary constant. Thus, as Stokes curves are crossed, the Stokes multiplier changes in value by 
\begin{equation*}\label{Stokes JUMP}
\big[\mathcal{S}\big]^+_-\sim \frac{i\pi}{\sqrt{\epsilon}H(|\chi| )},
\end{equation*}
and hence the exponential contribution, $R_N$, which experiences Stokes switching, changes by
\begin{equation*}\label{RnVALUEJUMP}
\big[R_N\big]^+_-\sim \frac{i\pi}{\sqrt{\epsilon}H(|\chi| )}F(s)e^{-\chi/\epsilon},
\end{equation*}
as Stokes curves are crossed.

\end{appendices}

\bibliographystyle{plain}
\bibliography{StokesdPIIbibliography}

\end{document}